\documentclass[aip,pop,reprint]{revtex4-2}

\usepackage{bm}
\usepackage{graphicx}
\usepackage{mathtools}
\usepackage{mathtools}
\usepackage{longtable}

\DeclarePairedDelimiterX\braket[2]{\langle}{\rangle}{#1\,\delimsize\vert\,\mathopen{}#2}

\usepackage[final]{changes}

\newcommand{\bea}{\begin{eqnarray}}
	\newcommand{\eea}{\end{eqnarray}}

\newcommand{\tsf}[1]{\textsf{#1}}
\newcommand{\trm}[1]{\textrm{#1}}

\newcommand{\mbf}[1]{\mathbf{#1}}

\newcommand{\figref}[1]{Fig. \ref{#1}}

\newcommand{\eqnref}[1]{Eq. (\ref{#1})}

\newcommand{\be}{\begin{equation}}
	\newcommand{\ee}{\end{equation}}
\newcommand{\bi}{\begin{itemize}}
	\newcommand{\ei}{\end{itemize}}

\newcommand{\mbi}{M_{\tsf{BI}}}
\newcommand{\cbi}{c_{\tsf{BI}}}

\begin{document}


\title{On Measuring Stimulated Photon-photon Scattering using Multiple Ultraintense Lasers}


\author{Hans G. Rinderknecht}
\email{hrin@lle.rochester.edu}
\affiliation{Laboratory for Laser Energetics, University of Rochester, Rochester, New York 14623, USA}

\author{E. Dill}
\affiliation{University of Rochester, Rochester, NY 14620}
\author{A. J.~MacLeod}
\affiliation{ELI Beamlines Facility, The Extreme Light Infrastructure ERIC, Za Radnic\'{i} 835, 25241 Doln\'{i} B\v{r}e\v{z}any, Czech Republic}
\author{B. King}
\affiliation{Centre for Mathematical Sciences, Plymouth University, Plymouth, PL4 8AA, United 
Kingdom}
\author{K. Sow}
\affiliation{University of Rochester, Rochester, NY 14620}

\author{S.-W. Bahk}
\affiliation{Laboratory for Laser Energetics, University of Rochester, Rochester, New York 14623, USA}
\author{I. A. Begishev}
\affiliation{Laboratory for Laser Energetics, University of Rochester, Rochester, New York 14623, USA}
\author{F. Karbstein}
\affiliation{Helmholtz-Institut Jena, Fr\"obelstieg 3, 07743 Jena, Germany}
\affiliation{GSI Helmholtzzentrum f\"ur Schwerionenforschung, Planckstrasse 1, 64291 Darmstadt, Germany}
\affiliation{Faculty of Physics and Astronomy, Friedrich-Schiller-Universit\"at Jena, 07743 Jena, Germany}
\author{J\"org Schreiber}
\affiliation{Ludwig-Maximilians-Universitat Munich, D-80539 Munich, Germany}
\author{M. Zepf}
\affiliation{Helmholtz-Institut Jena, Fr\"obelstieg 3, 07743 Jena, Germany}
\affiliation{GSI Helmholtzzentrum f\"ur Schwerionenforschung, Planckstrasse 1, 64291 Darmstadt, Germany}
\affiliation{Faculty of Physics and Astronomy, Friedrich-Schiller-Universit\"at Jena, 07743 Jena, Germany}

\author{A. Di Piazza}
\affiliation{Department of Physics and Astronomy, University of Rochester, Rochester, New York 14627, USA}
\affiliation{Laboratory for Laser Energetics, University of Rochester, Rochester, New York 14623, USA}

\date{}


\date{\today, work in progress}

\begin{abstract}
	Stimulated photon-photon scattering is a predicted consequence of quantum electrodynamics that has yet to be measured directly.
	Measuring the cross-section for stimulated photon-photon scattering is the aim of a flagship experiment for NSF OPAL, a proposed laser user facility with two, 25-PW beamlines.
	We present optimized experimental designs for achieving this challenging and canonical measurement.
	A family of experimental geometries is identified that satisfies the momentum- and energy-matching conditions for two selected laser frequency options.
	Numerical models predict a maximum signal exceeding 1000 scattered photons per shot at the experimental conditions envisaged at NSF OPAL.
	Experimental requirements on collision geometry, polarization, cotiming and copointing, background suppression, and diagnostic technologies are investigated numerically.
	These results confirm that a beam cotiming shorter than the pulse duration and control of the copointing on a scale smaller than the shortest laser wavelength are needed to robustly scatter photons on a per-shot basis.
    Finally, we assess the bounds that a successful execution of this experiment may place on the mass scale of Born-Infeld nonlinear electrodynamics beyond the Standard Model of physics.
\end{abstract}


\maketitle


\section{Introduction}
\label{sec:intro}
Quantum electrodynamics (QED) predicts that electromagnetic (EM) fields interact in vacuum, with the interaction mediated by virtual pairs of charged particles and antiparticles. 
This so-called `vacuum nonlinearity' is a purely quantum effect:  the classical Maxwell's equations in vacuum are strictly linear. 
The idea that the existence of particle/antiparticle fields gives rise to nonlinear effects in the propagation of EM fields in vacuum was developed in Refs.~\cite{Heisenberg_1936,Weisskopf_1936}, where the  Lagrangian density of a slowly-varying EM field was determined including the quantum effects of the electron-positron “vacuum fluctuations.” 
This is the well-studied Euler-Heisenberg Lagrangian density, which was re-computed later by Schwinger using quantum electrodynamics in Ref.\cite{Schwinger_1951}. 

The space (time) scale characterizing the rapidity of variation of the electromagnetic field is determined by the reduced Compton wavelength (Compton time) $\lambda_C=\hbar/mc\approx 3.9\times 10^{-11}\;\text{cm}$ ($\lambda_C/c=\hbar/mc^2\approx1.3\times 10^{-21}\;\text{s}$) \cite{Heisenberg_1936,Weisskopf_1936,Schwinger_1951}, with $m$ indicating the electron mass. 
The Euler-Heisenberg Lagrangian density was computed assuming uniform and constant fields, i.e., it does not contain  spacetime derivatives of the EM fields and depends only  
on the two Lorentz invariants \cite{Jackson_b_1975}:  $\mathcal{F}=-2(E^2-B^2)$ and $\mathcal{G}=-4\bm{E}\cdot\bm{B}$. 
The fact that the Euler-Heisenberg Lagrangian density depends nonlinearly on $\mathcal{F}$ and $\mathcal{G}$ implies that the resulting equations of motion of the electromagnetic field $(\bm{E},\bm{B})$ are also nonlinear\cite{Heisenberg_1936,Weisskopf_1936,Schwinger_1951}.\footnote{The leading-order derivative corrections to the Euler-Heisenberg Lagrangian density have been computed in Ref. \cite{JMP:Gusynin:1999}
} 

The importance of these nonlinear terms is determined by the strength of the electromagnetic field as compared to the so-called critical electric and magnetic fields of QED \cite{Heisenberg_1936,Weisskopf_1936,Schwinger_1951}: $E_{cr}=m^2c^3/\hbar|e|\approx 1.3\times 10^{16}\;\text{V/cm}$ and $B_{cr}=m^2c^3/\hbar|e|\approx 4.4\times 10^{13}\;\text{G}$ in our units where the vacuum permittivity $\epsilon_0$ is set equal to unity and where $e<0$ denotes the electron charge.
The critical fields of QED exceed by several orders of magnitude the most intense electromagnetic fields produced in the laboratory by high-power lasers, with the current record being\cite{Yoon_2019} around
$1.1\times 10^{23}\;\text{W/cm$^2$}$ corresponding to an electric field amplitude of approximately $6.4\times 10^{12}\;\text{V/cm}$. 
Several multipetawatt facilities are under construction or planned \cite{APOLLON_10P,ELI,CoReLS,Bromage_2019,XCELS}, which are expected to overcome the present record intensity by a factor five or more (see also the report of the recent Multi-Petawatt Physics Prioritization (MP3) Workshop \cite{Di_Piazza_2022}), but remain orders of magnitude below the critical field strength. 
This explains why vacuum-polarization effects are typically very small and challenging to measure.

A wide variety of experiments have been proposed to observe various consequences of vacuum nonlinearity. These include: vacuum polarization effects and the related process of photon-photon scattering, the cross-section of which was computed in Refs.~\cite{Euler_1936_a,Akhiezer_1937,Karplus_1951,De_Tollis_1964} (see also Refs. \cite{Ahmadiniaz_2023,Ahmadiniaz_2023_b}); birefringence and dichroic effects in the propagation of an EM wave through a strong laser field \cite{Aleksandrov_1985,Heinzl_2006,Di_Piazza_2006,Tommasini_2010,King_2010_b,Homma_2011,Dinu_2014,Dinu_2014_b,King_2016,Bragin_2017,Gies_2018,Karbstein_2018,King_2018_b,Pegoraro_2019_b,Bulanov_2020,Ahmadiniaz_2020,Karbstein_2021,Ataman_2021,PRD:Mosman:2021,Ahmadiniaz_2021,Jin_2022,Sainte-Marie_2022,Karbstein_2022,Aleksandrov_2023,Macleod_2023,Formanek_2024_b}; harmonic generation and photon splitting in intense laser fields \cite{Di_Piazza_2005,Narozhny_2007,Brodin_2007,Di_Piazza_2007,Di_Piazza_2013,Gies_2014_b,Sundqvist_2023}; vacuum Bragg scattering and Cherenkov radiation \cite{Kryuchkyan_2011,Macleod_2019,Bulanov_2019,Jirka_2023}; and vacuum-polarization effects in plasmas \cite{Di_Piazza_2007_a,Pegoraro_2019,Zhang_2020,Bret_2021}. 
Photon-photon scattering and related experimental proposals were analyzed, among others, in Refs. \cite{Varfolomeev_1966,Lundstroem_2006,Lundin_2007,King_2010,Jeong_2020,Karbstein_2021_c,Dumlu_2022,Berezin_2024}. 
The above list of works is not exhaustive; we refer the reader to the reviews \cite{Di_Piazza_2012,Dunne_2014,King_2016_c,Karbstein_2020,Gonoskov_2022,Fedotov_2023} for a more complete list of proposals.
Recently, it has been claimed that vacuum birefringence was observed in the presence of the strong magnetic field surrounding a neutron star;\cite{MNRAS:Mignani:2017} however, those conclusions have been criticized.\cite{EPJ:Capparelli:2017}
Scattering of GeV photons by a virtual photon field (Delbr{\"u}ck scattering) has been measured,\cite{PRD:Jarlskog:1973,Akhmadaliev:1998zz} and evidence for scattering of two GeV-scale virtual photons has also been reported in  relativistic heavy-ion collisions.\cite{NP:Aaboud:2017,CMS:2018erd,ATLAS:2019azn}.
However, the direct measurement of the predicted vacuum nonlinearity effects in experiments with sufficient statistics to constrain the theory remains to be done.

The lowest-order interaction between two photons requires a closed fermion loop with four vertices, making it highly suppressed with respect to, e.g., electron-photon scattering processes. The cross-section for scattering of two photons with energy $\hbar\omega$ in their center-of-momentum frame is calculated to be $\sigma_{\gamma\gamma} = \left[7.265\times10^{-66}~\textrm{cm}^2\right] \left(\hbar\omega/\textrm{eV}\right)^6$.\cite{Landau_b_4_1982}
The highest power laser system as of this writing \cite{HPLSE:Radier:2022} produces on the order of $10^{21}$ photons: if concentrating two such beams into a diffraction-limited f/\# 1 focus, the probability of photon-photon scattering remains negligibly low ($N^2\sigma/\pi R^2 \approx 2\times10^{-8}$).\cite{NJP:King:2012}  While upper-bound results exist in the literature,\cite{EPJD:Bernard:2000} no attempts to measure direct photon-photon scattering to date have recorded a significant signal.

In this manuscript, we study the experimental requirements to measure real photon-photon scattering for the first time, using the stimulated photon-photon scattering (SPPS) concept at a proposed 2$\times$ 25-PW laser facility, NSF OPAL.
 In this design, which was first proposed in Ref.\cite{Varfolomeev_1966}, three laser beams collide, one of which acts as a ``stimulating'' beam along which one of the two scattered photons is emitted. 
The SPPS process is analogous to non-linear 4-wave mixing in the quantum vacuum, and has two advantages for measuring photon-photon scattering: first, that the scattered photon signal propagates in a known direction that is distinct from the incident lasers; and second, that the presense of the stimulating beam significantly increases the cross-section for scattering, which in this design scales\cite{Heinzl:2024cia} as $\sigma \propto \omega^4$.
Prior theoretical work has investigated 3-beam stimulated photon-photon scattering, showing that potentially viable solutions exist at the level of a three, 10~PW beam capability.\cite{PRL:Lundstrom:2006,PRA:Lundin:2006,PRA:King:2018,PRD:Gies:2018b}

The National Science Foundation is currently funding the design of NSF OPAL, a laser user facility that will deliver two, 25~PW beams \added{with central wavelength of 920~nm and bandwidth in the range 830~nm to 1010~nm}.
If built, NSF OPAL will be the first facility that can potentially measure stimulated photon-photon scattering at optical frequencies with more than one scattered photon per shot.
Stimulated photon-photon scattering has been identified as a potential flagship experiment for the facility, motivating a rigorous examination of the expected signal and experimental requirements.
To achieve the SPPS experiment described above, we propose splitting and (optionally) frequency-doubling one of the two, 25~PW beams (Alpha-1) to provide the two scattering beams, and collide them with the second 25~PW beam (Alpha-2) as the stimulating beam.
We predict that over 1000 scattered photons can be generated per shot: high enough to avoid reliance on statistical methods to interpret the result and to permit a detailed study of the SPPS interaction over a range of parameters.
If successful, this experiment will provide a direct measurement of nonlinear effects in the quantum vacuum. 

We examine the stimulated photon-photon scattering experiment based on the expected performance of the NSF OPAL facility, and present a novel, optimized scattering geometry to maximize the scattering signal.
The theory underlying stimulated photon scattering is discussed in Section~\ref{sec:theory}.
Section~\ref{sec:models} describes the numerical models used to assess a more realistic photon scattering signal, taking into account beam focusing and pulse shaping, and presents results of these models.
Section~\ref{sec:design} describes an optimized experimental design for the stimulated photon-photon scattering experiments on NSF OPAL, discusses requirements to achieve success, and how those requirements might be met.
Finally, Section~\ref{sec:born-infeld} assesses the bounds that the proposed experiment can put on Born-Infeld nonlinear electrodynamics i.e. contributing to beyond-the-Standard-Model (BSM) physics.


\section{Theory}
\label{sec:theory}
Scattering of two real photons must simultaneously conserve momentum and energy:

\begin{eqnarray}
	\bm{k}_1 + \bm{k}_2 = \bm{k}_3 + \bm{k}_4 \\  \label{eq:k}
	\omega_1 + \omega_2 = \omega_3 + \omega_4   \label{eq:omega}
\end{eqnarray}
\noindent where $\bm{k}$, $\omega$ are the wave-vector and frequency of the initial (1,2) and scattered photons (3, 4), respectively.
For stimulated scattering, the frequencies and wave vectors of the three input beams (1, 2, 3) enforce a particular solution for the scattered photon (4).
(Throughout this work we will use normalized units in which $c = \hbar = 1$, unless otherwise stated.)

In the particular case in which the three input beams share the same frequency ($\omega_1 = \omega_2 = \omega_3 = \omega$), the scattered photon also has the same frequency.
Equation~\ref{eq:k} then admits a family of solutions:

\begin{equation}
	\begin{array}{lclcccl}
	\bm{k}_1 &=& \omega\left[\right.&\cos\phi,&\sin\phi,&0&\left.\right], \\
	\bm{k}_2 &=& \omega\left[\right.&\cos\phi,&-\sin\phi,&0&\left.\right], \\
	\bm{k}_3 &=& \omega\left[\right.&\cos\phi,&0,&\sin\phi&\left.\right], \\
	\bm{k}_4 &=& \omega\left[\right.&\cos\phi,&0,&-\sin\phi&\left.\right], \\	
	\end{array}
	\label{eq:1omega_family}
\end{equation}
\noindent where $\phi$ is the half-angle between beams 1 and 2, and without loss of generality we have made beams (1,2) symmetric with respect to the $(y,z)$ plane.

\begin{figure}
	\includegraphics[width=0.8\columnwidth]{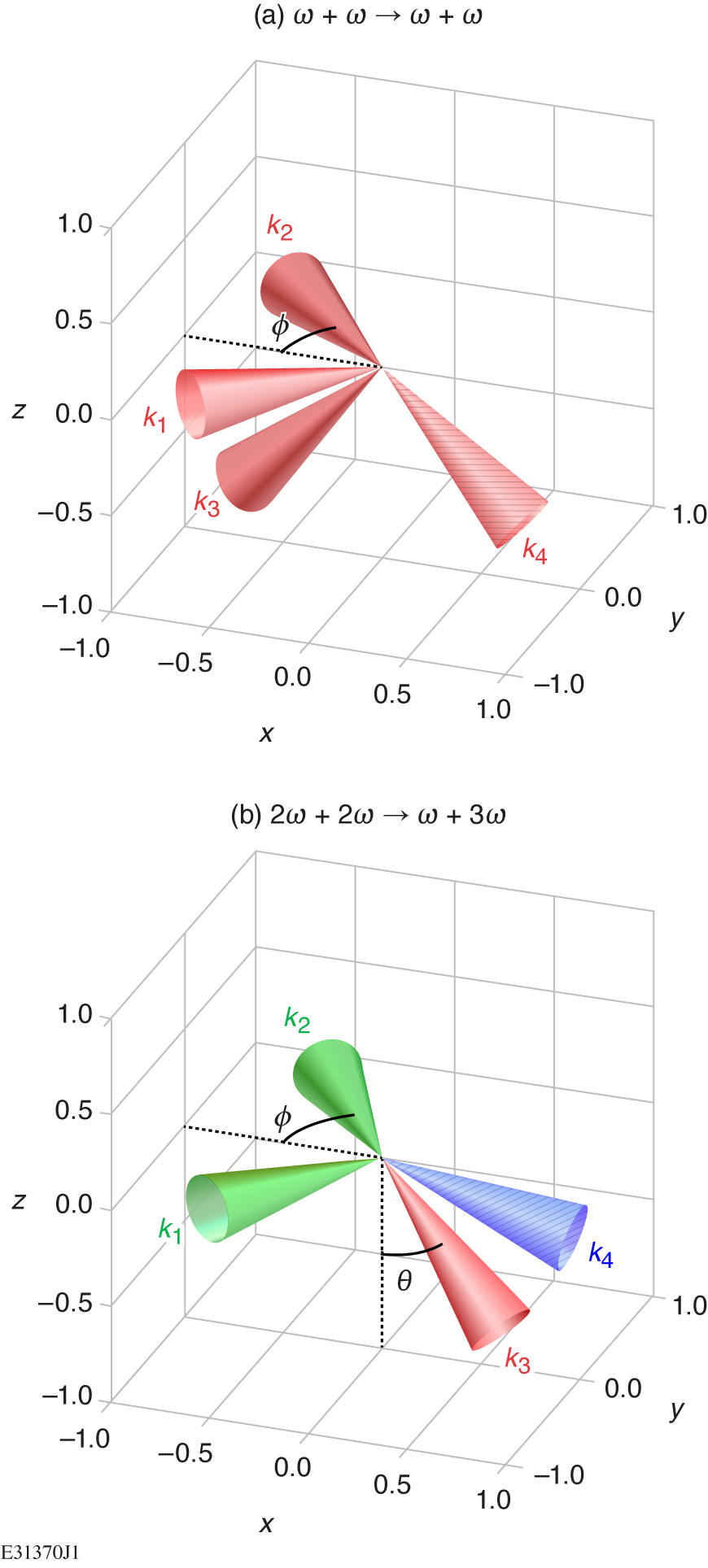}
	\caption{Cartoon of scattering geometry for (a) one-color scattering (Eq.~\ref{eq:1omega_family}); (b) three-color scattering (Eq.~\ref{eq:2omega_family}).  Incident lasers (beams 1--3) are shown focusing to the interaction point, whereas scattered photons (beam 4, dashed) are shown exiting the interaction.\label{fig:layouts}}
\end{figure}

Prior works have focused on cases in which the scattered photon has a distinct frequency from the incident beams, because this can increase detectability of the signal.
A common strategy\cite{PRL:Lundstrom:2006,PRA:Lundin:2006,PRA:King:2018,PRD:Gies:2018b} is to use $\omega_1 = \omega_2 = 2\omega$,  $\omega_3 = \omega$, for which Eq.~\ref{eq:omega} requires $\omega_4 = 3\omega$.
Such a scheme can be produced in the lab from a single laser source by frequency doubling beams 1 and 2.
With these frequency choices, Eq.~\ref{eq:k} admits the following family of solutions:

\begin{equation}
	\begin{array}{lcrcccl}
		\bm{k}_1 &=& 2\omega\left[\right.&\cos\phi,&\sin\phi,&0&\left.\right], \\
		\bm{k}_2 &=& 2\omega\left[\right.&\cos\phi,&-\sin\phi,&0&\left.\right], \\
		\bm{k}_3 &=& \omega\left[\right.&\sin\theta,&0,&\cos\theta&\left.\right], \\
		\bm{k}_4 &=& 3\omega\left[\right.&\frac{4\cos\phi-\sin\theta}{3},&0,&-\frac{\cos\theta}{3}&\left.\right], \\	
	\end{array}
	\label{eq:2omega_family}
\end{equation}
\noindent where additionally $\theta$ is the angle between beam 3 and the z-axis, and is a unique function of $\phi$ satisfying:

\begin{equation}
	\theta = \arcsin\left(2\cos\phi - \frac{1}{\cos\phi}\right).
	\label{eq:2w_theta}
\end{equation}
\noindent The particular solution studied in prior works is equivalent to e.g. $\mbf{k}_{1}=2\omega\,\mbf{e}_{x}$, $\mbf{k}_{2}=2\omega\,\mbf{e}_{y}$ and $\mbf{k}_{3}=\omega\,\mbf{e}_{z}$ where $\mbf{e}_{j}$ indicates a unit vector parallel to the $j$-axis. Figure~\ref{fig:layouts} shows the geometry for these two families of equations.  

The center-of-momentum frequency $\omega_{cm}$ is calculated using the 4-vector dot product as: 

\begin{equation}
	\omega_{cm}^2 = \frac{\omega_1\omega_2}{2}\left(1 - \hat{\bm{k}}_1\cdot\hat{\bm{k}}_2\right) =  \omega_1\omega_2\left(\sin\phi\right)^2
    \label{eq:cm_frequency}
\end{equation}
\noindent where $\hat{\bm{k}}$ indicates the direction of the vector $\bm{k}$ only.
In general, the scattering cross-section will be maximized with maximally-opposing beams.
For the one-color family of solutions (Eq.~\ref{eq:1omega_family}), fully counter-propagating collisions are admissible ($\phi = 90^\circ$), whereas for the three-color family of solutions (Eq.~\ref{eq:2omega_family}), impact angle is maximized at $\phi = 60^\circ, \theta = -90^\circ$.
In these maximal solutions, all three beams lie in a plane, and the scattered photon travels directly opposite beam 3, which complicates detection.
However, most of the benefit of increased scattering probability is attained with near-planar solutions.
For the three-color solution, assuming the final laser focusing subtends an opening angle of less than $28^\circ$ ($f/\#\geq2$), the cones of beams 3 and 4 will not overlap for values of $\phi < 57.1^\circ$ ($\theta > -49^\circ$).
At this limit, the cross-section scaling with $\omega_{cm}$ predicts an increase in scattering by $2.8\times$ compared to the previously-considered $\phi = 45^\circ$ solution.


To more fully assess the performance of photon-photon scattering experiments from this family of solutions, a numerical model is required to account for realistic focusing, temporal pulse shaping, and polarization effects.  
This model is described in the following section.

\section{Numerical Modeling}
\label{sec:models}
A full calculation of the stimulated photon-photon scattering signal from intense focused laser fields is achieved by starting from the lowest-order Euler-Heisenberg Lagrangian density \cite{Euler_1935,Heisenberg_1936,Weisskopf_1936,Schwinger_1951}
\begin{equation}
\mathcal{L}=\frac{1}{2}(E^2-B^2)+\frac{2\alpha^2}{45m^4}[(E^2-B^2)^2+7(\bm{E}\cdot\bm{B})^2],
\end{equation}
where $\alpha=e^2/4\pi$ is the fine-structure constant. Following the method presented in Ref. \cite{Di_Piazza_2006} (see also Ref. \cite{PR:McKenna:1963}), one can derive the following inhomogeneous wave equation for the electric field $\bm{E}(x)$, with $x=(t,\bm{r})$:
\begin{equation}
\nabla^2\bm{E}-\partial_t^2\bm{E}=\bm{\nabla}\times(\partial_t\bm{M})+\partial_t^2\bm{P}-\bm{\nabla}(\bm{\nabla}\cdot\bm{P}),
\end{equation}
where
\begin{align}
\bm{P}(x)&=\frac{4\alpha^2}{45m^4}[2(E^2-B^2)\bm{E}+7(\bm{E}\cdot\bm{B})\bm{B}], \label{eq:P} \\
\bm{M}(x)&=-\frac{4\alpha^2}{45m^4}[2(E^2-B^2)\bm{B}-7(\bm{E}\cdot\bm{B})\bm{E}], \label{eq:M}
\end{align}
are the vacuum-induced polarization and magnetization vectors.

By writing the total electromagnetic field as the sum $(\bm{E}_0(x),\bm{B}_0(x))$ of the electromagnetic fields of the colliding beams and the vacuum-induced electromagnetic field $(\bm{E}_1(x),\bm{B}_1(x))$, one can easily demonstrate that the energy $d\mathcal{E}/d\omega d\Omega$ radiated per unit frequency $\omega$ and unit solid angle $\Omega$ can be written as
\begin{align}
	\frac{d\mathcal{E}}{d\omega d\Omega} &= \frac{\omega^4}{16\pi^3}\left|\bm{n}\times\left[\bm{M}(k) + \bm{n}\times\bm{P}(k)\right]\right|^2, \label{eq:diff_energy} \\ 
	\bm{P}(k) &= \int d^3x dt\, e^{i\omega\left(t-\bm{n}\cdot\bm{x}\right)} \bm{P}(x),  \label{eq:Pprime}\\ 
	\bm{M}(k) &= \int d^3x dt\, e^{i\omega\left(t-\bm{n}\cdot\bm{x}\right)}\bm{M}(x), \label{eq:Mprime} 
\end{align}
where $k=(\omega,\bm{k})=\omega(1,\bm{n})$ and where from now on it is understood that the polarization and magnetization vectors $\bm{P}(x)$ and $\bm{M}(x)$ are computed with the total electromagnetic field $(\bm{E}_0(x),\bm{B}_0(x))$ of the colliding beams.

To fully calculate the radiated signal requires integrating the interacting laser fields over all relevant positions and times (4 dimensions), as well as over all relevant radiation angles (2 dimensions). 
A range of frequencies must also be assessed, as the spectral content of the short laser pulses allows a range of scattered photon frequencies.
The laser fields are defined using the paraxial approximation for Gaussian beams, and are fully defined by the set of parameters: wavelength, incident angle, polarization angle, phase offset, focal radius, pulse duration, and peak intensity.
The convention used for polarization angle \added{$\xi_k$} of a beam propagating along  $\bm{k}$ is that the polarization direction is $\hat{\bm{E}}=\cos(\added{\xi_k})\hat{\bm{e}}_1+\sin(\added{\xi_k})\hat{\bm{e}}_2$, with
\begin{align}
\hat{\bm{e}}_1=&\frac{(\hat{\bm{z}}\times\bm{k})\times\bm{k}}{|(\hat{\bm{z}}\times\bm{k})\times\bm{k}|},\\
\hat{\bm{e}}_2=&\frac{\hat{\bm{z}}\times\bm{k}}{|\hat{\bm{z}}\times\bm{k}|}.
\end{align}
For each timestep and location, the total field produced by the three lasers is determined and the scattering amplitude calculated.
The norm of the total scattering amplitude for each scattering vector and frequency is then calculated to determine the differential scattered spectrum.
By integrating over the spectrum and solid angle of emission, a total number of scattered photons is determined.  

\added{
The integration was performed using MATLAB\cite{vendor:Mathworks:MATLAB} and processed in parallel on GPUs.
Typical simulations used a 3-dimensional cartesian spatial grid with step sizes of $0.1\times$ the minimum wavelength and range of $\pm2\times$ the maximum beam focal width; a time axis with step size of $0.1\times$ the minimum period and range of $\pm2\times$ the maximum pulse duration; a scattered photon frequency band of 101 points covering a range up to $\pm 8\%$ of the nominal scattered frequency; and a quiver of scattering k-vectors evenly covering up to $20^\circ$ from the nominal $\hat{k}_4$ in 10 rings.
The highest-resolution simulations calculated of the order of $10^{13}$ grid points, and required 38 minutes to complete on two GPUs. 
}

\added{
The total number of scattered photons can be assessed in a simplified integration.
However, that calculation includes many photons scattered along the three incident beams, and is thus not appropriate for assessing the experimental signal in the present configuration.
We include it for completeness in Appendix~\ref{app:total_photon_number}.
}

\begin{figure}
	\includegraphics[width=\columnwidth]{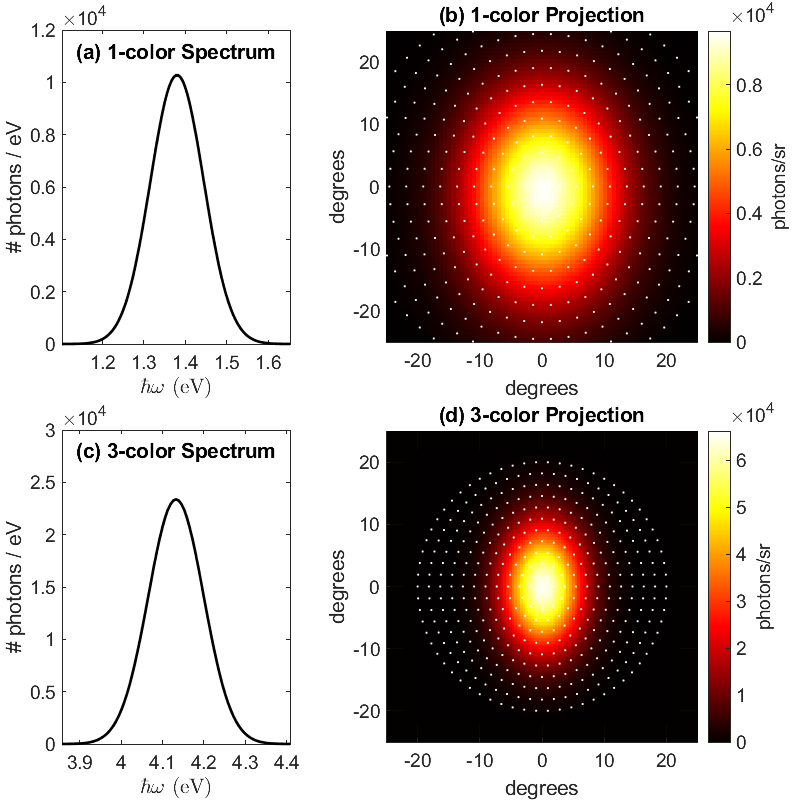}
	\caption{Simulation results for (a,b) one-color geometry ($\phi = 71.5^\circ$), (c,d) three-color planar geometry ($\phi = 57^\circ$).  (a,c) Scattered photon spectra; (b,d) scattered light profiles; white points indicate calculated scattering vectors.\label{fig:resultoptimal}}
\end{figure}

\subsection{Numerical integration results}

Simulation results for near-optimal \added{designs of the SPPS experiment that are experimentally realizable using the proposed NSF OPAL laser} are shown in Figure~\ref{fig:resultoptimal}.
We calculated scattering using a paraxial Gaussian beam model with f/2 focusing, 20~fs FWHM pulses, and optimal polarization choices as described in more detail in Sec.~\ref{sec:polarization}.
\added{The current optimistic performance estimate is that splitting a beam will reduce its peak power by $0.9\times$ (the beam must be masked to protect the splitting optic from damage).}
For the one-color solution, \added{$\phi=71.5^\circ$} and peak powers of \added{[11.25, 11.25, 25]} PW were assumed for beams [1, 2, 3], respectively, accounting for the loss of power due to beam splitting.  
A total signal of \added{2097} photons was calculated, with 90\% of the photons scattering within 19.4$^\circ$ of the beam-4 vector.
\added{Detailed simulation settings and results are presented in Table~\ref{table:resultoptimal_1w}.}

\begin{table}[tb]
	\begin{tabular}{|l|c|c|c|}
		\hline 
		Lasers & L1 & L2 & L3 \\
		\hline
		$\lambda$~($\mu$m) & 0.90 & 0.90 & 0.9 \\
		($\theta, \phi$) & (90$^\circ$, 71.5$^\circ$) & (90$^\circ$, -71.5$^\circ$) & (18.5$^\circ$, 0$^\circ$) \\
		Polarization & 315$^\circ$ & 135$^\circ$ & 45$^\circ$ \\
		Peak Power (PW) & 11.25 & 11.25 & 25 \\
		Pulse (fs, FWHM) & 20 & 20 & 20 \\
		Peak intensity (W/cm$^2$) & 4.85e23 & 4.85e23 & 1.21e24 \\
		beam FWHM ($\mu$m) & 1.349 & 1.349 & 1.349 \\
		Energy in beam (J) & 119.7  & 119.7 & 266.0 \\
		\hline
		\hline
		Domain & Range & Step & \# points \\
		\hline
		$[x, y, z]$ ($\mu$m) & $\pm$2.292 & 0.09 & 51$^3$ \\
		time (fs) & $\pm$ 24.02 & 0.30 & 161 \\
		$\omega_4$ & $\pm$ 20\% & 0.4\% & 101 \\
		$\bm{k}_4$ & 0$^\circ$--30$^\circ$ & 2.73$^\circ$ & 397 \\
		~  & ~ & TOTAL & 8.56e11 \\
		\hline
		\hline
		\multicolumn{2}{|l|}{Results} & \multicolumn{2}{|c|}{} \\
		\hline
		\multicolumn{2}{|l|}{Total scattered photons} & \multicolumn{2}{|c|}{2097} \\
		\multicolumn{2}{|l|}{Energy \& Bandwidth (eV)} & \multicolumn{2}{|c|}{1.383 $\pm$ 0.064} \\
		\multicolumn{2}{|l|}{Angle containing 50\% (90\%)} & \multicolumn{2}{|c|}{9.8$^\circ$ (19.4$^\circ$)} \\
		\hline		
	\end{tabular}
	\caption{Simulation parameters and results for the one-color point design.  Pulse duration, beam waist and energy in beam are numerically evaluated from the beam model.\label{table:resultoptimal_1w}}
\end{table}

For the three-color solution, \added{a scattering geometry of $\phi=57^\circ$ ($\theta=48.3^\circ$) was used.
The frequency conversion of beams 1 and 2 will reduce their peak power by an estimated $0.6\times$ (achieving this high value adds some technology risk\cite{HPLSE:Parker:2018}).
Therefore, peak powers of [6.75, 6.75, 25] PW were selected to account for losses from both frequency doubling and splitting.}
A total signal of \added{3163} photons was calculated, with 90\% of the photons scattering within 11.6$^\circ$ of the nominal direction.
\added{Detailed simulation settings and results are shown in Table~\ref{table:resultoptimal}.}
These signal levels are high enough to robustly measure stimulated photon-photon scattering in a single shot, and allow detailed characterization of scattering probability under various input conditions.  

The bandwidth in the scattered light signal is produced by the 20~fs pulse duration of the interaction, which introduces an uncertainty in the scattered frequency due to bandwidth of $h/(20~\textrm{fs})\approx 0.207$~eV.
This results in a larger fractional wavelength variation in the one-color scattering ($\lambda = 896.0\pm~41.5$~nm) as compared to the three-color scattering ($299.9\pm~4.8$~nm).
In both cases, the signal is scattered into roughly Gaussian ellipsoidal beams, which are elongated along the vertical axis, although the beam is more collimated in the three-color case.

The produced spectrum and beam angle will impact detector design.
Here we note that, to robustly collect the scattered photons, the incident beam 3 cone angle must not overlap with the detector collection angle.  
The f/2 focusing optics subtend a half-angle of 14$^\circ$. 
To collect 95\% of the scattered signal the projected photons must be centered at least 36.6$^\circ$ (27.6$^\circ$) from the center of the beam 3 axis, for the one-color (three-color) cases, respectively.
This in turn sets the maximal value for $\phi$ in both cases as 71.7$^\circ$ (57.2$^\circ$).
The simulations described here are thus close to the optimum achievable for this experimental concept.

\begin{table}[tb]
	\begin{tabular}{|l|c|c|c|}
		\hline 
		Lasers & L1 & L2 & L3 \\
		\hline
		$\lambda$~($\mu$m) & 0.45 & 0.45 & 0.9 \\
		($\theta, \phi$) & (90$^\circ$, 57$^\circ$) & (90$^\circ$, -57$^\circ$) & (48.31$^\circ$, 180$^\circ$) \\
		Polarization & 68.36$^\circ$ & 21.64$^\circ$ & 45$^\circ$ \\
		Peak Power (PW) & 6.75 & 6.75 & 25 \\
		Pulse (fs, FWHM) & 20 & 20 & 20 \\
		Peak intensity (W/cm$^2$) & 1.45e24 & 1.45e24 & 1.21e24 \\
		beam FWHM ($\mu$m) & 0.675 & 0.675 & 1.349 \\
		Energy in beam (J) & 71.7  & 71.7 & 266.0 \\
		\hline
		\hline
		Domain & Range & Step & \# points \\
		\hline
		$[x, y, z]$ ($\mu$m) & $\pm$2.292 & 0.045 & 101$^3$ \\
		time (fs) & $\pm$ 24.02 & 0.15 & 321 \\
		$\omega_4$ & $\pm$ 8\% & 0.16\% & 101 \\
		$\bm{k}_4$ & 0$^\circ$--20$^\circ$ & 1.82$^\circ$ & 397 \\
		~  & ~ & TOTAL & 1.33e13 \\
		\hline
		\hline
		\multicolumn{2}{|l|}{Results} & \multicolumn{2}{|c|}{} \\
		\hline
		\multicolumn{2}{|l|}{Total scattered photons} & \multicolumn{2}{|c|}{3163} \\
		\multicolumn{2}{|l|}{Energy \& Bandwidth (eV)} & \multicolumn{2}{|c|}{4.134 $\pm$ 0.067} \\
		\multicolumn{2}{|l|}{Angle containing 50\% (90\%)} & \multicolumn{2}{|c|}{5.7$^\circ$ (11.6$^\circ$)} \\
		\hline		
	\end{tabular}
	\caption{Simulation parameters and results for the three-color point design.  Pulse duration, beam waist and energy in beam are numerically evaluated from the beam model.\label{table:resultoptimal}}
\end{table}


Using the numerical model, we \added{proceed in the following sections to} assess the effects of various input parameters on the scattering performance.  

\subsection{Polarization}
\label{sec:polarization} 
\begin{figure}
	\includegraphics[width=\columnwidth]{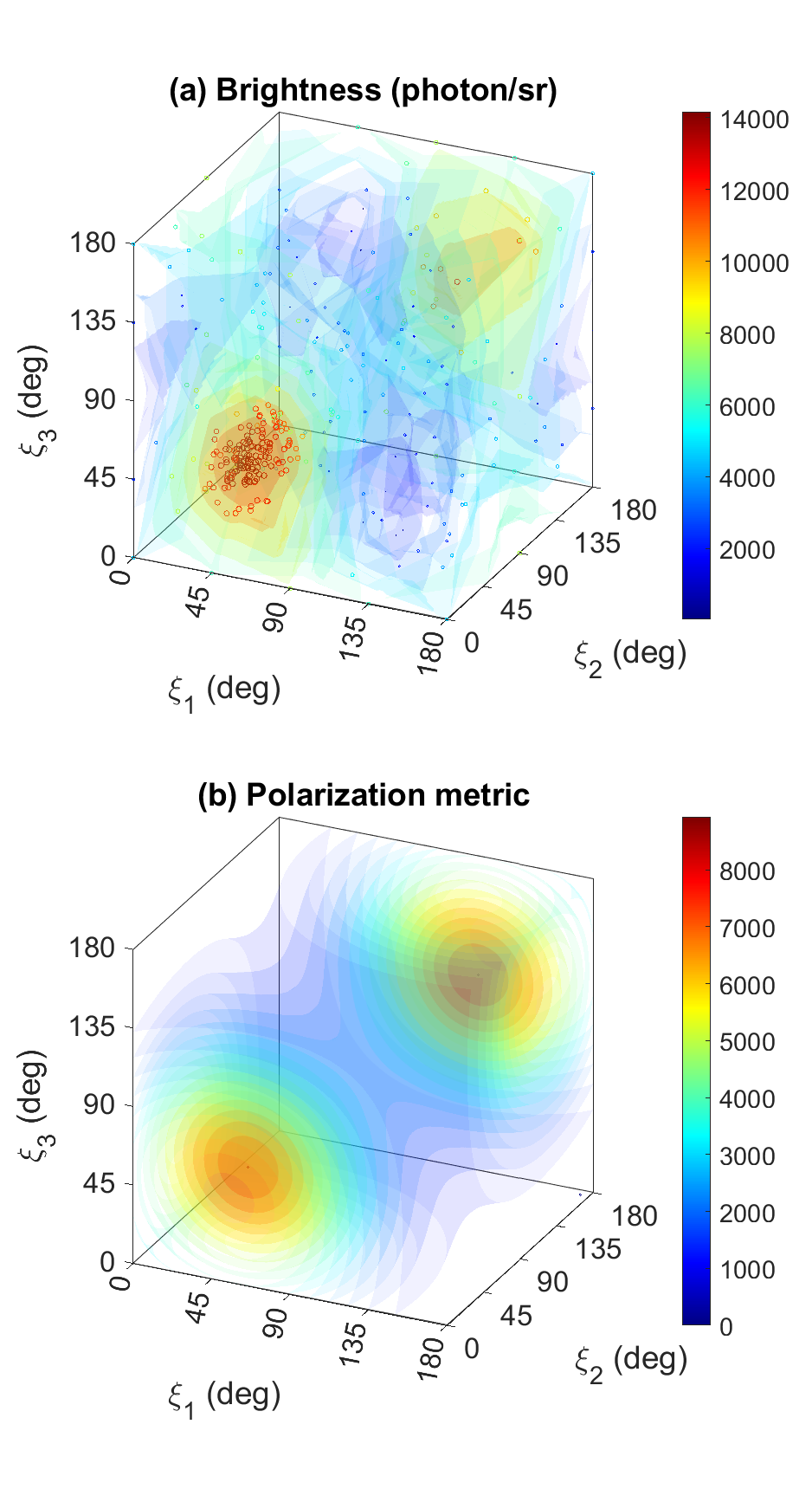}
	\caption{Effect of polarization in the planar three-color geometry. (a) Brightness of scattered photons calculated along $\bm{k}_4$ for 318 simulations (points); interpolation over points (surfaces). (b) Evaluation of polarization metric for this case, Eq.~\ref{eq:polmetric}.\label{fig:poltest}}
\end{figure}

Numerical studies demonstrated that the scattered photon \added{signal} depends on the relative polarization of all three incident beams.
\added{In this study, we calculate the photon brightness (in units of photons/sr) along the central vector $\bm{k}_4$ as a proxy for total scattering.
(This is done in order to reduce the size of each individual calculation by $\sim$400$\times$, allowing for a numerical study covering a wide range of input conditions.)}
For the planar three-color case, a series of random sets of polarizations for the three beams $\left(\xi_1, \xi_2, \xi_3\right)$ was generated. 
The peak photon brightness from this study was found to occur near two polarization nodes: $\left(45^\circ, 45^\circ, 45^\circ\right)$ and $\left(135^\circ, 135^\circ, 135^\circ\right)$, as shown in Figure~\ref{fig:poltest}a (including concentrated sampling points near the first node).

For a selected laser geometry, it was found that the optimal polarization can be identified by evaluating just the vector-dependent terms of Equation~\ref{eq:diff_energy}.
Since we are considering linearly-polarized colliding beams, the sum of the three polarization directions \added{($\vec{E_d} = \sum_i \hat{E}_i$)} and cross-polarization directions \added{($\vec{B_d} = \sum_i \hat{B}_i$)} define a polarization-dependent metric $L$ for scattering effectiveness:
\begin{align}
	L &\equiv \left|\bm{n}\times\left[\added{\vec{M_d}} + \bm{n}\times\added{\vec{P_d}}\right]\right|^2, \label{eq:polmetric} 
\end{align}
\added{\noindent where $\vec{P_d}, \vec{M_d}$ are the results of evaluating Eq.~\ref{eq:P},\ref{eq:M} using the summed field directions $\vec{E_d}, \vec{B_d}$.}
Since the metric $L$ depends only on the direction and polarization of the three interacting lasers, it can be quickly evaluated for all polarization combinations and maximized.
Calculations verify that the metric $L$ varies with polarization (modulo 180$^\circ$) in the same way as the scattered photon brightness, as shown in Figure~\ref{fig:poltest}b.

\begin{figure}
	\includegraphics[width=\columnwidth]{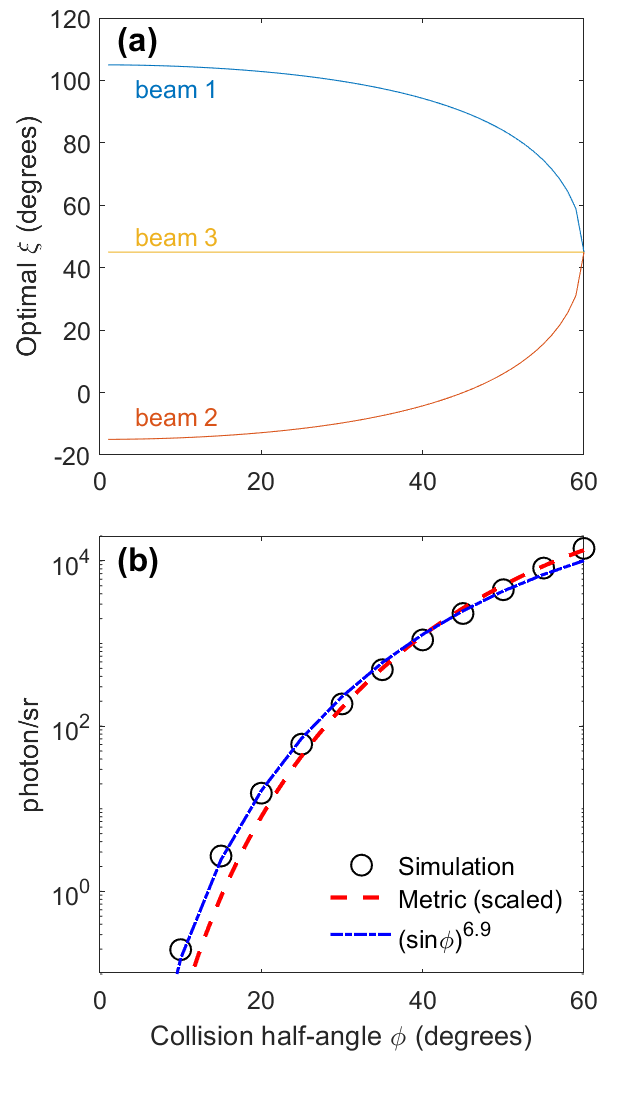}
	\caption{Effect of varying collision angle $\phi$ in three-color family of solutions.  (a) Optimum beam polarization angles used in calculating peak photon brightness. (b) Photon brightness along $\hat{k}_4$ with optimum polarization (points); amplitude of polarization metric, scaled to fit (red); best fit model (blue). \label{fig:2wpolarization}}
\end{figure}

The polarization metric predicts that the optimal polarization in the one-color family is always $\left(45^\circ, 45^\circ, 45^\circ\right)$; this result is confirmed by numerical simulations.
However, the optimal polarization in the three-color family depends on the scattering geometry, as shown in Figure~\ref{fig:2wpolarization}a.
The optimal polarization of beams 1 and 2 vary smoothly with the collision angle $\phi$, while beam 3 remains consistent at $45^\circ$. 
Using these optima, the peak photon brightness was calculated for each value of $\phi$.
The results, shown in Figure~\ref{fig:2wpolarization}b, demonstrate reasonable agreement between the trend in photon brightness and the scaled amplitude of the polarization metric.
Notably, this falls off much faster than the center-of-momentum frequency scaling of the scattering cross section alone ($\propto\omega_{cm}^4$, see Eq.~\ref{eq:cm_frequency}).
A reasonable fit to the simulation outputs is given by a scaling $\propto \left(\sin \phi\right)^{6.9}$ (blue dashed line).
The maximum amplitude of the polarization metric (red) is found to fall off even faster than the simulation outputs, as $\left(\sin \phi\right)^8$.
While the polarization metric captures the effects of beam geometry and polarization on scattering efficiency, the numerical scaling additionally takes into account the details of the finite beam focus, which result in an elongated scattering volume as angle decreases (roughly $\propto \sin \phi^{-1}$).  
A table describing the vectors of laser propagation and optimal polarization for the one-color and three-color experiment families is included in Appendix~\ref{app:tables}.

In particular, we note that, for the same laser power and optimal polarizations, the three-axis ($\phi =  45^\circ$) experiment previously studied in the literature (see Ref.~\cite{PRL:Lundstrom:2006}, e.g.) is calculated to produce $6.1\times$ less signal than the planar case.
The maximum collision angle consistent with non-overlapping beams 3 and 4 ($\phi \approx 57^\circ$ for f/2 focusing) provides a $\approx4\times$ boost in signal compared to the 3-axis case.  
Although arbitrary polarization angles may be challenging to achieve experimentally, the increased signal for higher collision angles is significant enough to merit serious consideration.

\subsection{Sensitivity studies}
Given that the amount of scattering scales sensitively with the overlapped peak field amplitude of the three beams, the  inaccurate cotiming and copointing of the beams constitutes the main risk of shot-to-shot signal variability and loss.  
To assess the sensitivity of the scattering signal to these metrics, a series of simulations was run with timing and pointing variations introduced into the beams.

\begin{figure}
	\includegraphics[width=\columnwidth]{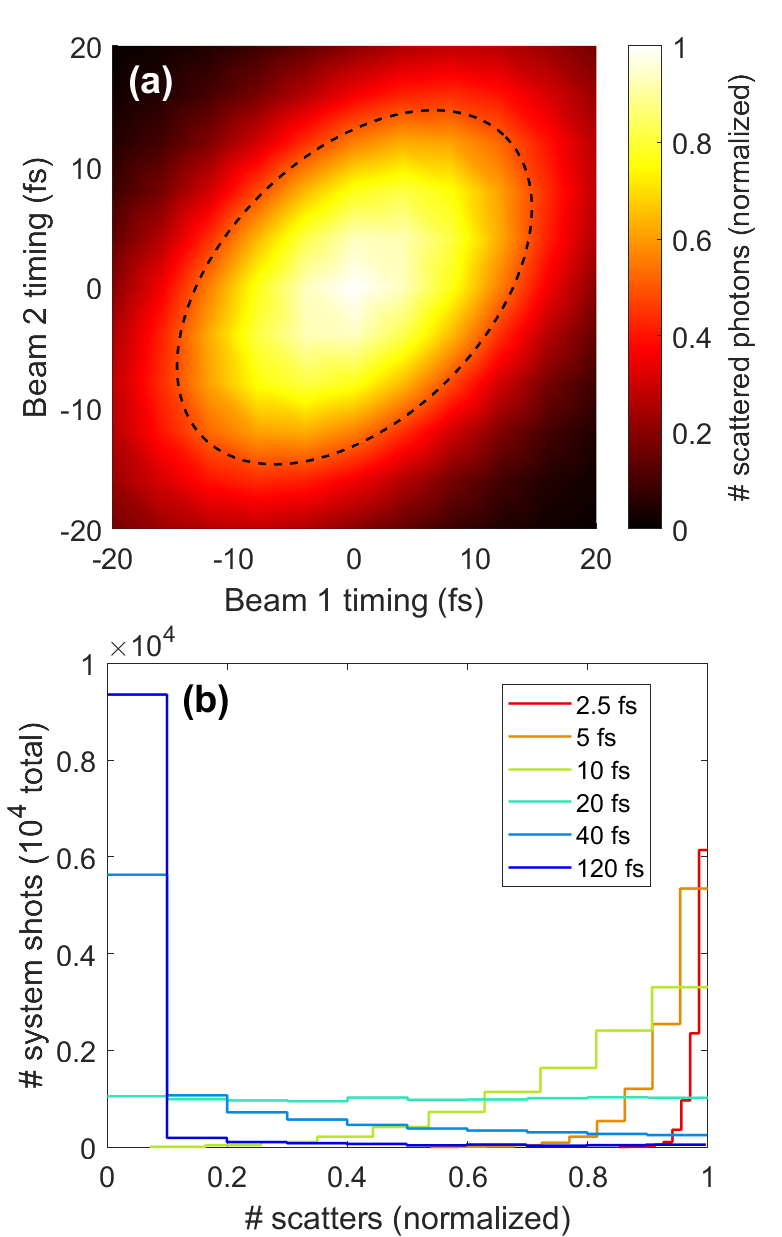}
	\caption{(a) Normalized scattered photon emission vs mistiming of beams 1 and 2 for the one-color point design.  (black dashed line) Contour of 50\% emission. (b) Statistical estimate of scattered photon probability for various values of random (Gaussian FWHM) beam timing uncertainty from 2.5~fs (red) to 120~fs (blue).\label{fig:cotiming}}
\end{figure}

The result of a timing sensitivity study is shown in Figure~\ref{fig:cotiming}a.
This study was based on the point design for the one-color experimental layout (the results of the three-color layout for this timing study are almost identical). 
The time of peak intensity for beams 1 and 2 were varied over a range of 20~fs (the pulse duration full width at half maximum); beam 3 was held constant without loss of generality. 
The degradation in the signal is correlated with respect to the direction of timing delay in the two beams: 50\% degradation is observed when the beams are delayed by  17.7~fs in the same direction, or 10.9~fs in opposite directions.
This result was used to simulate a campaign of 10,000 system shots, with each beam's timing randomly sampled from a normal distribution with various full-widths at half maximum (FWHM).
The results of this sampling study are shown in Figure~\ref{fig:cotiming}b.
When beams have random timing uncertainties of 10~fs FWHM  (green curve) or less, the distribution of results is weighted towards observation of the scattering signal on most shots.
However, when the timing uncertainty exceeds 20~fs FWHM (cyan curve), the scattering signal is dominated by mistiming.  
As expected, this threshold is comparable to the pulse duration of the laser pulses themselves.
To limit the dependence of the experimental results on mistiming, the relative beam cotiming should be equal to or better than the pulse duration. 

\begin{figure}
	\includegraphics[width=\columnwidth]{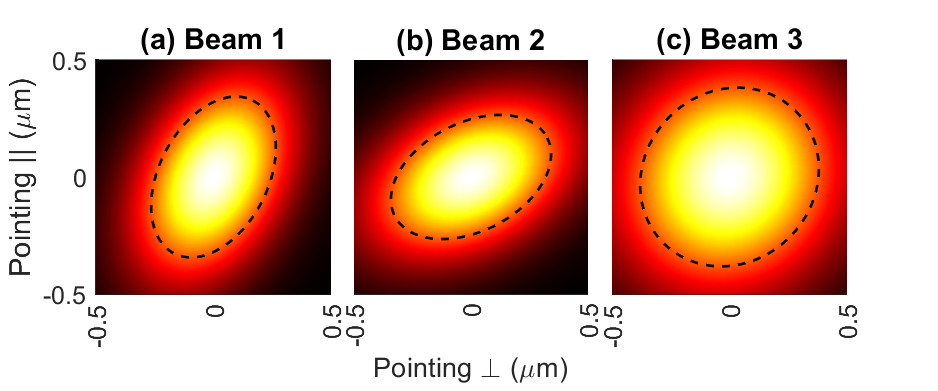}
	\caption{Normalized scattered photon brightness along $\hat{k}_4$ versus single-beam mispointings for each beam in three-color point design. (black dashed) Contours of 50\% emission; best-fit ellipse parameters are given in Table~\ref{table:pointing_50pc_axes}.\label{fig:copointing}}
\end{figure}

The sensitivity of the photon-photon scattering signal to beam mispointing is more challenging to rigorously evaluate due to the increased dimensionality of the problem, as each beam can be mispointed in two directions.
To assess this sensitivity, various 2-dimensional planes of mispointing were evaluated in isolation.
For this study, we evaluate only the scattered photon brightness along the $\hat{k}_4$ vector, rather than performing the full angular integral, to reduce computational time.
(This `peak brightness' was found to scale linearly with the total photon number to better than 10\%.)
The sensitivity of the three-color point design to single-beam mispointing is shown in Figure~\ref{fig:copointing}.
The single-beam pointing sensitivity is well modeled as elliptical Gaussians with half-widths and angles given in Table~\ref{table:pointing_50pc_axes}.
The minor axes are smaller for the frequency-doubled beams (1, 2) than for beam 3, which is due to the 0.5$\times$ smaller wavelength and focal spot of these beams with fixed focal geometry (f/2): the beam waist in the laser model has a FWHM of $\sigma_{\textrm{FWHM}} = 2\lambda{f}\sqrt{2 \ln 2}/\pi \approx 1.5\lambda$ for f/2 focusing.
On a relative scale, the mispointing tolerance FWHM is roughly equal to the wavelength of the frequency-doubled beams (0.45~$\mu$m).

\begin{figure}
	\includegraphics[width=\columnwidth]{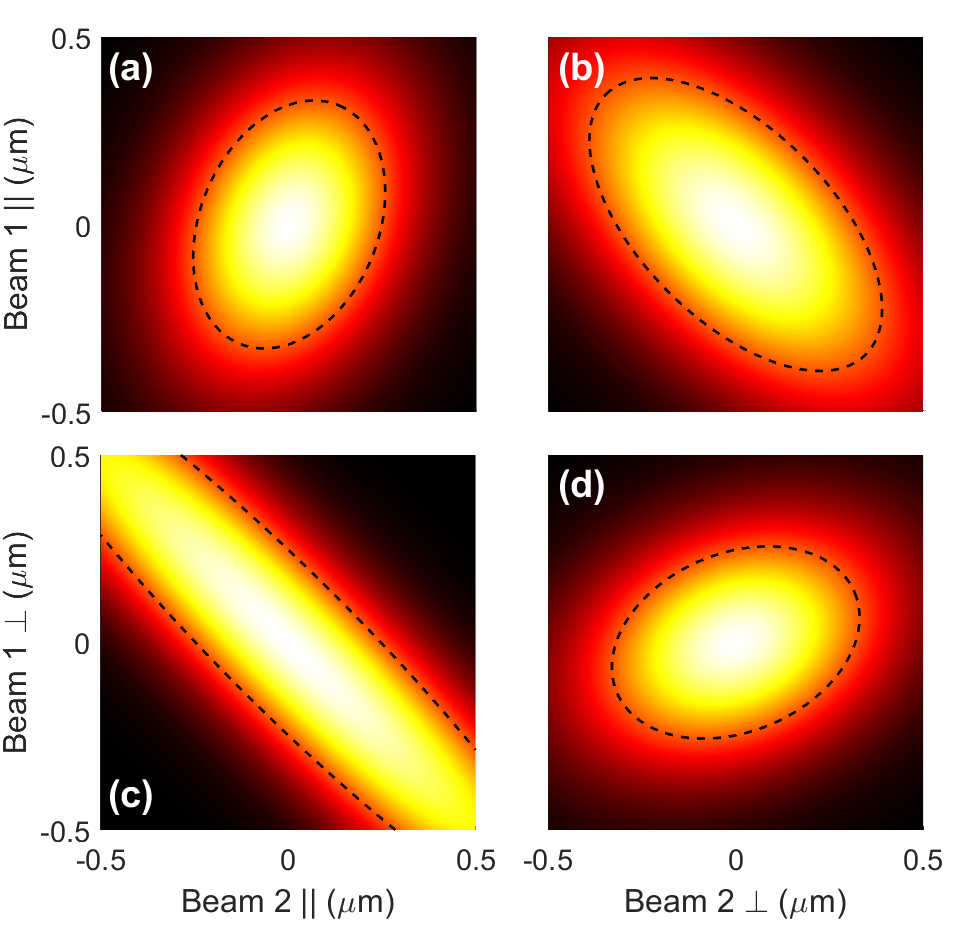}
	\caption{Normalized scattered photon brightness along $\hat{k}_4$ versus selected mispointings of beams 1 and 2 in the three-color point design: (a) $\parallel_1$, $\parallel_2$; (b) $\parallel_1$, $\perp_2$; (c) $\perp_1$, $\parallel_2$; (d) $\perp_1$, $\perp_2$. (black dashed) Contours of 50\% emission.\label{fig:copointing_beam1v2}}
\end{figure}

\begin{figure}
	\includegraphics[width=\columnwidth]{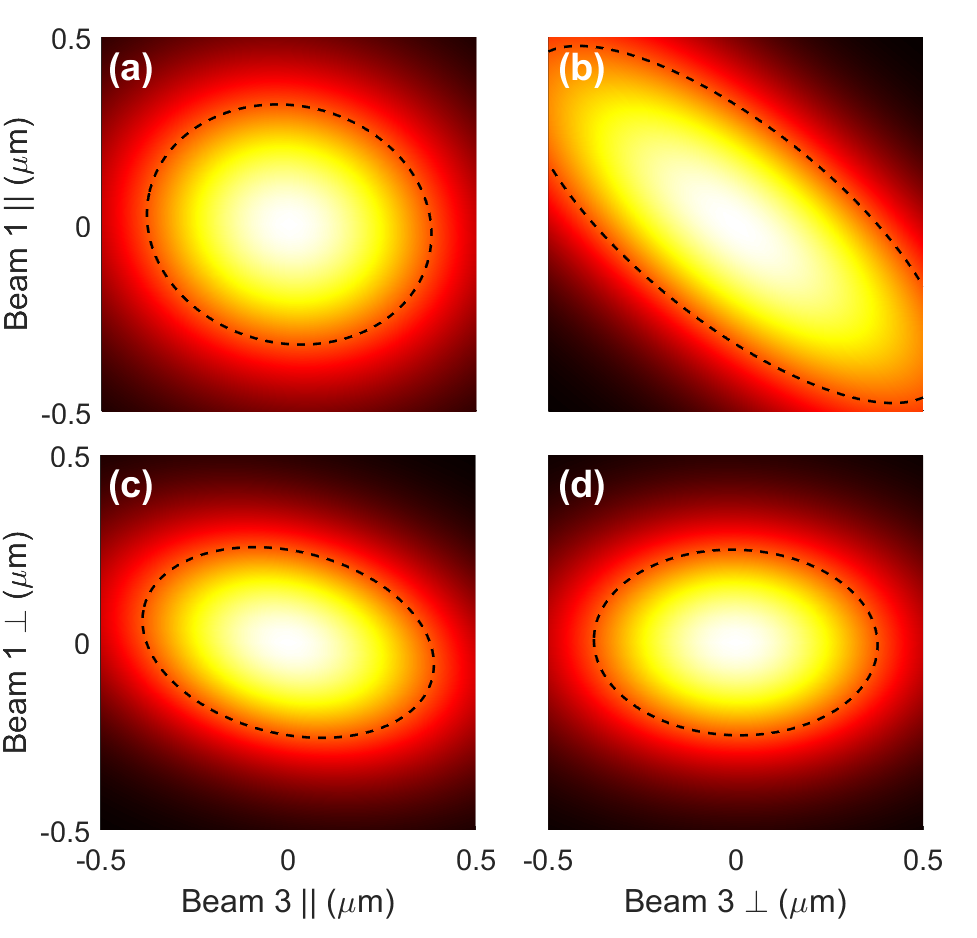}
	\caption{Normalized scattered photon brightness along $\hat{k}_4$ versus selected mispointings of beams 1 and 3 in the three-color point design: (a) $\parallel_1$, $\parallel_3$; (b) $\parallel_1$, $\perp_3$; (c) $\perp_1$, $\parallel_3$; (d) $\perp_1$, $\perp_3$. (black dashed) Contours of 50\% emission.\label{fig:copointing_beam1v3}}
\end{figure}

Correlations between the mispointing of the three beams exist in the cases where they are mispointed in the same direction.
Results of selected multi-beam mispointing studies are shown in Figures~\ref{fig:copointing_beam1v2} and \ref{fig:copointing_beam1v3}.
Each figure shows correlations in the scattered photon yield with mispointing of beam 1 in the (polarization, cross-polarization) directions versus mispointing of beam 2 or 3 in the (polarization, cross-polarization) directions, resulting in four studies for each pair of beams.
Each result is again well-fit with an elliptical Gaussian, with major/minor axes (FWHM) given in Table~\ref{table:pointing_50pc_axes}.
Due to the three-dimensional nature of the interaction volume, large variance is observed in the sensitivity with mispointing direction: the FWHM is seen to varies from 0.355~$\mu$m to 2.12~$\mu$m; or, 0.78$\times$ to 4.7$\times$ the shortest wavelength.
However, due to the random nature of the pointing variance in real experiments, the smaller value will limit the statistical performance of the experimental campaign.

\begin{table}
	\caption{Elliptical best-fit parameters for 50\% degradation in single-beam mispointing sensitivity study (black contours) in Figure~\ref{fig:copointing}.\label{table:pointing_50pc_axes}}
	\begin{tabular}{|c|c|c|c|}
		\hline
		Beam axes & major axis & minor axis & angle \\
		\hline
		1 &  0.732~$\mu$m & 0.463~$\mu$m & 62.7~deg \\
		2 &  0.732~$\mu$m & 0.463~$\mu$m & 27.3~deg \\
		3 &  0.795~$\mu$m & 0.724~$\mu$m & 45~deg \\
		\hline
		$1_{||}, 2_{||}$ &  0.688~$\mu$m & 0.475~$\mu$m & 67.3~deg \\
		$1_{||}, 2_{\perp}$ &  0.981~$\mu$m & 0.508~$\mu$m & 135~deg \\
		$1_{\perp}, 2_{||}$ &  2.123~$\mu$m & 0.355~$\mu$m & 135~deg \\
		$1_{\perp}, 2_{\perp}$ &  0.688~$\mu$m & 0.475~$\mu$m & 22.7~deg \\
		\hline
		$1_{||}, 3_{||}$ &  0.768~$\mu$m & 0.633~$\mu$m & 166.9~deg \\
		$1_{||}, 3_{\perp}$ &  1.382~$\mu$m & 0.522~$\mu$m & 141.5~deg \\
		$1_{\perp}, 3_{||}$ &  0.792~$\mu$m & 0.486~$\mu$m & 166.1~deg \\
		$1_{\perp}, 3_{\perp}$ &  0.757~$\mu$m & 0.495~$\mu$m & 178.2~deg \\
		\hline
		
	\end{tabular}
\end{table}

\begin{figure}
	\includegraphics[width=\columnwidth]{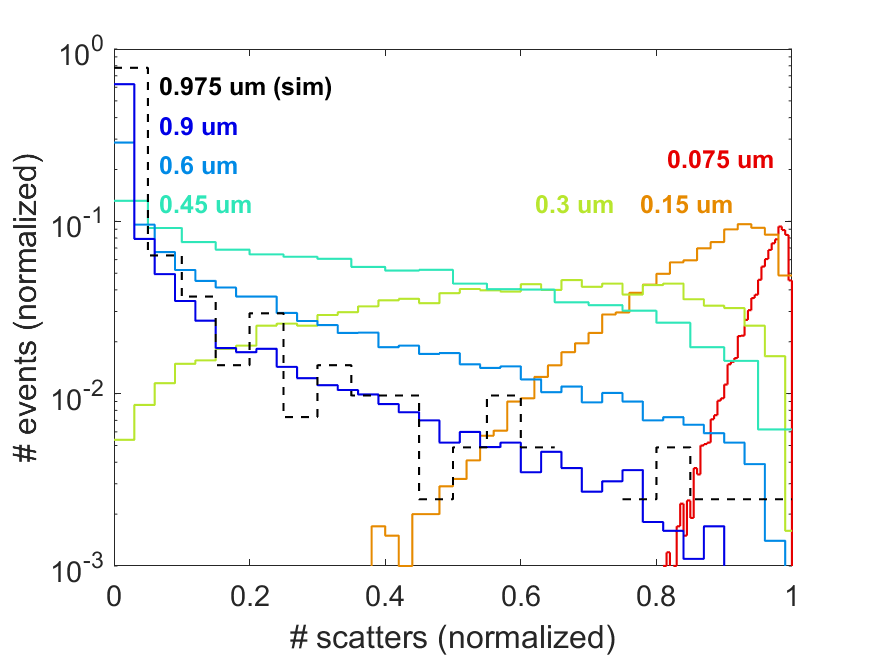}
	\caption{(color) Estimates of total photon scattering distribution from 10$^4$ shots with beam mispointing selected from a normal distribution (FWHM in legend), based on a fit to Figures~\ref{fig:copointing}--\ref{fig:copointing_beam1v3}. (black dashed) Calculated total photon scattering for 410 shots with mispointing randomly selected from a normal distribution with FWHM 0.975~$\mu$m. \label{fig:copointing_statistics}}
\end{figure}

To assess the combined effect of mispointing, a combined elliptical Gaussian model in six dimensions was fit to the mispointing results shown in Figures~\ref{fig:copointing}--\ref{fig:copointing_beam1v3} and a sampling of 410 randomly-selected off-plane calculations.  
This model was then sampled 10,000 times with random beam mispointings selected from a normal distribution with various FWHM widths.
The results of this statistical mispointing study are shown in Figure~\ref{fig:copointing_statistics}.
As with the statistical mistiming study (Figure~\ref{fig:cotiming}b), a transition point is observed between 0.3 and 0.45~$\mu$m FWHM mispointing: below the transition point, the number of scatters is weighted towards the maximum value achieved at optimal pointing, but above the transition point the most likely number of scatters reaches zero, and the tail of scattering events drops with increased mispointing.
The result of the 410 full simulations is also included, and matches the trend of the sampled fits.
This study confirms that to reliably detect scattering events on a shot-by-shot basis, a pointing stability of better than the shortest laser wavelength (here, 0.45~$\mu$m) is required.
The statistical reduction in the signal from mispointing and mistiming are expected to be uncorrelated and compound multiplicatively.

Other numerical scalings were tested to verify the expected theoretical behavior.
Simulations confirm that the peak brightness of scattering signal scales proportionally to the product of the three beam powers, $(P_1P_2P_3)$, as expected.
Doubling the laser frequencies while keeping the energy, focusing (f/\#) and pulse duration of the lasers constant was observed to increase the scattering by a factor of 8.6$\times$.  
Taking into account the effect on beam photon density and overlap volume, this is inferred to denote a change in cross-section of $\sigma \propto N\omega = 17.6$, approximately consistent with the expected $\omega^4$ scaling of stimulated photon-photon scattering.
Lastly, simulations varying the relative phase of the three laser pulses \added{($\varphi_0$ in Eq.~\ref{eq:beammodel})} do not predict a change in scattering signal above numerical noise.

\section{Implications for Experiments}
\label{sec:design}



\subsection{Geometry}
While the planar solution is expected to produce the greatest scattering signal, measurement of the signal in that geometry is complicated by the fact that it counter-propagates relative to beam 3.
This situation would require the signal to share optics with beam 3 at some point: either the final focusing optic or an optic further upstream must be able to separate the (backwards-propagating) scattered light from the (forward-propagating) beam. 
Given the relative intensity of these two signals, a discrimination of the order of 10$^{19}$ would be required.  
For comparison, the reflectivity of optics used to guide the compressed beams is estimated to be 98\%.  
The distribution of the remaining 2\% of the beam energy in the bandwidth range of the signal is not known, however even very low probabilities ($\sim10^{-17}$) of scattering and, in the case of the three-color solution, third-harmonic conversion, would be sufficient to dominate the signal of interest.
Given the challenge of the proposed measurement, it is prudent to avoid shared beam paths as much as possible.
\added{
For this reason we propose separating the signal path from the path of beam 3 by selecting $\phi < 57^\circ$ for the three-color solution.}

\added{
Laser pointing stability is usually reported as angular variation, as defined by the angular mispointing of the final focusing optic.
For the NSF OPAL point design, the full-aperture 25-PW beam has a square profile with a side length of 86~cm, and we presume for this experiment it will be focused with f/2 OAPs (172~cm focal length).
Absolute pointing variance of 0.45~$\mu$m with that system would then require pointing stability of 0.26~$\mu$rad.
This presents a significant challenge for the facility, and will be addressed in its design, including active pointing stabilization.
This risk may be mitigated by splitting the 2$\omega$ beams immediately prior to the final focusing optics, such that the mispointing of the two most-sensitive beams is substantially correlated.}

\added{
Arbitrary beam polarization is difficult to achieve in high-power laser experiments, as optics are often designed to work with particular polarization orientations.  
For this reason, the optimal polarizations described in Section~\ref{sec:polarization} may not be experimentally achievable.
In particular, we note that a poor choice of polarization can nearly or completely eliminate the SPPS signal, and these nodes include choices that are otherwise reasonable from a hardware perspective: a choice of $[\xi_1, \xi_2, \xi_3] = [90^\circ, 90^\circ, 0^\circ]$ reduces the scattering signal by a multiplier of $\times0.003$.
If we select from designs that are `simple' in the sense that all beams are either horizontally or vertically polarized, we find that a choice of all vertical ($[0^\circ, 0^\circ, 0^\circ]$) or all horizontal ($[90^\circ, 90^\circ, 90^\circ]$) polarization both produce a reduction of approximately $\times0.3$ in the total scattered signal as compared to the optimal choice, which may be tolerable.
Polarization is a critical element that must be considered in selecting an experimental chamber design.  
}

\subsection{Background}
Despite the loss of power associated with frequency doubling, we anticipate that measurement of the scattered light will be significantly more challenging in the one-color family of solutions, for which the measured signal has the same frequency as the lasers.
The presence of a single free electron intersecting with a laser pulse in the field of view of the scattered light detector would result in significant scattered light onto the detector.\cite{NJP:Doyle:2022}
As a first estimate, the cross-section for Thomson scattering is $\sigma_{t} \approx 0.665~$b, resulting in an estimate of the number of scattered photons per electron: $N_{s} \approx \sigma_{t} I\tau/\hbar\omega \approx 3.8\times10^{4}$ for a 25~PW NSF OPAL beam.
In reality, the relativistic motion of electrons will radiate photons over a range of energies.
Because of this, free electrons must not be allowed within the intersection of the incident lasers and the field-of-view of the detector.  
With f/2 focusing, the 10~PW-scale beams are above ionizing intensity within several centimeters of best focus, a volume over which it is likely impossible to achieve perfect vacuum.

We propose to collimate the detection angle using optics matched to the scattering angle of the light.  
Assuming an f/2 optic with spatial filter can be used to collect the scattered light, and light produced outside this collection angle can be rejected by light-absorbing baffles, then the region from which light can access the detector is limited to the intersection of the four f/2 focal regions. 
Additionally, an opposing f/2 `black box' will block light scattering up the detector line of sight from beyond the interaction region.
If less than one molecule in the interaction region is required and a vacuum of 10$^{-9}$ torr is achievable, then the observed interaction region must have a volume smaller than $3\times10^{4}$~$\mu$m$^3$.
This will likely require tight spatial filtering on the detection axis.
A numerical estimate using lasers as defined in Table~\ref{table:resultoptimal} and $\phi = 45^\circ$ predicts that the volume containing above 10$^{14}$ W/cm$^2$ reaches this size when observed by a collection optic with f/2 focusing and spatial filtering with an acceptance FWHM of about 36~$\mu$m.
Notably, as the angle increases and beam 3 more closely approaches the collection line of sight, the overlapping volume grows rapidly (as 1/sin of the relative angle).
At $\phi = 57^\circ$, the detection of scattering from beam 3 is depth-of-focus limited, and a volume of 3$\times10^{4}$~$\mu$m$^3$ is observed with spatial filtering FWHM of 21~$\mu$m.  
However, this value is very sensitive to small changes in the geometry.
If 10~$\mu$m FWHM spatial filtering could be achieved, the observed volume in this limiting case drops by an order of magnitude to $3.4\times10^{3}$~$\mu$m$^{3}$.
This could relax the pressure requirements to roughly 10$^{-8}$~torr.


The details of the expected radiation field will continue to be studied in future work, but we can infer the general principles that the detector must be isolated spatially, temporally, and spectrally as much as possible.
These principles significantly benefit from the proposed three-color, off-planar solution.
Additionally, after the interaction point, the lasers should be transported away from the experimental chamber and dumped, to reduce the scattering of free laser light into the detector.

\subsection{Detector Technologies}
The primary requirements of the diagnostic system is to accurately count the number $N$ of photons radiated into the scattering solid angle and frequency band on a per-shot basis.  
The fundamental statistical uncertainty of the number of scattered photons is Poisson distributed.
This establishes that the necessary counting accuracy for the detector system should be comparable to the Poisson uncertainty ($N^{1/2}$).
For signals of the order of 1000 photons, the desired counting accuracy is then approximately $\pm30$ photons.

Photomultiplier tubes (PMTs) are a well-developed technology for amplifying and detecting low light signals. 
Many commercial products exist, covering the full range of required wavelengths and with quantum efficiency exceeding 25\%.
Statistically, the width of a PMT signal distribution scales as the square root of the signal charge,\cite{MS:Macleod:2007} indicating that the uncertainty in inferred photon number scales as $N^{1/2}$.
As this is the same scaling as the expected statistical uncertainty in the photon signal, a PMT detector would likely suffice for this experiment, but detectors with better statistical uncertainty will benefit the measurement.

Single-photon counting technologies, such as superconducting nanowire single-photon detectors (SNSPD)\cite{APL:Goltsman:2001} and single-photon avalanche diodes (SPAD)\cite{Optica:Cova:1996}, offer the highest detection efficiency for low-light photon signals and picosecond-scale time resolution.
However, each detection element in these systems must recover from a ``dead time'' after the detection event during which subsequent photons are not detected, which typically takes of the order of nanoseconds.  
To detect the very short (20~fs) expected pulses of scattered photons, the signal would need to be collimated onto an array of detection elements with average incidence well below one photon per element.
To achieve this, arrays with more than $10^4$ elements would be required.
Commercial SNSPD products are currently optimized for efficiency at optical and near-infrared frequencies,\cite{SingleQuantum:SNSPD} which is appropriate for the one-color scattering signal. 
Avalanche photodiodes using 4H-SiC have been demonstrated to achieve quantum efficiency above 50\% in the ultraviolet ($300~$nm) range required for three-color scattering, with high visible light rejection ratio ($>10^3$).\cite{IEEE:Zhou:2018}
For either of these technologies, a sufficiently-large detection array will likely require custom development. 

A secondary goal of the detector includes measuring the polarization distribution.
The polarization distribution of the scattered photons potentially contains information about the coupling processes, in particular the fundamental low-energy constants of the Euler-Heisenberg Lagrangian.\cite{Karbstein:2022uwf, PRA:Macleod:2024, PRD:Schutze:2024}
Measuring the polarization content can be done by splitting the polarizations of the collimated scattered photon beam using a Wollaston prism and separately detecting the two channels.
Measuring polarization would increase the uncertainty by at least $\sqrt{2}$ compared to the full signal counting.

The angular distribution of the scattered photons is primarily determined by the spatial overlap of the beams, and contains information about the on-shot beam copointing that may be useful in statistical assessment of the dataset.
Measuring the angular distribution may then be useful to evaluate otherwise hard-to-assess correlations between on-shot beam pointing and scattered signal level.  
This could be done by collimating the signal onto a low-noise imaging array, such as a qCMOS.
Commercial technologies offer readout noise as low as 0.3~electrons per pixel,\cite{OrcaQUEST2} and a few examples with even lower readout noise have been demonstrated in the literature.
The low noise level allows photon counting imaging, although the noise is increased compared to true single-photon-counting detectors.  
To effectively use such a detector, the angularly distributed signal would be collimated onto a small region of the detector array.  
Assuming Poisson statistics in the read noise, to keep the statistical variation in the noise below $N^{1/2}$, the number of pixels in the region-of-interest $N_p$ must be less than the number of detected photons divided by the read noise. 
Taking the quantum efficiency of detection into account, resolution of the signal with of the order of 100 pixels is viable.
Skipper-CCDs read the charge levels of the CCD array non-destructively, and can thus reduce effective read noise to levels as low as desired with multiple reads.\cite{SPIE:Janesick:1990, PRL:Tiffenberg:2017} 
While read noises below 0.1~e/pixel may require minutes of acquisition time, this delay is well matched to the planned 5-minute shot rate for the NSF OPAL facility.

\section{Projected Bounds on Born-Infeld}
\label{sec:born-infeld}
Stimulated photon-photon scattering can be sensitive to contributions from beyond-the-Standard-Model (BSM) physics. 
As an example, we include here an assessment of the expected bounds that the proposed experiment could place on Born-Infeld nonlinear electrodynamics.
\added{The Born-Infeld (BI) Lagrangian density was originally suggested to avoid the appearance of the infinite electron self-energy due to its Coulomb field, and would contribute a signal to photon-photon scattering.
The structure of the Lagrangian density was proposed in analogy with the modification of the relation between energy and velocity when passing from non-relativistic to relativistic mechanics.}
The Born-Infeld Lagrangian:
\bea
\mathcal{L}_{\tsf{BI}} = b^{2}\left[1-\sqrt{1+\frac{\mathcal{F}}{2b^{2}}-\frac{\mathcal{G}^{2}}{16b^{4}}}\right],
\eea
with $\mathcal{F}$ and $\mathcal{G}$ as defined in Sec.~\ref{sec:intro}, depends on a phenomenological parameter, $b$, with units of mass-squared, i.e. $\mbi = \sqrt{b}$, which controls the strength of the interaction.   
In Born and Infeld's work \cite{Born:1934gh}, $b$ was chosen \added{to equal the strength of the Coulomb field at the classical electron radius. 
Written in terms of a field strength, Born and Infeld's value corresponds to $b \approx 1.19\times 10^{18}\,\rm{Vcm}^{-1}$ (approximately 100 times the QED critical field strength), indicating a mass scale of $\mbi \approx 10~m_ec^2 \approx 5\,\trm{MeV}$.
Interest has been renewed in Born-Infeld electrodynamics in recent decades: the structure is the same as the low-energy limit of the effective action of D-branes in string theory, where the parameter $b$ can take any value not currently ruled out by experiment\cite{FRADKIN1985123}.}
One of the strongest bounds on $\mbi$ \cite{Ellis:2017edi} was calculated using the ATLAS photon scattering results in ultra-peripheral heavy ion collisions \cite{NP:Aaboud:2017}, which bounded $\mbi>100\,\trm{GeV}$ (see also \cite{Ellis:2022uxv}). 
However, this bound was derived assuming that the Born-Infeld mass scale was much larger than the center-of-momentum (CM) energy of the experiment $\sim O(\trm{GeV})$ (such a high mass scale is required if the physical mechanism arises from string theory). If the Born-Infeld mass scale is of the order of or lower than the energy scale probed by ATLAS, further analysis would be required to verify if Born-Infeld is excluded by the experimental results. 
Furthermore, the ATLAS and CMS\cite{CMS:2018erd} experiments probed the high-energy limit of the QED photon-photon scattering cross-section; the CM energy of the NSF OPAL experiment is $\sim O(\trm{eV})$ and so probes the low-energy behavior. 
Since this energy scale is certainly much less than the Born-Infeld mass scale, we can approximate the Born-Infeld Lagrangian by its low-energy expansion, and add it to the low-energy QED Lagrangian to give:
\bea \label{eqn:Lagrangian}
	\mathcal{L} = c_{1}\mathcal{F}^{2}+c_{2}\mathcal{G}^{2},
\eea
with the combined constants:
\bea 
	c_{1} =  \frac{8 \alpha^{2}}{720 m^{4}} + \frac{1}{32 \mbi^4}; \qquad 
	c_{2} =  \frac{14 \alpha^{2}}{720 m^{4}} + \frac{1}{32 \mbi^4},
	\label{eqn:csQEDBI1}
\eea
in which the first terms are the QED low-energy constants and the seconds terms arise from Born-Infeld.
%

The coupling to the Born-Infeld mass scale can therefore be viewed as a BSM correction to the QED predicted low-energy constants.
As such, it is useful to express the Born-Infeld mass analogously to the electron mass in \eqnref{eqn:csQEDBI1}, writing:
\begin{align}
	\mbi = 	\Big(\frac{720}{4\alpha^{2} \cbi}\Big)^{1/4} m .
\end{align}
This results in a simplified version of Eq.~\ref{eqn:csQEDBI1} as:
\begin{align}\label{eqn:csQEDBI2}
	c_{1} =	&  \frac{2 \alpha^{2}}{720 m^{4}} \big(4 + \cbi\big)	\quad	&  \
	c_{2}	=	&	\frac{2 \alpha^{2}}{720 m^{4}}	\big(7 + \cbi\big),
\end{align}
where $\cbi$ is a dimensionless parameter controlling the strength of the coupling to Born-Infeld BSM physics.
As an estimate, the range $20 \text{~MeV} \lesssim \mbi \lesssim 200 \text{~MeV}$ corresponds to  $10^{-4} < \cbi < 10^{-1}$.
Given that the dimensionless numerical constants of QED [first terms in brackets in \eqnref{eqn:csQEDBI2}] are of order unity, the values of the mass scale determined by $\cbi$ that NSF OPAL will be sensitive to will largely be dictated by the statistics of the experiment.

It is well-known \cite{Fouche:2016qqj} that the forward-scattered vacuum birefringence signal is proportional to $c_{1}-c_{2}$, so experiments that measure vacuum birefringence are not sensitive to the Born-Infeld interaction (unless the non-birefringent part is also measured, but this requires a more developed set-up \cite{Karbstein:2022uwf}). 
However, if the momentum change of the scattered photon is measured, then the constants $c_{1}$ and $c_{2}$ can be inferred independently \cite{Macleod:2024jxl} and Born-Infeld can make a contribution. 
For example, in the three-color SPPS experimental design, if all the beam polarizations are parallel, the cross-section scales with \cite{king24toappear}:
\bea
	\sigma_{3\omega} \propto (6c_{1})^{2} + (c_{1}-5c_{2})^{2}.
\eea
Inserting the dependence from \eqnref{eqn:csQEDBI1} or \eqnref{eqn:csQEDBI2} then gives:
\begin{align}
	\sigma_{3\omega} &= \left[ \sigma_{3\omega}^{\tsf{QED}} + \sigma_{3\omega}^{\tsf{QED+BI}} + \sigma_{3\omega}^{\tsf{BI}} \right] \\
	\quad &\propto \left[1 + 0.349 \cbi + 0.034 \cbi^2\right] 
\end{align}
Therefore the cross-section picks up an interference term \cite{Davila:2013wba} between QED and Born-Infeld, which will produce the strongest signal of Born-Infeld.

\begin{figure}[t!!]
	\includegraphics[width=0.45\textwidth,trim={0.0cm 0.0cm 0.0cm 0.0cm},clip=true]{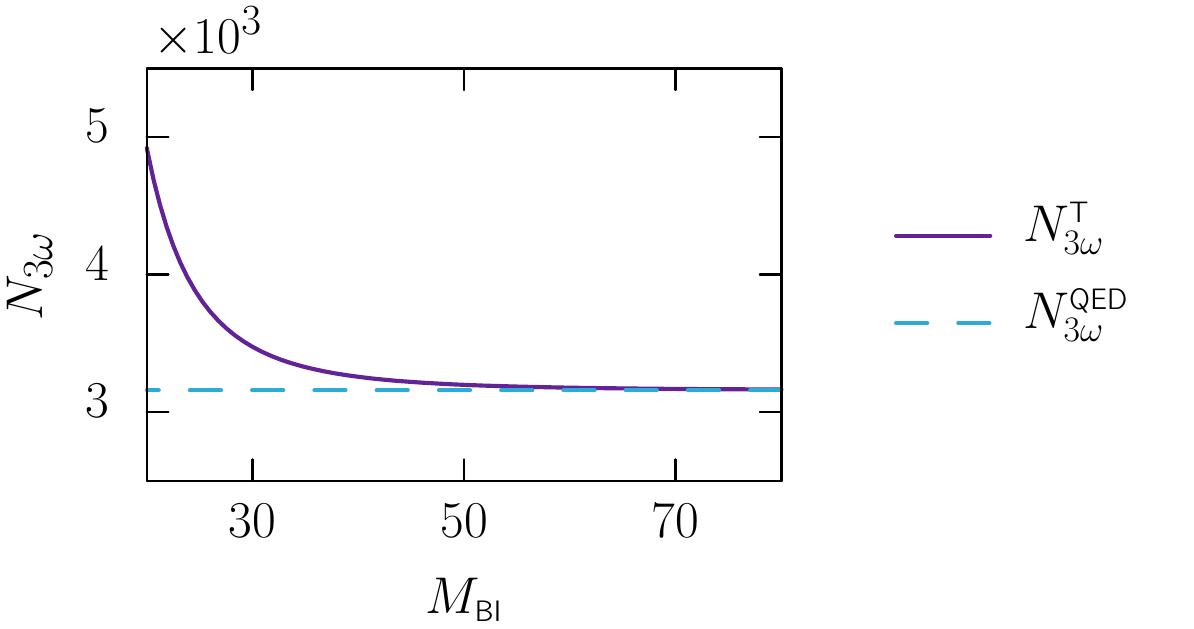}
	
	\caption{\label{fig:SingleShot} 
		Number of signal photons per optimal shot in the three-colour set-up as a function of the Born-Infeld mass scale $\mbi$.
		(Purple solid) Total number of signal photons, $N_{3\omega}^{\tsf{T}}$, including both QED contribution and Born-Infeld; (blue dashed) signal due to QED, $N_{3\omega}^{\tsf{QED}} \approx 3.2 \times 10^{3}$, determined from \eqnref{eqn:Lagrangian} with \eqnref{eqn:csQEDBI2} and $\cbi = 0$.
	}
\end{figure}

If the Born-Infeld interaction exists in nature, then for every shot of the NSF OPAL campaign there will be a total number of signal photons per shot, $N^{\tsf{T}} = N^{\tsf{QED}} + N^{\tsf{BI}}$, where we denote the contribution from the pure QED interaction [i.e. from \eqnref{eqn:Lagrangian} with \eqnref{eqn:csQEDBI2} and $c_{\tsf{BI}} = 0$] as $N^{\tsf{QED}}$ and the Born-Infeld signal as $N^{\tsf{BI}}$.
The Born-Infeld signal includes both the pure Born-Infeld interaction, where the QED low-energy constants are set to zero, and the interference term.
The signal of BSM physics due to the Born-Infeld interaction will require an excess of photons above the QED prediction to be measured.

\figref{fig:SingleShot} shows the estimated number of photons scattered per optimal collision in the NSF OPAL three-color experiment as a function of the Born-Infeld mass scale $\mbi$. 
The solid purple line shows the total number of signal photons including Born-Infeld effects ($N_{3\omega}^{\tsf{T}}$) while the blue dashed line corresponds to the pure QED result ($N_{3\omega}^{\tsf{QED}} \approx 3.2 \times 10^{3}$).
In this light, we can see that the QED contribution $N^{\tsf{QED}}$ acts as a minimum background for the measurement of Born-Infeld effects.

\begin{figure}[h!!]
	\includegraphics[width=0.45\textwidth,trim={0.0cm 0.0cm 0.0cm 0.0cm},clip=true]{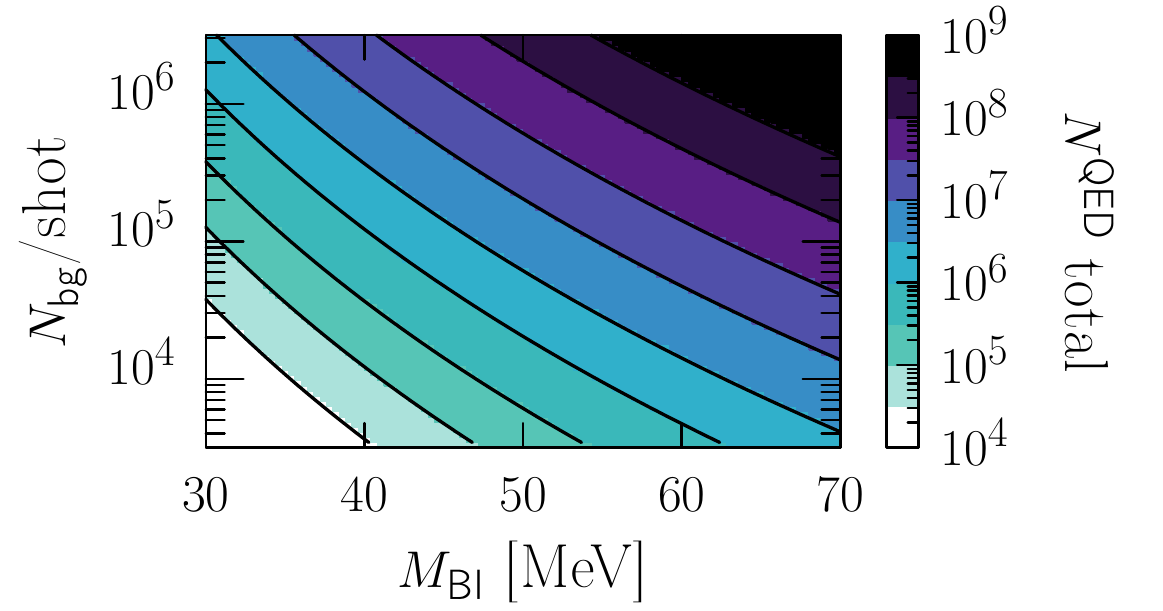}
	\caption{\label{fig:MassShots1} 
		Total number of measured QED photons required to place bound on Born-Infeld mass scale, $\mbi$, to $5\sigma$ significance for a particular number of background photons per shot $N_{\tsf{bg}}$ in the three-colour set-up, see \eqnref{eqn:Shots}.
	}
\end{figure}

To determine whether the Born-Infeld mass scale $\mbi$ could be measured, or otherwise bounded, at NSF OPAL will require both estimates on the number of Born-Infeld induced signal photons, $N^{\tsf{BI}} = N^{\tsf{T}} - N^{\tsf{QED}}$, and background photons, $N_{\tsf{bg}}$, per shot.
We consider the signal as a Poisson process, where the minimum number of optimal shots required to measure the Born-Infeld mass scale to a statistical significance of $n\sigma$ is given by \cite{Cowan:2010js},
\begin{align}\label{eqn:Shots}
	\mathsf{N}_{\tsf{shots}}^{n\sigma}
	\gtrsim
	&
	\frac{n^{2}}{2}
	\Big[
	(N^{\tsf{BI}} + N_{\tsf{bg}}) 
	\ln \Big(
	1 + \frac{N^{\tsf{BI}}}{N_{\tsf{bg}}}
	\Big) 
	- 
	N^{\tsf{BI}}
	\Big]^{-1}
	\,.
\end{align}
For a given number of shots there will be a total of $\mathsf{N}_{\tsf{shots}}^{n\sigma} N^{\tsf{QED}}$ photons measured due to the pure QED interaction.
Thus, if the Born-Infeld interaction does not occur in nature, the detection of $\mathsf{N}_{\tsf{shots}}^{n\sigma} N^{\tsf{QED}}$ total signal photons at a number of background photons per shot, $N_{\tsf{bg}}$, will place an $n\sigma$ significance bound on the mass scale $\mbi$ to a particular value.
This is shown in \figref{fig:MassShots1}, which plots the total number of QED signal photons that would be required to place a $5\sigma$ significance bound on a value of the mass scale $\mbi$ for a given number of background photons per shot.
Given the projected repetition rate of the NSF OPAL laser system ($\sim 5$~min/shot) and assuming optimal shots for which the number of QED photons per shot is $N^{\tsf{QED}} \approx 3.2 \times 10^{3}$, it is likely that a maximum number of signal photons of the order of $10^{7}$ (corresponding to $\approx 3000$ shots) could be observed in one experimental campaign.
Using these values, we predict that NSF OPAL could be used to place bounds on the Born-Infeld mass up to $M_{\tsf{BI}} \gtrsim 70$~MeV, depending on the control of the other sources of background in the measurement.

\section{Conclusions}
We present a design study for experiments to measure Stimulated Photon-Photon Scattering (SPPS) using NSF OPAL, \added{a multi-petawatt laser that is currently being designed}.
Two families of experimental geometries were derived that satisfy the necessary conservation of energy and momentum relations.
Designs with a half-angle of $71.5^\circ$ for the one-color case and $57^\circ$ for the three-color case optimize for maximum scattered photon yield while maintaining separation between the third laser beam and the detection solid angle.  
A numerical integration of the scattered photon signal from the Euler-Heisenberg Lagrangian was performed using Gaussian beams with bandwidth temporal pulse-shaping to match NSF OPAL parameters (25~PW peak power, 20~fs duration, focusing with f/2 optics), and predict a maximum value of 2097 (3163) scattered photons per shot in the one-color (three-color) cases.
Optimal polarization angles of the beams were determined.
Sensitivity studies demonstrate that co-timing of better than the pulse duration (20~fs FWHM) and co-pointing of better than the shortest wavelength (0.45~$\mu$m FWHM) is required to reliably scatter photons.
Detector designs and viable detector technologies were discussed.
Given the high background suppression required to successfully measure the scattered signal, we identify the three-color point design as the most likely experimental setup for this experiment to succeed on NSF OPAL.  
Using this design, we estimate that a campaign of the order of 3000 shots may bound the Born-Infeld mass scale to $\mbi\gtrsim70$~MeV.

Several areas of research will be pursued to support the development of the SPPS experiment and the NSF OPAL facility.
We will continue to develop our numerical simulations to include more realistic focused laser fields, \added{including the effect of a non-Gaussian near field, focal aberration, chromatic aberration, and spatiotemporal distortion.
We will also} validate the results in comparison with other numerical approaches.\cite{PRD:Blinne:2019}
Second-harmonic conversion of compressed short pulses remains a developmental technology, and is additionally complicated by the very large beam areas (\added{86}~cm)$^2$ anticipated at NSF OPAL.
To mitigate this risk, we will develop a large-aperture, high-aspect-ratio second harmonic conversion crystal technology.
This technology will be demonstrated using MTW-OPAL, a prototype laser for NSF OPAL \added{in operation at LLE}.\cite{HPLSE:Bromage:2021}
If second-harmonic conversion proves unworkable, the point design for the SPPS experiment will be changed to the one-color version.
We will also design and field a prototype detector on MTW-OPAL to measure in-situ the anticipated background from residual chamber gas and to test detector options and background mitigation strategies.
It is possible that a relativistic-scale prepulse on one of the beams (above $10^{18}~$W/cm$^2$) will sweep electrons from the interaction volume, mitigating the need for ultra-high vacuum at the interaction point: MTW-OPAL experiments using the prototype detector will test this hypothesis.
These ongoing research efforts will inform the NSF OPAL design in preparation for a successful measurement of stimulated photon-photon scattering when the facility is complete. 

\subsection*{Acknowledgments}
This material is based upon work supported by the Department of Energy [National Nuclear Security Administration] University of Rochester “National Inertial Confinement Fusion Program” under Award Number(s) DE-NA0004144.
This report was prepared as an account of work sponsored by an agency of the United States Government. Neither the United States Government nor any agency thereof, nor any of their employees, makes any warranty, express or implied, or assumes any legal liability or responsibility for the accuracy, completeness, or usefulness of any information, apparatus, product, or process disclosed, or represents that its use would not infringe privately owned rights. Reference herein to any specific commercial product, process, or service by trade name, trademark, manufacturer, or otherwise does not necessarily constitute or imply its endorsement, recommendation, or favoring by the United States Government or any agency thereof. The views and opinions of authors expressed herein do not necessarily state or reflect those of the United States Government or any agency thereof.

This material is based upon work supported by the U.S. National Science Foundation \added{Mid-scale Research Infrastructure Program under Award No.} PHY-2329970. Any opinions, findings and conclusions or recommendations expressed in this material are those of the author(s) and do not necessarily reflect the views of the National Science Foundation.

This work has been funded \added{also} by the Deutsche Forschungsgemeinschaft (DFG) under Grants No. 416607684, No. 416702141, and No. 416708866 within the Research Unit FOR2783/2.

BK acknowledges support from The Leverhulme Trust Grant RPG-2024-142. \added{ADP is partially supported by the U.S. National Science Foundation Mid-scale Research Infrastructure Program under Award No. PHY-2329970.}

\appendix

\section{\added{Total number of scattered photons}}
\label{app:total_photon_number}

The total number of scattered photons can be assessed in a simplified integration. By dividing the emitted energy by the scattered photon energy $\omega$, one can write the total number $N$ of scattered photons as
\begin{equation}
\label{N}
\begin{split}
N&=\int_0^{\infty}d\omega\int d\Omega\frac{\omega^3}{16\pi^3}|\bm{n}\times[\bm{M}(k)+\bm{n}\times\bm{P}(k)]|^2\\
&=\int\frac{d^3\bm{k}}{(2\pi)^3}\frac{1}{2\omega}|\bm{k}\times[\bm{M}(k)+\bm{n}\times\bm{P}(k)]|^2.
\end{split}
\end{equation}
By introducing a four-dimensional notation with the metric $\eta^{\mu\nu}=\text{diag}(+1,-1,-1,-1)$ and the four-dimensional scalar product $a\cdot b=a_0b_0-\bm{a}\cdot\bm{b}$ between two arbitrary four-vectors $a^{\mu}=(a_0,\bm{a})$ and $b^{\mu}=(b_0,\bm{b})$, the above expression of $N$ can written in a manifestly covariant form. Indeed, analogously to the electric and magnetic field, the magnetization and the polarization three-dimensional vectors are the elements of a second-rank, antisymmetric tensor \cite{Barut_b_1980}, 
\begin{align}
\Sigma^{\mu\nu}(x) &=
\begin{pmatrix}
0 & -P_x(x) & -P_y(x) &-P_z(x)\\
P_x(x) & 0 & M_z(x) & -M_y(x)\\
P_y(x) & -M_z(x) & 0 & M_x(x)\\
P_z(x) & M_y(x) & -M_x(x) & 0\\
\end{pmatrix} \nonumber \\
&=-\frac{\alpha}{45\pi F_{cr}^2}\left[\mathcal{F}(x)F^{\mu\nu}(x)+\frac{7}{4}\mathcal{G}(x)\tilde{F}^{\mu\nu}(x)\right],
\end{align}
where $F_{cr}=m^2/|e|$ indicates either the electric or the magnetic critical field of QED, $\mathcal{F}(x)=F_{\mu\nu}(x)F^{\mu\nu}(x)=-2[\bm{E}^2(x)-\bm{B}^2(x)]$ and $\mathcal{G}(x)=F_{\mu\nu}(x)\tilde{F}^{\mu\nu}(x)=-4\bm{E}(x)\cdot\bm{B}(x)$ are the two electromagnetic invariants, and $F^{\mu\nu}(x)$ and $\tilde{F}^{\mu\nu}(x)=(1/2)\varepsilon^{\mu\nu\lambda\rho}F_{\lambda\rho}(x)$ are the field tensor and its dual, with $\varepsilon^{\mu\nu\lambda\rho}$ being the fully antisymmetric tensor ($\varepsilon^{0123}=+1$).

In fact, one can easily show that Eq. (\ref{N}) can be written as
\begin{align}
\label{N_2}
N&=\int\frac{d^3\bm{k}}{(2\pi)^3}\frac{1}{2\omega}\int d^4x\,d^4y\,e^{ik\cdot(x-y)}k_{\mu}\Sigma^{\mu\nu}(x)\Sigma_{\nu\lambda}(y)k^{\lambda} \nonumber \\
&=\int\frac{d^3\bm{k}}{(2\pi)^3}\frac{1}{2\omega}\int d^4x\,d^4y\,e^{ik\cdot(x-y)}\nonumber\\
&\quad\times\partial_{x,\mu}\Sigma^{\mu\nu}(x)\partial_y{}^{\lambda}\Sigma_{\nu\lambda}(y),
\end{align}
where the indices $x$ and $y$ in the partial derivatives indicate the corresponding four-dimensional variables (notice that, since at the considered order of perturbation theory the fields are assumed to fulfill the free Maxwell's equations in the expression of $\Sigma^{\mu\nu}(x)$, then $\partial_{x,\mu}F^{\mu\nu}(x)=\partial_{x,\mu}\tilde{F}^{\mu\nu}(x)=0$). 
This expression is manifestly covariant under proper, orthochronous Lorentz transformations because the identity
\begin{equation}
\int\frac{d^3\bm{k}}{(2\pi)^3}\frac{1}{2\omega}=\int \frac{d^4k}{(2\pi)^3}\,\theta(k^0)\delta(k^2),
\end{equation}
with $\theta(\cdot)$ being the Heaviside step function, holds.

Now, we notice that the integral in $d^3\bm{k}$ in Eq. (\ref{N_2}) can be taken analytically. By passing to spherical coordinates and by applying the usual $i0$ prescription, we obtain
\begin{equation}
\int\frac{d^3\bm{k}}{(2\pi)^3}\frac{1}{2\omega}e^{ik\cdot(x-y)}=-\frac{1}{(2\pi)^2}\frac{1}{(t_x-t_y+i0)^2-|\bm{x}-\bm{y}|^2}.
\end{equation}
By introducing the four-vector $S^{\nu}(x)=\partial_{x,\mu}\Sigma^{\mu\nu}(x)$, we can rewrite $N$ as
\begin{align}
N&=\frac{1}{8\pi^2}\int \frac{d^4x\,d^4y}{|\bm{x}-\bm{y}|} S_{\mu}(x)S^{\mu}(y)  \nonumber \\
&\times\left(\frac{1}{t_x-t_y-|\bm{x}-\bm{y}|+i0}-\frac{1}{t_x-t_y+|\bm{x}-\bm{y}|+i0}\right).
\end{align}
We can now apply the identity
\begin{equation}
\frac{1}{x+i0}=\mathcal{P}\,\frac{1}{x}-i\pi\delta(x),
\end{equation}
to be understood inside an integral in $x$, where $\mathcal{P}$ indicates the principal value of the integral and observe that due to the symmetry of the integrand in the change of variable $x_{-}\to-x_{-}$, the terms proportional to the delta functions cancel out. By passing to the variables $x_+^{\mu}=(x^{\mu}+y^{\mu})/2$ and $x_-^{\mu}=x^{\mu}-y^{\mu}$, the result can be written as
\begin{align}
N=&\frac{1}{8\pi^2}\int d^4x_+\int\frac{d^3\bm{x}_-}{|\bm{x}_-|}\mathcal{P}\int dt_- \ \nonumber \\
&\times\left(\frac{1}{t_--|\bm{x}_-|}-\frac{1}{t_-+|\bm{x}_-|}\right)\nonumber\\
&\times S_{\mu}\left(x_++\frac{x_-}{2}\right)S^{\mu}\left(x_+-\frac{x_-}{2}\right).
\end{align}
By exploiting again the symmetry of the integrand, 
one can show that the two contributions to $N$ are equal to each other:
\begin{align}
N=&\frac{1}{4\pi^2}\int d^4x_+\int\frac{d^3\bm{x}_-}{|\bm{x}_-|}\mathcal{P}\int dt_-  \frac{1}{t_--|\bm{x}_-|}\nonumber \\
&\times S_{\mu}\left(x_++\frac{x_-}{2}\right)S^{\mu}\left(x_+-\frac{x_-}{2}\right).
\end{align}
Finally, by shifting the variable $t_-$ according to $t_-\to t_-+|\bm{x}_-|$ and, again, by exploiting the symmetry properties of the integrand, one obtains
\begin{align}
N=&\frac{1}{8\pi^2}\int d^4x_+\int\frac{d^3\bm{x}_-}{|\bm{x}_-|}\int \frac{dt_-}{t_-}  \nonumber \\
&\times\left[ S_{\mu}\left(t_++\frac{t_-+|\bm{x}_-|}{2},\bm{x}_++\frac{\bm{x}_-}{2}\right) \right. \nonumber \\
&\left.\times S^{\mu}\left(t_+-\frac{t_-+|\bm{x}_-|}{2},\bm{x}_+-\frac{\bm{x}_-}{2}\right)-(t_-\leftrightarrow -t_-)\right].
\end{align}
where the symbol $-(t_-\leftrightarrow -t_-)$ indicates that the previous expression in the square brackets has to be subtracted with $t_-$ replaced by $-t_-$ and where the principal value symbol has been removed as the integrand is now regular at $t_-=0$.

\section{Beam model}
\label{app:beam_model}
In this work, we use a beam model equivalent to a paraxial Gaussian travelling-beam model\cite{APB:Salamin:2007} to calculate the instantaneous electric and magnetic field vectors for the three colliding lasers.
Here we present formula for the beam model and the derived relationships between integrated quantities of peak intensity, peak power, and total energy.

The amplitude of the field as a function of space and time is given by:
\begin{align}
    E =& E_0 \exp\left[-\frac{\left(t-r_{\parallel}\right)^2}{2\tau^2}\right] \exp\left[-\frac{r_\perp^2}{\sigma^2\left(1+\frac{r_\parallel^2}{z_R^2}\right)}\right]\times \nonumber \\
    &\sin\left[\varphi_0 + \omega t-\omega r_\parallel\left(1+\frac{1}{2}\frac{r_\perp^2}{r_\parallel^2+z_R^2}\right)+\tan^{-1}\left(\frac{r_\parallel}{z_R}\right)\right] \nonumber \\
    & \times\left[1+\frac{r_\parallel^2}{z_R^2}\right]^{-1/2}, 
    \label{eq:beammodel}
\end{align}
where $t$ is the normalized time and ($r_\parallel, r_\perp$) the normalized distance along and perpendicular to the axis, respectively, with units of (1/eV).
The Rayleigh length $z_R = \pi \sigma^2/\lambda$ is a function of the beam waist $\sigma$ and the wavelength $\lambda$.
Additionally, we define the beam waist $\sigma = 2 \lambda f/\pi$ in terms of the focusing optic f-stop $f$, so the spatial geometry of the fields is in practice determined entirely by $f$.
The scalar field amplitude is entirely determined by the peak field constant $E_0$, including all units, and the instantaneous intensity $I \propto E^2$.

By integrating the instantaneous intensity over ($t, r_\perp$) at $r_\parallel = 0$, we derive an approximate algebraic relationship between peak intensity $I_0$ and peak power $P_0$; and by subsequently integrating over time, we obtain a relationship between these quantities and total beam energy $U$, as follows:
\begin{align}
    P_{0} &= \frac{\pi \sigma^2}{2} I_0 \\
    U &= \frac{\pi^{3/4}}{4}\sigma^2\tau I_0 = \frac{\sqrt{\pi}}{2} P_0 \tau.
    \label{eq:IPU}
\end{align}
Experimental quantities typically will refer to the temporal duration and spot radius as full-width at half maximum intensity (FWHM). 
The relation between these quantities and $\tau, \sigma$ is as follows:
\begin{align}
    t_{\textrm{FWHM}} &= 2\tau\sqrt{\ln 2} \\
    r_{\textrm{FWHM}} &= \sigma \sqrt{2 \ln 2}
\end{align}

In practice, instead of algebraically introducing a Gaussian temporal envelope (the first exponential in Eq.~\ref{eq:beammodel}), we obtained temporal pulse shaping by adding together multiple sub-beams (typically 21) with evenly-spaced frequencies around the fundamental frequency.
This was done to more realistically represent a bandwidth-limited compressed laser pulse, and enable future studies of pulse imperfections such as chromatic chirp.
When using bandwidth pulse shaping, a Gaussian spectral amplitude was used, and the bandwidth was selected to produce the desired value of $t_{\textrm{FWHM}}$.
In testing, the bandwidth-shaped model produced nearly identical results to the Gaussian algebraic temporal pulse shaping described in Eq.~\ref{eq:beammodel}, and the intensity/power/energy relations (Eq.~\ref{eq:IPU}) were numerically validated.

\section{Optimal polarization vectors}
\label{app:tables}
Here we include the optimal polarization vectors $(\bm{k}_{i,\parallel})$ for the three beams as a function of the collision angle $\phi$.
The optimal polarization for the one-color family of solutions is given in Table~\ref{table:polarization1w}, and for the three-color family of solutions in Table~\ref{table:polarization2w}.
The propagation directions $\bm{k_i}$ of the three beams can be calculated directly from the vector components of Eqs.~\ref{eq:1omega_family},\ref{eq:2omega_family}. 

\begingroup
\renewcommand{\arraystretch}{0.6}
\begin{table*}[t]
   \caption{Optimal polarization directions for one-color family of solutions.  Propagation directions are as described in Eq.~\ref{eq:1omega_family}.\label{table:polarization1w}} 
   \begin{tabular}{|c|c|c|c|c|c|c|c|c|c|}
        \hline
        $\phi$ & \multicolumn{3}{|c|}{$\bm{k}_{1,\parallel}$} & \multicolumn{3}{|c|}{$\bm{k}_{2,\parallel}$} & \multicolumn{3}{|c|}{$\bm{k}_{3,\parallel}$} \\
        (deg) & x & y & z & x & y & z & x & y & z \\
        \hline 
        90 & -0.7071 & 0 & -0.7071 & 0.7071 & 0 & -0.7071 & 0 & -0.7071 & -0.7071 \\ 
        85 & -0.7071 & 0.0616 & -0.7044 & 0.7071 & -0.0616 & -0.7044 & -0.0616 & -0.7071 & -0.7044 \\ 
        80 & -0.7071 & 0.1228 & -0.6964 & 0.7071 & -0.1228 & -0.6964 & -0.1228 & -0.7071 & -0.6964 \\ 
        75 & -0.7071 & 0.183 & -0.683 & 0.7071 & -0.183 & -0.683 & -0.183 & -0.7071 & -0.683 \\ 
        70 & -0.7071 & 0.2418 & -0.6645 & 0.7071 & -0.2418 & -0.6645 & -0.2418 & -0.7071 & -0.6645 \\ 
        65 & -0.7071 & 0.2988 & -0.6409 & 0.7071 & -0.2988 & -0.6409 & -0.2988 & -0.7071 & -0.6409 \\ 
        60 & -0.7071 & 0.3536 & -0.6124 & 0.7071 & -0.3536 & -0.6124 & -0.3536 & -0.7071 & -0.6124 \\ 
        55 & -0.7071 & 0.4056 & -0.5792 & 0.7071 & -0.4056 & -0.5792 & -0.4056 & -0.7071 & -0.5792 \\ 
        50 & -0.7071 & 0.4545 & -0.5417 & 0.7071 & -0.4545 & -0.5417 & -0.4545 & -0.7071 & -0.5417 \\ 
        45 & -0.7071 & 0.5 & -0.5 & 0.7071 & -0.5 & -0.5 & -0.5 & -0.7071 & -0.5 \\ 
        40 & -0.7071 & 0.5417 & -0.4545 & 0.7071 & -0.5417 & -0.4545 & -0.5417 & -0.7071 & -0.4545 \\ 
        35 & -0.7071 & 0.5792 & -0.4056 & 0.7071 & -0.5792 & -0.4056 & -0.5792 & -0.7071 & -0.4056 \\ 
        30 & -0.7071 & 0.6124 & -0.3536 & 0.7071 & -0.6124 & -0.3536 & -0.6124 & -0.7071 & -0.3536 \\ 
        25 & -0.7071 & 0.6409 & -0.2988 & 0.7071 & -0.6409 & -0.2988 & -0.6409 & -0.7071 & -0.2988 \\ 
        20 & -0.7071 & 0.6645 & -0.2418 & 0.7071 & -0.6645 & -0.2418 & -0.6645 & -0.7071 & -0.2418 \\ 
        15 & -0.7071 & 0.683 & -0.183 & 0.7071 & -0.683 & -0.183 & -0.683 & -0.7071 & -0.183 \\ 
        10 & -0.7071 & 0.6964 & -0.1228 & 0.7071 & -0.6964 & -0.1228 & -0.6964 & -0.7071 & -0.1228 \\ 
        5 & -0.7071 & 0.7044 & -0.0616 & 0.7071 & -0.7044 & -0.0616 & -0.7044 & -0.7071 & -0.0616 \\ 
        \hline
    \end{tabular}
\end{table*}
\endgroup

\begingroup
\renewcommand{\arraystretch}{0.6}
\begin{table*}[t]
   \caption{Optimal polarization directions for three-color family of solutions.  Propagation directions are as described in Eq.~\ref{eq:2omega_family}.\label{table:polarization2w}} 
   \begin{tabular}{|c|c|c|c|c|c|c|c|c|c|}
        \hline
        $\phi$ & \multicolumn{3}{|c|}{$\bm{k}_{1,\parallel}$} & \multicolumn{3}{|c|}{$\bm{k}_{2,\parallel}$} & \multicolumn{3}{|c|}{$\bm{k}_{3,\parallel}$} \\
        (deg) & x & y & z & x & y & z & x & y & z \\
        \hline 
        60 & -0.6124 & 0.3536 & -0.7071 & 0.6124 & 0.3536 & -0.7071 & 0 & -0.7071 & -0.7071 \\ 
        59.75 & -0.6809 & 0.3971 & -0.6154 & 0.5316 & 0.31 & -0.7882 & -0.1493 & -0.7071 & -0.6912 \\ 
        59.5 & -0.7048 & 0.4152 & -0.5752 & 0.4956 & 0.2919 & -0.818 & -0.2092 & -0.7071 & -0.6754 \\ 
        59.25 & -0.7213 & 0.4291 & -0.5437 & 0.4673 & 0.278 & -0.8393 & -0.254 & -0.7071 & -0.6599 \\ 
        59 & -0.7338 & 0.4409 & -0.5168 & 0.443 & 0.2662 & -0.8561 & -0.2908 & -0.7071 & -0.6445 \\ 
        58 & -0.7644 & 0.4777 & -0.4329 & 0.3672 & 0.2294 & -0.9014 & -0.3973 & -0.7071 & -0.5849 \\ 
        57 & -0.7796 & 0.5062 & -0.3688 & 0.3093 & 0.2009 & -0.9295 & -0.4703 & -0.7071 & -0.5281 \\ 
        56 & -0.7867 & 0.5306 & -0.3156 & 0.2617 & 0.1765 & -0.9489 & -0.525 & -0.7071 & -0.4737 \\ 
        55 & -0.7887 & 0.5523 & -0.2699 & 0.2211 & 0.1548 & -0.9629 & -0.5676 & -0.7071 & -0.4216 \\ 
        54 & -0.7874 & 0.5721 & -0.2298 & 0.1859 & 0.135 & -0.9732 & -0.6015 & -0.7071 & -0.3717 \\ 
        53 & -0.7835 & 0.5904 & -0.1939 & 0.1549 & 0.1167 & -0.981 & -0.6286 & -0.7071 & -0.3239 \\ 
        52 & -0.7776 & 0.6076 & -0.1617 & 0.1274 & 0.0995 & -0.9868 & -0.6502 & -0.7071 & -0.2779 \\ 
        51 & -0.7703 & 0.6238 & -0.1324 & 0.1029 & 0.0833 & -0.9912 & -0.6674 & -0.7071 & -0.2336 \\ 
        50 & -0.7618 & 0.6392 & -0.1057 & 0.0809 & 0.0679 & -0.9944 & -0.6808 & -0.7071 & -0.191 \\ 
        49 & -0.7522 & 0.6539 & -0.0811 & 0.0612 & 0.0532 & -0.9967 & -0.691 & -0.7071 & -0.15 \\ 
        48 & -0.7419 & 0.668 & -0.0585 & 0.0434 & 0.0391 & -0.9983 & -0.6984 & -0.7071 & -0.1105 \\ 
        47 & -0.7308 & 0.6815 & -0.0375 & 0.0274 & 0.0256 & -0.9993 & -0.7034 & -0.7071 & -0.0723 \\ 
        46 & -0.7192 & 0.6945 & -0.0181 & 0.013 & 0.0126 & -0.9998 & -0.7062 & -0.7071 & -0.0355 \\ 
        45 & -0.7071 & 0.7071 & 0 & 0 & 0 & -1 & 0.7071 & 0.7071 & 0 \\ 
        40 & -0.641 & 0.7639 & 0.0742 & -0.0477 & -0.0568 & -0.9972 & 0.6887 & 0.7071 & -0.1603 \\ 
        35 & -0.5688 & 0.8124 & 0.1285 & -0.0737 & -0.1053 & -0.9917 & 0.6425 & 0.7071 & -0.2952 \\ 
        30 & -0.4928 & 0.8536 & 0.1691 & -0.0846 & -0.1464 & -0.9856 & 0.5774 & 0.7071 & -0.4082 \\ 
        25 & -0.4141 & 0.8881 & 0.1997 & -0.0844 & -0.181 & -0.9799 & 0.4985 & 0.7071 & -0.5015 \\ 
        20 & -0.3335 & 0.9161 & 0.2225 & -0.0761 & -0.209 & -0.9749 & 0.4095 & 0.7071 & -0.5764 \\ 
        15 & -0.2513 & 0.9379 & 0.239 & -0.0619 & -0.2308 & -0.971 & 0.3132 & 0.7071 & -0.634 \\ 
        10 & -0.1681 & 0.9535 & 0.2502 & -0.0434 & -0.2464 & -0.9682 & 0.2116 & 0.7071 & -0.6747 \\ 
        5 & -0.0842 & 0.9628 & 0.2567 & -0.0224 & -0.2557 & -0.9665 & 0.1066 & 0.7071 & -0.699 \\ 
        \hline
    \end{tabular}
\end{table*}
\endgroup


\begin{thebibliography}{119}%
	\makeatletter
	\providecommand \@ifxundefined [1]{%
		\@ifx{#1\undefined}
	}%
	\providecommand \@ifnum [1]{%
		\ifnum #1\expandafter \@firstoftwo
		\else \expandafter \@secondoftwo
		\fi
	}%
	\providecommand \@ifx [1]{%
		\ifx #1\expandafter \@firstoftwo
		\else \expandafter \@secondoftwo
		\fi
	}%
	\providecommand \natexlab [1]{#1}%
	\providecommand \enquote  [1]{``#1''}%
	\providecommand \bibnamefont  [1]{#1}%
	\providecommand \bibfnamefont [1]{#1}%
	\providecommand \citenamefont [1]{#1}%
	\providecommand \href@noop [0]{\@secondoftwo}%
	\providecommand \href [0]{\begingroup \@sanitize@url \@href}%
	\providecommand \@href[1]{\@@startlink{#1}\@@href}%
	\providecommand \@@href[1]{\endgroup#1\@@endlink}%
	\providecommand \@sanitize@url [0]{\catcode `\\12\catcode `\$12\catcode
		`\&12\catcode `\#12\catcode `\^12\catcode `\_12\catcode `\%12\relax}%
	\providecommand \@@startlink[1]{}%
	\providecommand \@@endlink[0]{}%
	\providecommand \url  [0]{\begingroup\@sanitize@url \@url }%
	\providecommand \@url [1]{\endgroup\@href {#1}{\urlprefix }}%
	\providecommand \urlprefix  [0]{URL }%
	\providecommand \Eprint [0]{\href }%
	\providecommand \doibase [0]{https://doi.org/}%
	\providecommand \selectlanguage [0]{\@gobble}%
	\providecommand \bibinfo  [0]{\@secondoftwo}%
	\providecommand \bibfield  [0]{\@secondoftwo}%
	\providecommand \translation [1]{[#1]}%
	\providecommand \BibitemOpen [0]{}%
	\providecommand \bibitemStop [0]{}%
	\providecommand \bibitemNoStop [0]{.\EOS\space}%
	\providecommand \EOS [0]{\spacefactor3000\relax}%
	\providecommand \BibitemShut  [1]{\csname bibitem#1\endcsname}%
	\let\auto@bib@innerbib\@empty
	\bibitem [{\citenamefont {Heisenberg}\ and\ \citenamefont
		{Euler}(1936)}]{Heisenberg_1936}%
	\BibitemOpen
	\bibfield  {author} {\bibinfo {author} {\bibfnamefont {W.}~\bibnamefont
			{Heisenberg}}\ and\ \bibinfo {author} {\bibfnamefont {H.}~\bibnamefont
			{Euler}},\ }\href@noop {} {\bibfield  {journal} {\bibinfo  {journal} {Z.
				Phys.}\ }\textbf {\bibinfo {volume} {98}},\ \bibinfo {pages} {714} (\bibinfo
		{year} {1936})}\BibitemShut {NoStop}%
	\bibitem [{\citenamefont {Weisskopf}(1936)}]{Weisskopf_1936}%
	\BibitemOpen
	\bibfield  {author} {\bibinfo {author} {\bibfnamefont {V.}~\bibnamefont
			{Weisskopf}},\ }\href@noop {} {\bibfield  {journal} {\bibinfo  {journal} {K.
				Dan. Vidensk. Selsk. Mat. Fys. Medd.}\ }\textbf {\bibinfo {volume} {14}},\
		\bibinfo {pages} {1} (\bibinfo {year} {1936})}\BibitemShut {NoStop}%
	\bibitem [{\citenamefont {Schwinger}(1951)}]{Schwinger_1951}%
	\BibitemOpen
	\bibfield  {author} {\bibinfo {author} {\bibfnamefont {J.}~\bibnamefont
			{Schwinger}},\ }\href@noop {} {\bibfield  {journal} {\bibinfo  {journal}
			{Phys. Rev.}\ }\textbf {\bibinfo {volume} {82}},\ \bibinfo {pages} {664}
		(\bibinfo {year} {1951})}\BibitemShut {NoStop}%
	\bibitem [{\citenamefont {Jackson}(1975)}]{Jackson_b_1975}%
	\BibitemOpen
	\bibfield  {author} {\bibinfo {author} {\bibfnamefont {J.~D.}\ \bibnamefont
			{Jackson}},\ }\href@noop {} {\emph {\bibinfo {title} {Classical
				Electrodynamics}}}\ (\bibinfo  {publisher} {John Wiley \& Sons, New York},\
	\bibinfo {year} {1975})\BibitemShut {NoStop}%
	\bibitem [{Note1()}]{Note1}%
	\BibitemOpen
	\bibinfo {note} {The leading-order derivative corrections to the
		Euler-Heisenberg Lagrangian density have been computed in Ref. \cite
		{JMP:Gusynin:1999}}\BibitemShut {NoStop}%
	\bibitem [{\citenamefont {Yoon}\ \emph {et~al.}(2019)\citenamefont {Yoon},
		\citenamefont {Jeon}, \citenamefont {Shin}, \citenamefont {Lee},
		\citenamefont {Lee}, \citenamefont {Choi}, \citenamefont {Kim}, \citenamefont
		{Sung},\ and\ \citenamefont {Nam}}]{Yoon_2019}%
	\BibitemOpen
	\bibfield  {author} {\bibinfo {author} {\bibfnamefont {J.~W.}\ \bibnamefont
			{Yoon}}, \bibinfo {author} {\bibfnamefont {C.}~\bibnamefont {Jeon}}, \bibinfo
		{author} {\bibfnamefont {J.}~\bibnamefont {Shin}}, \bibinfo {author}
		{\bibfnamefont {S.~K.}\ \bibnamefont {Lee}}, \bibinfo {author} {\bibfnamefont
			{H.~W.}\ \bibnamefont {Lee}}, \bibinfo {author} {\bibfnamefont {I.~W.}\
			\bibnamefont {Choi}}, \bibinfo {author} {\bibfnamefont {H.~T.}\ \bibnamefont
			{Kim}}, \bibinfo {author} {\bibfnamefont {J.~H.}\ \bibnamefont {Sung}},\ and\
		\bibinfo {author} {\bibfnamefont {C.~H.}\ \bibnamefont {Nam}},\ }\href@noop
	{} {\bibfield  {journal} {\bibinfo  {journal} {Opt. Express}\ }\textbf
		{\bibinfo {volume} {27}},\ \bibinfo {pages} {20412} (\bibinfo {year}
		{2019})}\BibitemShut {NoStop}%
	\bibitem [{\citenamefont {Papadopoulos}\ \emph {et~al.}(2016)\citenamefont
		{Papadopoulos}, \citenamefont {Zou}, \citenamefont {Le~Blanc}, \citenamefont
		{Ch\'{e}riaux}, \citenamefont {Georges}, \citenamefont {Druon}, \citenamefont
		{Mennerat}, \citenamefont {Ramirez}, \citenamefont {Martin}, \citenamefont
		{Fr\'{e}neaux}, \citenamefont {Beluze}, \citenamefont {Lebas}, \citenamefont
		{Monot}, \citenamefont {Mathieu},\ and\ \citenamefont
		{Audebert}}]{APOLLON_10P}%
	\BibitemOpen
	\bibfield  {author} {\bibinfo {author} {\bibfnamefont {D.~N.}\ \bibnamefont
			{Papadopoulos}}, \bibinfo {author} {\bibfnamefont {J.~P.}\ \bibnamefont
			{Zou}}, \bibinfo {author} {\bibfnamefont {C.}~\bibnamefont {Le~Blanc}},
		\bibinfo {author} {\bibfnamefont {G.}~\bibnamefont {Ch\'{e}riaux}}, \bibinfo
		{author} {\bibfnamefont {P.}~\bibnamefont {Georges}}, \bibinfo {author}
		{\bibfnamefont {F.}~\bibnamefont {Druon}}, \bibinfo {author} {\bibfnamefont
			{G.}~\bibnamefont {Mennerat}}, \bibinfo {author} {\bibfnamefont
			{P.}~\bibnamefont {Ramirez}}, \bibinfo {author} {\bibfnamefont
			{L.}~\bibnamefont {Martin}}, \bibinfo {author} {\bibfnamefont
			{A.}~\bibnamefont {Fr\'{e}neaux}}, \bibinfo {author} {\bibfnamefont
			{A.}~\bibnamefont {Beluze}}, \bibinfo {author} {\bibfnamefont
			{N.}~\bibnamefont {Lebas}}, \bibinfo {author} {\bibfnamefont
			{P.}~\bibnamefont {Monot}}, \bibinfo {author} {\bibfnamefont
			{F.}~\bibnamefont {Mathieu}},\ and\ \bibinfo {author} {\bibfnamefont
			{P.}~\bibnamefont {Audebert}},\ }\href@noop {} {\bibfield  {journal}
		{\bibinfo  {journal} {High Power Laser Sci. Eng.}\ }\textbf {\bibinfo
			{volume} {4}},\ \bibinfo {pages} {e34} (\bibinfo {year} {2016})}\BibitemShut
	{NoStop}%
	\bibitem [{ELI()}]{ELI}%
	\BibitemOpen
	\href@noop {} {\enquote {\bibinfo {title} {{Extreme Light Infrastructure
					(ELI)}},}\ }\bibinfo {howpublished} {\url{https://eli-laser.eu/}}\BibitemShut
	{NoStop}%
	\bibitem [{CoR()}]{CoReLS}%
	\BibitemOpen
	\href@noop {} {\enquote {\bibinfo {title} {{Center for Relativistic Laser
					Science (CoReLS)}},}\ }\bibinfo {howpublished}
	{\url{https://www.ibs.re.kr/eng/sub02_03_05.do}}\BibitemShut {NoStop}%
	\bibitem [{\citenamefont {Bromage}\ \emph {et~al.}(2019)\citenamefont
		{Bromage}, \citenamefont {Bahk}, \citenamefont {Begishev}, \citenamefont
		{Dorrer}, \citenamefont {Guardalben}, \citenamefont {Hoffman}, \citenamefont
		{Oliver}, \citenamefont {Roides}, \citenamefont {Schiesser}, \citenamefont
		{Shoup~III}, \citenamefont {Spilatro}, \citenamefont {Webb}, \citenamefont
		{Weiner},\ and\ \citenamefont {Zuegel}}]{Bromage_2019}%
	\BibitemOpen
	\bibfield  {author} {\bibinfo {author} {\bibfnamefont {J.}~\bibnamefont
			{Bromage}}, \bibinfo {author} {\bibfnamefont {S.-W.}\ \bibnamefont {Bahk}},
		\bibinfo {author} {\bibfnamefont {I.~A.}\ \bibnamefont {Begishev}}, \bibinfo
		{author} {\bibfnamefont {C.}~\bibnamefont {Dorrer}}, \bibinfo {author}
		{\bibfnamefont {M.~J.}\ \bibnamefont {Guardalben}}, \bibinfo {author}
		{\bibfnamefont {B.~N.}\ \bibnamefont {Hoffman}}, \bibinfo {author}
		{\bibfnamefont {J.~B.}\ \bibnamefont {Oliver}}, \bibinfo {author}
		{\bibfnamefont {R.~G.}\ \bibnamefont {Roides}}, \bibinfo {author}
		{\bibfnamefont {E.~M.}\ \bibnamefont {Schiesser}}, \bibinfo {author}
		{\bibfnamefont {M.~J.}\ \bibnamefont {Shoup~III}}, \bibinfo {author}
		{\bibfnamefont {M.}~\bibnamefont {Spilatro}}, \bibinfo {author}
		{\bibfnamefont {B.}~\bibnamefont {Webb}}, \bibinfo {author} {\bibfnamefont
			{D.}~\bibnamefont {Weiner}},\ and\ \bibinfo {author} {\bibfnamefont {J.~D.}\
			\bibnamefont {Zuegel}},\ }\href@noop {} {\bibfield  {journal} {\bibinfo
			{journal} {High Power Laser Sci. Eng.}\ }\textbf {\bibinfo {volume} {7}},\
		\bibinfo {pages} {e4} (\bibinfo {year} {2019})}\BibitemShut {NoStop}%
	\bibitem [{XCE()}]{XCELS}%
	\BibitemOpen
	\href@noop {} {\enquote {\bibinfo {title} {{Exawatt Center for Extreme Light
					Studies (XCELS)}},}\ }\bibinfo {howpublished}
	{\url{https://www.cremlinplus.eu/collaboration/russian_megascience_projects/xcels/}}\BibitemShut
	{NoStop}%
	\bibitem [{\citenamefont {Di~Piazza}, \citenamefont {Willingale},\ and\
		\citenamefont {Zuegel}(2022)}]{Di_Piazza_2022}%
	\BibitemOpen
	\bibfield  {author} {\bibinfo {author} {\bibfnamefont {A.}~\bibnamefont
			{Di~Piazza}}, \bibinfo {author} {\bibfnamefont {L.}~\bibnamefont
			{Willingale}},\ and\ \bibinfo {author} {\bibfnamefont {J.~D.}\ \bibnamefont
			{Zuegel}},\ }\href@noop {} {} (\bibinfo {year} {2022}),\ \Eprint
	{https://arxiv.org/abs/2211.13187} {arXiv:2211.13187} \BibitemShut {NoStop}%
	\bibitem [{\citenamefont {Euler}(1936)}]{Euler_1936_a}%
	\BibitemOpen
	\bibfield  {author} {\bibinfo {author} {\bibfnamefont {H.}~\bibnamefont
			{Euler}},\ }\href@noop {} {\bibfield  {journal} {\bibinfo  {journal} {Ann.
				Phys. (Leipzig)}\ }\textbf {\bibinfo {volume} {26}},\ \bibinfo {pages} {398}
		(\bibinfo {year} {1936})}\BibitemShut {NoStop}%
	\bibitem [{\citenamefont {Akhiezer}(1937)}]{Akhiezer_1937}%
	\BibitemOpen
	\bibfield  {author} {\bibinfo {author} {\bibfnamefont {A.~I.}\ \bibnamefont
			{Akhiezer}},\ }\href@noop {} {\bibfield  {journal} {\bibinfo  {journal}
			{Phys. Z. Sowjetunion}\ }\textbf {\bibinfo {volume} {11}},\ \bibinfo {pages}
		{263} (\bibinfo {year} {1937})}\BibitemShut {NoStop}%
	\bibitem [{\citenamefont {Karplus}\ and\ \citenamefont
		{Neuman}(1950)}]{Karplus_1951}%
	\BibitemOpen
	\bibfield  {author} {\bibinfo {author} {\bibfnamefont {R.}~\bibnamefont
			{Karplus}}\ and\ \bibinfo {author} {\bibfnamefont {M.}~\bibnamefont
			{Neuman}},\ }\href@noop {} {\bibfield  {journal} {\bibinfo  {journal} {Phys.
				Rev.}\ }\textbf {\bibinfo {volume} {80}},\ \bibinfo {pages} {380} (\bibinfo
		{year} {1950})}\BibitemShut {NoStop}%
	\bibitem [{\citenamefont {De~Tollis}(1964)}]{De_Tollis_1964}%
	\BibitemOpen
	\bibfield  {author} {\bibinfo {author} {\bibfnamefont {B.}~\bibnamefont
			{De~Tollis}},\ }\href@noop {} {\bibfield  {journal} {\bibinfo  {journal}
			{Nuovo Cim.}\ }\textbf {\bibinfo {volume} {32}},\ \bibinfo {pages} {757}
		(\bibinfo {year} {1964})}\BibitemShut {NoStop}%
	\bibitem [{\citenamefont {Ahmadiniaz}\ \emph
		{et~al.}(2023{\natexlab{a}})\citenamefont {Ahmadiniaz}, \citenamefont
		{Lopez-Arcos}, \citenamefont {Lopez-Lopez},\ and\ \citenamefont
		{Schubert}}]{Ahmadiniaz_2023}%
	\BibitemOpen
	\bibfield  {author} {\bibinfo {author} {\bibfnamefont {N.}~\bibnamefont
			{Ahmadiniaz}}, \bibinfo {author} {\bibfnamefont {C.}~\bibnamefont
			{Lopez-Arcos}}, \bibinfo {author} {\bibfnamefont {M.~A.}\ \bibnamefont
			{Lopez-Lopez}},\ and\ \bibinfo {author} {\bibfnamefont {C.}~\bibnamefont
			{Schubert}},\ }\href@noop {} {\bibfield  {journal} {\bibinfo  {journal}
			{Nucl. Phys. B}\ }\textbf {\bibinfo {volume} {991}},\ \bibinfo {pages}
		{116216} (\bibinfo {year} {2023}{\natexlab{a}})}\BibitemShut {NoStop}%
	\bibitem [{\citenamefont {Ahmadiniaz}\ \emph
		{et~al.}(2023{\natexlab{b}})\citenamefont {Ahmadiniaz}, \citenamefont
		{Lopez-Arcos}, \citenamefont {Lopez-Lopez},\ and\ \citenamefont
		{Schubert}}]{Ahmadiniaz_2023_b}%
	\BibitemOpen
	\bibfield  {author} {\bibinfo {author} {\bibfnamefont {N.}~\bibnamefont
			{Ahmadiniaz}}, \bibinfo {author} {\bibfnamefont {C.}~\bibnamefont
			{Lopez-Arcos}}, \bibinfo {author} {\bibfnamefont {M.~A.}\ \bibnamefont
			{Lopez-Lopez}},\ and\ \bibinfo {author} {\bibfnamefont {C.}~\bibnamefont
			{Schubert}},\ }\href@noop {} {\bibfield  {journal} {\bibinfo  {journal}
			{Nucl. Phys. B}\ }\textbf {\bibinfo {volume} {991}},\ \bibinfo {pages}
		{116217} (\bibinfo {year} {2023}{\natexlab{b}})}\BibitemShut {NoStop}%
	\bibitem [{\citenamefont {Aleksandrov}, \citenamefont {Ansel’m},\ and\
		\citenamefont {Moskalev}(1985)}]{Aleksandrov_1985}%
	\BibitemOpen
	\bibfield  {author} {\bibinfo {author} {\bibfnamefont {E.}~\bibnamefont
			{Aleksandrov}}, \bibinfo {author} {\bibfnamefont {A.}~\bibnamefont
			{Ansel’m}},\ and\ \bibinfo {author} {\bibfnamefont {A.}~\bibnamefont
			{Moskalev}},\ }\bibfield  {title} {\enquote {\bibinfo {title} {Vacuum
				birefringence in an intense laser radiation field},}\ }\href@noop {}
	{\bibfield  {journal} {\bibinfo  {journal} {Sov. Phys. JETP}\ }\textbf
		{\bibinfo {volume} {62}},\ \bibinfo {pages} {680} (\bibinfo {year}
		{1985})}\BibitemShut {NoStop}%
	\bibitem [{\citenamefont {Heinzl}\ \emph {et~al.}(2006)\citenamefont {Heinzl},
		\citenamefont {Liesfeld}, \citenamefont {Amthor}, \citenamefont {Schwoerer},
		\citenamefont {Sauerbrey},\ and\ \citenamefont {Wipf}}]{Heinzl_2006}%
	\BibitemOpen
	\bibfield  {author} {\bibinfo {author} {\bibfnamefont {T.}~\bibnamefont
			{Heinzl}}, \bibinfo {author} {\bibfnamefont {B.}~\bibnamefont {Liesfeld}},
		\bibinfo {author} {\bibfnamefont {K.-U.}\ \bibnamefont {Amthor}}, \bibinfo
		{author} {\bibfnamefont {H.}~\bibnamefont {Schwoerer}}, \bibinfo {author}
		{\bibfnamefont {R.}~\bibnamefont {Sauerbrey}},\ and\ \bibinfo {author}
		{\bibfnamefont {A.}~\bibnamefont {Wipf}},\ }\href@noop {} {\bibfield
		{journal} {\bibinfo  {journal} {Opt. Commun.}\ }\textbf {\bibinfo {volume}
			{267}},\ \bibinfo {pages} {318} (\bibinfo {year} {2006})}\BibitemShut
	{NoStop}%
	\bibitem [{\citenamefont {Di~Piazza}, \citenamefont {Hatsagortsyan},\ and\
		\citenamefont {Keitel}(2006)}]{Di_Piazza_2006}%
	\BibitemOpen
	\bibfield  {author} {\bibinfo {author} {\bibfnamefont {A.}~\bibnamefont
			{Di~Piazza}}, \bibinfo {author} {\bibfnamefont {K.~Z.}\ \bibnamefont
			{Hatsagortsyan}},\ and\ \bibinfo {author} {\bibfnamefont {C.~H.}\
			\bibnamefont {Keitel}},\ }\href@noop {} {\bibfield  {journal} {\bibinfo
			{journal} {Phys. Rev. Lett.}\ }\textbf {\bibinfo {volume} {97}},\ \bibinfo
		{pages} {083603} (\bibinfo {year} {2006})}\BibitemShut {NoStop}%
	\bibitem [{\citenamefont {Tommasini}\ and\ \citenamefont
		{Michinel}(2010)}]{Tommasini_2010}%
	\BibitemOpen
	\bibfield  {author} {\bibinfo {author} {\bibfnamefont {D.}~\bibnamefont
			{Tommasini}}\ and\ \bibinfo {author} {\bibfnamefont {H.}~\bibnamefont
			{Michinel}},\ }\href@noop {} {\bibfield  {journal} {\bibinfo  {journal}
			{Phys. Rev. A}\ }\textbf {\bibinfo {volume} {82}},\ \bibinfo {pages} {011803}
		(\bibinfo {year} {2010})}\BibitemShut {NoStop}%
	\bibitem [{\citenamefont {King}, \citenamefont {Di~Piazza},\ and\ \citenamefont
		{Keitel}(2010{\natexlab{a}})}]{King_2010_b}%
	\BibitemOpen
	\bibfield  {author} {\bibinfo {author} {\bibfnamefont {B.}~\bibnamefont
			{King}}, \bibinfo {author} {\bibfnamefont {A.}~\bibnamefont {Di~Piazza}},\
		and\ \bibinfo {author} {\bibfnamefont {C.~H.}\ \bibnamefont {Keitel}},\
	}\href@noop {} {\bibfield  {journal} {\bibinfo  {journal} {Phys. Rev. A}\
		}\textbf {\bibinfo {volume} {82}},\ \bibinfo {pages} {032114} (\bibinfo
		{year} {2010}{\natexlab{a}})}\BibitemShut {NoStop}%
	\bibitem [{\citenamefont {Homma}, \citenamefont {Habs},\ and\ \citenamefont
		{Tajima}(2011)}]{Homma_2011}%
	\BibitemOpen
	\bibfield  {author} {\bibinfo {author} {\bibfnamefont {K.}~\bibnamefont
			{Homma}}, \bibinfo {author} {\bibfnamefont {D.}~\bibnamefont {Habs}},\ and\
		\bibinfo {author} {\bibfnamefont {T.}~\bibnamefont {Tajima}},\ }\href@noop {}
	{\bibfield  {journal} {\bibinfo  {journal} {Appl. Phys. B}\ }\textbf
		{\bibinfo {volume} {104}},\ \bibinfo {pages} {769} (\bibinfo {year}
		{2011})}\BibitemShut {NoStop}%
	\bibitem [{\citenamefont {Dinu}\ \emph
		{et~al.}(2014{\natexlab{a}})\citenamefont {Dinu}, \citenamefont {Heinzl},
		\citenamefont {Ilderton}, \citenamefont {Marklund},\ and\ \citenamefont
		{Torgrimsson}}]{Dinu_2014}%
	\BibitemOpen
	\bibfield  {author} {\bibinfo {author} {\bibfnamefont {V.}~\bibnamefont
			{Dinu}}, \bibinfo {author} {\bibfnamefont {T.}~\bibnamefont {Heinzl}},
		\bibinfo {author} {\bibfnamefont {A.}~\bibnamefont {Ilderton}}, \bibinfo
		{author} {\bibfnamefont {M.}~\bibnamefont {Marklund}},\ and\ \bibinfo
		{author} {\bibfnamefont {G.}~\bibnamefont {Torgrimsson}},\ }\href@noop {}
	{\bibfield  {journal} {\bibinfo  {journal} {Phys. Rev. D}\ }\textbf {\bibinfo
			{volume} {89}},\ \bibinfo {pages} {125003} (\bibinfo {year}
		{2014}{\natexlab{a}})}\BibitemShut {NoStop}%
	\bibitem [{\citenamefont {Dinu}\ \emph
		{et~al.}(2014{\natexlab{b}})\citenamefont {Dinu}, \citenamefont {Heinzl},
		\citenamefont {Ilderton}, \citenamefont {Marklund},\ and\ \citenamefont
		{Torgrimsson}}]{Dinu_2014_b}%
	\BibitemOpen
	\bibfield  {author} {\bibinfo {author} {\bibfnamefont {V.}~\bibnamefont
			{Dinu}}, \bibinfo {author} {\bibfnamefont {T.}~\bibnamefont {Heinzl}},
		\bibinfo {author} {\bibfnamefont {A.}~\bibnamefont {Ilderton}}, \bibinfo
		{author} {\bibfnamefont {M.}~\bibnamefont {Marklund}},\ and\ \bibinfo
		{author} {\bibfnamefont {G.}~\bibnamefont {Torgrimsson}},\ }\href@noop {}
	{\bibfield  {journal} {\bibinfo  {journal} {Phys. Rev. D}\ }\textbf {\bibinfo
			{volume} {90}},\ \bibinfo {pages} {045025} (\bibinfo {year}
		{2014}{\natexlab{b}})}\BibitemShut {NoStop}%
	\bibitem [{\citenamefont {King}\ and\ \citenamefont {Hu}(2016)}]{King_2016}%
	\BibitemOpen
	\bibfield  {author} {\bibinfo {author} {\bibfnamefont {B.}~\bibnamefont
			{King}}\ and\ \bibinfo {author} {\bibfnamefont {H.}~\bibnamefont {Hu}},\
	}\href@noop {} {\bibfield  {journal} {\bibinfo  {journal} {Phys. Rev. D}\
		}\textbf {\bibinfo {volume} {94}},\ \bibinfo {pages} {125010} (\bibinfo
		{year} {2016})}\BibitemShut {NoStop}%
	\bibitem [{\citenamefont {Bragin}\ \emph {et~al.}(2017)\citenamefont {Bragin},
		\citenamefont {Meuren}, \citenamefont {Keitel},\ and\ \citenamefont
		{Di~Piazza}}]{Bragin_2017}%
	\BibitemOpen
	\bibfield  {author} {\bibinfo {author} {\bibfnamefont {S.}~\bibnamefont
			{Bragin}}, \bibinfo {author} {\bibfnamefont {S.}~\bibnamefont {Meuren}},
		\bibinfo {author} {\bibfnamefont {C.~H.}\ \bibnamefont {Keitel}},\ and\
		\bibinfo {author} {\bibfnamefont {A.}~\bibnamefont {Di~Piazza}},\ }\href@noop
	{} {\bibfield  {journal} {\bibinfo  {journal} {Phys. Rev. Lett.}\ }\textbf
		{\bibinfo {volume} {119}},\ \bibinfo {pages} {250403} (\bibinfo {year}
		{2017})}\BibitemShut {NoStop}%
	\bibitem [{\citenamefont {Gies}, \citenamefont {Karbstein},\ and\ \citenamefont
		{Kohlf\"urst}(2018)}]{Gies_2018}%
	\BibitemOpen
	\bibfield  {author} {\bibinfo {author} {\bibfnamefont {H.}~\bibnamefont
			{Gies}}, \bibinfo {author} {\bibfnamefont {F.}~\bibnamefont {Karbstein}},\
		and\ \bibinfo {author} {\bibfnamefont {C.}~\bibnamefont {Kohlf\"urst}},\
	}\href@noop {} {\bibfield  {journal} {\bibinfo  {journal} {Phys. Rev. D}\
		}\textbf {\bibinfo {volume} {97}},\ \bibinfo {pages} {036022} (\bibinfo
		{year} {2018})}\BibitemShut {NoStop}%
	\bibitem [{\citenamefont {Karbstein}(2018)}]{Karbstein_2018}%
	\BibitemOpen
	\bibfield  {author} {\bibinfo {author} {\bibfnamefont {F.}~\bibnamefont
			{Karbstein}},\ }\href@noop {} {\bibfield  {journal} {\bibinfo  {journal}
			{Phys. Rev. D}\ }\textbf {\bibinfo {volume} {98}},\ \bibinfo {pages} {056010}
		(\bibinfo {year} {2018})}\BibitemShut {NoStop}%
	\bibitem [{\citenamefont {King}, \citenamefont {Hu},\ and\ \citenamefont
		{Shen}(2018{\natexlab{a}})}]{King_2018_b}%
	\BibitemOpen
	\bibfield  {author} {\bibinfo {author} {\bibfnamefont {B.}~\bibnamefont
			{King}}, \bibinfo {author} {\bibfnamefont {H.}~\bibnamefont {Hu}},\ and\
		\bibinfo {author} {\bibfnamefont {B.}~\bibnamefont {Shen}},\ }\href@noop {}
	{\bibfield  {journal} {\bibinfo  {journal} {Phys. Rev. A}\ }\textbf {\bibinfo
			{volume} {98}},\ \bibinfo {pages} {023817} (\bibinfo {year}
		{2018}{\natexlab{a}})}\BibitemShut {NoStop}%
	\bibitem [{\citenamefont {Pegoraro}\ and\ \citenamefont
		{Bulanov}(2019)}]{Pegoraro_2019_b}%
	\BibitemOpen
	\bibfield  {author} {\bibinfo {author} {\bibfnamefont {F.}~\bibnamefont
			{Pegoraro}}\ and\ \bibinfo {author} {\bibfnamefont {S.~V.}\ \bibnamefont
			{Bulanov}},\ }\href@noop {} {\bibfield  {journal} {\bibinfo  {journal} {Phys.
				Rev. D}\ }\textbf {\bibinfo {volume} {100}},\ \bibinfo {pages} {036004}
		(\bibinfo {year} {2019})}\BibitemShut {NoStop}%
	\bibitem [{\citenamefont {Bulanov}\ \emph {et~al.}(2020)\citenamefont
		{Bulanov}, \citenamefont {Sasorov}, \citenamefont {Pegoraro}, \citenamefont
		{Kadlecov\'a}, \citenamefont {Bulanov}, \citenamefont {Esirkepov},
		\citenamefont {Rosanov},\ and\ \citenamefont {Korn}}]{Bulanov_2020}%
	\BibitemOpen
	\bibfield  {author} {\bibinfo {author} {\bibfnamefont {S.~V.}\ \bibnamefont
			{Bulanov}}, \bibinfo {author} {\bibfnamefont {P.~V.}\ \bibnamefont
			{Sasorov}}, \bibinfo {author} {\bibfnamefont {F.}~\bibnamefont {Pegoraro}},
		\bibinfo {author} {\bibfnamefont {H.}~\bibnamefont {Kadlecov\'a}}, \bibinfo
		{author} {\bibfnamefont {S.~S.}\ \bibnamefont {Bulanov}}, \bibinfo {author}
		{\bibfnamefont {T.~Z.}\ \bibnamefont {Esirkepov}}, \bibinfo {author}
		{\bibfnamefont {N.~N.}\ \bibnamefont {Rosanov}},\ and\ \bibinfo {author}
		{\bibfnamefont {G.}~\bibnamefont {Korn}},\ }\href@noop {} {\bibfield
		{journal} {\bibinfo  {journal} {Phys. Rev. D}\ }\textbf {\bibinfo {volume}
			{101}},\ \bibinfo {pages} {016016} (\bibinfo {year} {2020})}\BibitemShut
	{NoStop}%
	\bibitem [{\citenamefont {Ahmadiniaz}\ \emph {et~al.}(2020)\citenamefont
		{Ahmadiniaz}, \citenamefont {Cowan}, \citenamefont {Sauerbrey}, \citenamefont
		{Schramm}, \citenamefont {Schlenvoigt},\ and\ \citenamefont
		{Sch\"utzhold}}]{Ahmadiniaz_2020}%
	\BibitemOpen
	\bibfield  {author} {\bibinfo {author} {\bibfnamefont {N.}~\bibnamefont
			{Ahmadiniaz}}, \bibinfo {author} {\bibfnamefont {T.~E.}\ \bibnamefont
			{Cowan}}, \bibinfo {author} {\bibfnamefont {R.}~\bibnamefont {Sauerbrey}},
		\bibinfo {author} {\bibfnamefont {U.}~\bibnamefont {Schramm}}, \bibinfo
		{author} {\bibfnamefont {H.-P.}\ \bibnamefont {Schlenvoigt}},\ and\ \bibinfo
		{author} {\bibfnamefont {R.}~\bibnamefont {Sch\"utzhold}},\ }\href@noop {}
	{\bibfield  {journal} {\bibinfo  {journal} {Phys. Rev. D}\ }\textbf {\bibinfo
			{volume} {101}},\ \bibinfo {pages} {116019} (\bibinfo {year}
		{2020})}\BibitemShut {NoStop}%
	\bibitem [{\citenamefont {Karbstein}\ \emph {et~al.}(2021)\citenamefont
		{Karbstein}, \citenamefont {Sundqvist}, \citenamefont {Schulze},
		\citenamefont {Uschmann}, \citenamefont {Gies},\ and\ \citenamefont
		{Paulus}}]{Karbstein_2021}%
	\BibitemOpen
	\bibfield  {author} {\bibinfo {author} {\bibfnamefont {F.}~\bibnamefont
			{Karbstein}}, \bibinfo {author} {\bibfnamefont {C.}~\bibnamefont
			{Sundqvist}}, \bibinfo {author} {\bibfnamefont {K.~S.}\ \bibnamefont
			{Schulze}}, \bibinfo {author} {\bibfnamefont {I.}~\bibnamefont {Uschmann}},
		\bibinfo {author} {\bibfnamefont {H.}~\bibnamefont {Gies}},\ and\ \bibinfo
		{author} {\bibfnamefont {G.~G.}\ \bibnamefont {Paulus}},\ }\href@noop {}
	{\bibfield  {journal} {\bibinfo  {journal} {New J. Phys.}\ }\textbf {\bibinfo
			{volume} {23}},\ \bibinfo {pages} {095001} (\bibinfo {year}
		{2021})}\BibitemShut {NoStop}%
	\bibitem [{\citenamefont {Ataman}(2018)}]{Ataman_2021}%
	\BibitemOpen
	\bibfield  {author} {\bibinfo {author} {\bibfnamefont {S.}~\bibnamefont
			{Ataman}},\ }\href@noop {} {\bibfield  {journal} {\bibinfo  {journal} {Phys.
				Rev. A}\ }\textbf {\bibinfo {volume} {97}},\ \bibinfo {pages} {063811}
		(\bibinfo {year} {2018})}\BibitemShut {NoStop}%
	\bibitem [{\citenamefont {Mosman}\ and\ \citenamefont
		{Karbstein}(2021)}]{PRD:Mosman:2021}%
	\BibitemOpen
	\bibfield  {author} {\bibinfo {author} {\bibfnamefont {E.~A.}\ \bibnamefont
			{Mosman}}\ and\ \bibinfo {author} {\bibfnamefont {F.}~\bibnamefont
			{Karbstein}},\ }\bibfield  {title} {\enquote {\bibinfo {title} {Vacuum
				birefringence and diffraction at an x-ray free-electron laser: {From}
				analytical estimates to optimal parameters},}\ }\href
	{https://doi.org/10.1103/PhysRevD.104.013006} {\bibfield  {journal} {\bibinfo
			{journal} {Physical Review D}\ }\textbf {\bibinfo {volume} {104}},\ \bibinfo
		{pages} {013006} (\bibinfo {year} {2021})}\BibitemShut {NoStop}%
	\bibitem [{\citenamefont {Ahmadiniaz}\ \emph {et~al.}(2021)\citenamefont
		{Ahmadiniaz}, \citenamefont {Bussmann}, \citenamefont {Cowan}, \citenamefont
		{Debus}, \citenamefont {Kluge},\ and\ \citenamefont
		{Sch\"utzhold}}]{Ahmadiniaz_2021}%
	\BibitemOpen
	\bibfield  {author} {\bibinfo {author} {\bibfnamefont {N.}~\bibnamefont
			{Ahmadiniaz}}, \bibinfo {author} {\bibfnamefont {M.}~\bibnamefont
			{Bussmann}}, \bibinfo {author} {\bibfnamefont {T.~E.}\ \bibnamefont {Cowan}},
		\bibinfo {author} {\bibfnamefont {A.}~\bibnamefont {Debus}}, \bibinfo
		{author} {\bibfnamefont {T.}~\bibnamefont {Kluge}},\ and\ \bibinfo {author}
		{\bibfnamefont {R.}~\bibnamefont {Sch\"utzhold}},\ }\href@noop {} {\bibfield
		{journal} {\bibinfo  {journal} {Phys. Rev. D}\ }\textbf {\bibinfo {volume}
			{104}},\ \bibinfo {pages} {L011902} (\bibinfo {year} {2021})}\BibitemShut
	{NoStop}%
	\bibitem [{\citenamefont {Jin}, \citenamefont {Shen},\ and\ \citenamefont
		{Xu}(2022)}]{Jin_2022}%
	\BibitemOpen
	\bibfield  {author} {\bibinfo {author} {\bibfnamefont {B.}~\bibnamefont
			{Jin}}, \bibinfo {author} {\bibfnamefont {B.}~\bibnamefont {Shen}},\ and\
		\bibinfo {author} {\bibfnamefont {D.}~\bibnamefont {Xu}},\ }\href@noop {}
	{\bibfield  {journal} {\bibinfo  {journal} {Phys. Rev. A}\ }\textbf {\bibinfo
			{volume} {106}},\ \bibinfo {pages} {013502} (\bibinfo {year}
		{2022})}\BibitemShut {NoStop}%
	\bibitem [{\citenamefont {Sainte-Marie}\ \emph {et~al.}(2022)\citenamefont
		{Sainte-Marie}, \citenamefont {Fedeli}, \citenamefont {Za\"{i}m},
		\citenamefont {Karbstein},\ and\ \citenamefont
		{Vincenti}}]{Sainte-Marie_2022}%
	\BibitemOpen
	\bibfield  {author} {\bibinfo {author} {\bibfnamefont {A.}~\bibnamefont
			{Sainte-Marie}}, \bibinfo {author} {\bibfnamefont {L.}~\bibnamefont
			{Fedeli}}, \bibinfo {author} {\bibfnamefont {N.}~\bibnamefont {Za\"{i}m}},
		\bibinfo {author} {\bibfnamefont {F.}~\bibnamefont {Karbstein}},\ and\
		\bibinfo {author} {\bibfnamefont {H.}~\bibnamefont {Vincenti}},\ }\href@noop
	{} {\bibfield  {journal} {\bibinfo  {journal} {New J. Phys.}\ }\textbf
		{\bibinfo {volume} {24}},\ \bibinfo {pages} {065005} (\bibinfo {year}
		{2022})}\BibitemShut {NoStop}%
	\bibitem [{\citenamefont {Karbstein}\ \emph
		{et~al.}(2022{\natexlab{a}})\citenamefont {Karbstein}, \citenamefont
		{Ullmann}, \citenamefont {Mosman},\ and\ \citenamefont
		{Zepf}}]{Karbstein_2022}%
	\BibitemOpen
	\bibfield  {author} {\bibinfo {author} {\bibfnamefont {F.}~\bibnamefont
			{Karbstein}}, \bibinfo {author} {\bibfnamefont {D.}~\bibnamefont {Ullmann}},
		\bibinfo {author} {\bibfnamefont {E.~A.}\ \bibnamefont {Mosman}},\ and\
		\bibinfo {author} {\bibfnamefont {M.}~\bibnamefont {Zepf}},\ }\href@noop {}
	{\bibfield  {journal} {\bibinfo  {journal} {Phys. Rev. Lett.}\ }\textbf
		{\bibinfo {volume} {129}},\ \bibinfo {pages} {061802} (\bibinfo {year}
		{2022}{\natexlab{a}})}\BibitemShut {NoStop}%
	\bibitem [{\citenamefont {Aleksandrov}\ and\ \citenamefont
		{Shabaev}()}]{Aleksandrov_2023}%
	\BibitemOpen
	\bibfield  {author} {\bibinfo {author} {\bibfnamefont {I.~A.}\ \bibnamefont
			{Aleksandrov}}\ and\ \bibinfo {author} {\bibfnamefont {V.~M.}\ \bibnamefont
			{Shabaev}},\ }\href@noop {} {}\Eprint {https://arxiv.org/abs/2303.16273}
	{arXiv:2303.16273} \BibitemShut {NoStop}%
	\bibitem [{\citenamefont {Macleod}\ \emph {et~al.}(2023)\citenamefont
		{Macleod}, \citenamefont {Edwards}, \citenamefont {Heinzl}, \citenamefont
		{King},\ and\ \citenamefont {Bulanov}}]{Macleod_2023}%
	\BibitemOpen
	\bibfield  {author} {\bibinfo {author} {\bibfnamefont {A.~J.}\ \bibnamefont
			{Macleod}}, \bibinfo {author} {\bibfnamefont {J.~P.}\ \bibnamefont
			{Edwards}}, \bibinfo {author} {\bibfnamefont {T.}~\bibnamefont {Heinzl}},
		\bibinfo {author} {\bibfnamefont {B.}~\bibnamefont {King}},\ and\ \bibinfo
		{author} {\bibfnamefont {S.~V.}\ \bibnamefont {Bulanov}},\ }\href@noop {}
	{\bibfield  {journal} {\bibinfo  {journal} {New J. Phys.}\ }\textbf {\bibinfo
			{volume} {25}},\ \bibinfo {pages} {093002} (\bibinfo {year}
		{2023})}\BibitemShut {NoStop}%
	\bibitem [{\citenamefont {Formanek}\ \emph {et~al.}(2024)\citenamefont
		{Formanek}, \citenamefont {Palastro}, \citenamefont {Ramsey}, \citenamefont
		{Weber},\ and\ \citenamefont {Di~Piazza}}]{Formanek_2024_b}%
	\BibitemOpen
	\bibfield  {author} {\bibinfo {author} {\bibfnamefont {M.}~\bibnamefont
			{Formanek}}, \bibinfo {author} {\bibfnamefont {J.~P.}\ \bibnamefont
			{Palastro}}, \bibinfo {author} {\bibfnamefont {D.}~\bibnamefont {Ramsey}},
		\bibinfo {author} {\bibfnamefont {S.}~\bibnamefont {Weber}},\ and\ \bibinfo
		{author} {\bibfnamefont {A.}~\bibnamefont {Di~Piazza}},\ }\href@noop {}
	{\bibfield  {journal} {\bibinfo  {journal} {Phys. Rev. D}\ }\textbf {\bibinfo
			{volume} {109}},\ \bibinfo {pages} {056009} (\bibinfo {year}
		{2024})}\BibitemShut {NoStop}%
	\bibitem [{\citenamefont {Di~Piazza}, \citenamefont {Hatsagortsyan},\ and\
		\citenamefont {Keitel}(2005)}]{Di_Piazza_2005}%
	\BibitemOpen
	\bibfield  {author} {\bibinfo {author} {\bibfnamefont {A.}~\bibnamefont
			{Di~Piazza}}, \bibinfo {author} {\bibfnamefont {K.~Z.}\ \bibnamefont
			{Hatsagortsyan}},\ and\ \bibinfo {author} {\bibfnamefont {C.~H.}\
			\bibnamefont {Keitel}},\ }\href@noop {} {\bibfield  {journal} {\bibinfo
			{journal} {Phys. Rev. D}\ }\textbf {\bibinfo {volume} {72}},\ \bibinfo
		{pages} {085005} (\bibinfo {year} {2005})}\BibitemShut {NoStop}%
	\bibitem [{\citenamefont {Narozhny}\ and\ \citenamefont
		{Fedotov}(2006)}]{Narozhny_2007}%
	\BibitemOpen
	\bibfield  {author} {\bibinfo {author} {\bibfnamefont {N.~B.}\ \bibnamefont
			{Narozhny}}\ and\ \bibinfo {author} {\bibfnamefont {A.~M.}\ \bibnamefont
			{Fedotov}},\ }\href@noop {} {\bibfield  {journal} {\bibinfo  {journal} {Laser
				Phys.}\ }\textbf {\bibinfo {volume} {17}},\ \bibinfo {pages} {350} (\bibinfo
		{year} {2006})}\BibitemShut {NoStop}%
	\bibitem [{\citenamefont {Brodin}\ \emph {et~al.}(2007)\citenamefont {Brodin},
		\citenamefont {Marklund}, \citenamefont {Eliasson},\ and\ \citenamefont
		{Shukla}}]{Brodin_2007}%
	\BibitemOpen
	\bibfield  {author} {\bibinfo {author} {\bibfnamefont {G.}~\bibnamefont
			{Brodin}}, \bibinfo {author} {\bibfnamefont {M.}~\bibnamefont {Marklund}},
		\bibinfo {author} {\bibfnamefont {B.}~\bibnamefont {Eliasson}},\ and\
		\bibinfo {author} {\bibfnamefont {P.~K.}\ \bibnamefont {Shukla}},\
	}\href@noop {} {\bibfield  {journal} {\bibinfo  {journal} {Phys. Rev. Lett.}\
		}\textbf {\bibinfo {volume} {98}},\ \bibinfo {pages} {125001} (\bibinfo
		{year} {2007})}\BibitemShut {NoStop}%
	\bibitem [{\citenamefont {Di~Piazza}, \citenamefont {Milstein},\ and\
		\citenamefont {Keitel}(2007)}]{Di_Piazza_2007}%
	\BibitemOpen
	\bibfield  {author} {\bibinfo {author} {\bibfnamefont {A.}~\bibnamefont
			{Di~Piazza}}, \bibinfo {author} {\bibfnamefont {A.~I.}\ \bibnamefont
			{Milstein}},\ and\ \bibinfo {author} {\bibfnamefont {C.~H.}\ \bibnamefont
			{Keitel}},\ }\href@noop {} {\bibfield  {journal} {\bibinfo  {journal} {Phys.
				Rev. A}\ }\textbf {\bibinfo {volume} {76}},\ \bibinfo {pages} {032103}
		(\bibinfo {year} {2007})}\BibitemShut {NoStop}%
	\bibitem [{\citenamefont {Di~Piazza}(2013)}]{Di_Piazza_2013}%
	\BibitemOpen
	\bibfield  {author} {\bibinfo {author} {\bibfnamefont {A.}~\bibnamefont
			{Di~Piazza}},\ }\href@noop {} {\bibfield  {journal} {\bibinfo  {journal}
			{Ann. Phys. (N. Y.)}\ }\textbf {\bibinfo {volume} {338}},\ \bibinfo {pages}
		{302} (\bibinfo {year} {2013})}\BibitemShut {NoStop}%
	\bibitem [{\citenamefont {Gies}, \citenamefont {Karbstein},\ and\ \citenamefont
		{Shaisultanov}(2014)}]{Gies_2014_b}%
	\BibitemOpen
	\bibfield  {author} {\bibinfo {author} {\bibfnamefont {H.}~\bibnamefont
			{Gies}}, \bibinfo {author} {\bibfnamefont {F.}~\bibnamefont {Karbstein}},\
		and\ \bibinfo {author} {\bibfnamefont {R.}~\bibnamefont {Shaisultanov}},\
	}\href@noop {} {\bibfield  {journal} {\bibinfo  {journal} {Phys. Rev. D}\
		}\textbf {\bibinfo {volume} {90}},\ \bibinfo {pages} {033007} (\bibinfo
		{year} {2014})}\BibitemShut {NoStop}%
	\bibitem [{\citenamefont {Sundqvist}\ and\ \citenamefont
		{Karbstein}(2023)}]{Sundqvist_2023}%
	\BibitemOpen
	\bibfield  {author} {\bibinfo {author} {\bibfnamefont {C.}~\bibnamefont
			{Sundqvist}}\ and\ \bibinfo {author} {\bibfnamefont {F.}~\bibnamefont
			{Karbstein}},\ }\href@noop {} {\bibfield  {journal} {\bibinfo  {journal}
			{Phys. Rev. D}\ }\textbf {\bibinfo {volume} {108}},\ \bibinfo {pages}
		{056028} (\bibinfo {year} {2023})}\BibitemShut {NoStop}%
	\bibitem [{\citenamefont {Kryuchkyan}\ and\ \citenamefont
		{Hatsagortsyan}(2011)}]{Kryuchkyan_2011}%
	\BibitemOpen
	\bibfield  {author} {\bibinfo {author} {\bibfnamefont {G.~{\relax Yu}.}\
			\bibnamefont {Kryuchkyan}}\ and\ \bibinfo {author} {\bibfnamefont {K.~Z.}\
			\bibnamefont {Hatsagortsyan}},\ }\href@noop {} {\bibfield  {journal}
		{\bibinfo  {journal} {Phys. Rev. Lett.}\ }\textbf {\bibinfo {volume} {107}},\
		\bibinfo {pages} {053604} (\bibinfo {year} {2011})}\BibitemShut {NoStop}%
	\bibitem [{\citenamefont {Macleod}, \citenamefont {Noble},\ and\ \citenamefont
		{Jaroszynski}(2019)}]{Macleod_2019}%
	\BibitemOpen
	\bibfield  {author} {\bibinfo {author} {\bibfnamefont {A.~J.}\ \bibnamefont
			{Macleod}}, \bibinfo {author} {\bibfnamefont {A.}~\bibnamefont {Noble}},\
		and\ \bibinfo {author} {\bibfnamefont {D.~A.}\ \bibnamefont {Jaroszynski}},\
	}\href@noop {} {\bibfield  {journal} {\bibinfo  {journal} {Phys. Rev. Lett.}\
		}\textbf {\bibinfo {volume} {122}},\ \bibinfo {pages} {161601} (\bibinfo
		{year} {2019})}\BibitemShut {NoStop}%
	\bibitem [{\citenamefont {Bulanov}\ \emph {et~al.}(2019)\citenamefont
		{Bulanov}, \citenamefont {Sasorov}, \citenamefont {Bulanov},\ and\
		\citenamefont {Korn}}]{Bulanov_2019}%
	\BibitemOpen
	\bibfield  {author} {\bibinfo {author} {\bibfnamefont {S.~V.}\ \bibnamefont
			{Bulanov}}, \bibinfo {author} {\bibfnamefont {P.}~\bibnamefont {Sasorov}},
		\bibinfo {author} {\bibfnamefont {S.~S.}\ \bibnamefont {Bulanov}},\ and\
		\bibinfo {author} {\bibfnamefont {G.}~\bibnamefont {Korn}},\ }\href@noop {}
	{\bibfield  {journal} {\bibinfo  {journal} {Phys. Rev. D}\ }\textbf {\bibinfo
			{volume} {100}},\ \bibinfo {pages} {016012} (\bibinfo {year}
		{2019})}\BibitemShut {NoStop}%
	\bibitem [{\citenamefont {Jirka}, \citenamefont {Sasorov},\ and\ \citenamefont
		{Bulanov}(2023)}]{Jirka_2023}%
	\BibitemOpen
	\bibfield  {author} {\bibinfo {author} {\bibfnamefont {M.}~\bibnamefont
			{Jirka}}, \bibinfo {author} {\bibfnamefont {P.}~\bibnamefont {Sasorov}},\
		and\ \bibinfo {author} {\bibfnamefont {S.~V.}\ \bibnamefont {Bulanov}},\
	}\href@noop {} {\bibfield  {journal} {\bibinfo  {journal} {Phys. Rev. A}\
		}\textbf {\bibinfo {volume} {107}},\ \bibinfo {pages} {052805} (\bibinfo
		{year} {2023})}\BibitemShut {NoStop}%
	\bibitem [{\citenamefont {Di~Piazza}, \citenamefont {Hatsagortsyan},\ and\
		\citenamefont {Keitel}(2007)}]{Di_Piazza_2007_a}%
	\BibitemOpen
	\bibfield  {author} {\bibinfo {author} {\bibfnamefont {A.}~\bibnamefont
			{Di~Piazza}}, \bibinfo {author} {\bibfnamefont {K.~Z.}\ \bibnamefont
			{Hatsagortsyan}},\ and\ \bibinfo {author} {\bibfnamefont {C.~H.}\
			\bibnamefont {Keitel}},\ }\href@noop {} {\bibfield  {journal} {\bibinfo
			{journal} {Phys. Plasmas}\ }\textbf {\bibinfo {volume} {14}},\ \bibinfo
		{pages} {032102} (\bibinfo {year} {2007})}\BibitemShut {NoStop}%
	\bibitem [{\citenamefont {Pegoraro}(2019)}]{Pegoraro_2019}%
	\BibitemOpen
	\bibfield  {author} {\bibinfo {author} {\bibfnamefont {F.}~\bibnamefont
			{Pegoraro}},\ }\href@noop {} {\bibfield  {journal} {\bibinfo  {journal}
			{Rend. Fis. Acc. Lincei}\ }\textbf {\bibinfo {volume} {30}} (\bibinfo {year}
		{2019})},\ \bibinfo {note} {11}\BibitemShut {NoStop}%
	\bibitem [{\citenamefont {Zhang}\ \emph {et~al.}(2020)\citenamefont {Zhang},
		\citenamefont {Bulanov}, \citenamefont {Seipt}, \citenamefont {Arefiev},\
		and\ \citenamefont {Thomas}}]{Zhang_2020}%
	\BibitemOpen
	\bibfield  {author} {\bibinfo {author} {\bibfnamefont {P.}~\bibnamefont
			{Zhang}}, \bibinfo {author} {\bibfnamefont {S.~S.}\ \bibnamefont {Bulanov}},
		\bibinfo {author} {\bibfnamefont {D.}~\bibnamefont {Seipt}}, \bibinfo
		{author} {\bibfnamefont {A.~V.}\ \bibnamefont {Arefiev}},\ and\ \bibinfo
		{author} {\bibfnamefont {A.~G.~R.}\ \bibnamefont {Thomas}},\ }\href@noop {}
	{\bibfield  {journal} {\bibinfo  {journal} {Phys. Plasmas}\ }\textbf
		{\bibinfo {volume} {27}} (\bibinfo {year} {2020})},\ \bibinfo {note}
	{050601}\BibitemShut {NoStop}%
	\bibitem [{\citenamefont {Bret}(2021)}]{Bret_2021}%
	\BibitemOpen
	\bibfield  {author} {\bibinfo {author} {\bibfnamefont {A.}~\bibnamefont
			{Bret}},\ }\href@noop {} {\bibfield  {journal} {\bibinfo  {journal}
			{Europhys. Lett.}\ }\textbf {\bibinfo {volume} {135}},\ \bibinfo {pages}
		{35001} (\bibinfo {year} {2021})}\BibitemShut {NoStop}%
	\bibitem [{\citenamefont {Varfolomeev}(1966)}]{Varfolomeev_1966}%
	\BibitemOpen
	\bibfield  {author} {\bibinfo {author} {\bibfnamefont {A.~A.}\ \bibnamefont
			{Varfolomeev}},\ }\href@noop {} {\bibfield  {journal} {\bibinfo  {journal}
			{Sov. Phys. JETP}\ }\textbf {\bibinfo {volume} {23}},\ \bibinfo {pages} {681}
		(\bibinfo {year} {1966})}\BibitemShut {NoStop}%
	\bibitem [{\citenamefont {Lundstr\"om}\ \emph {et~al.}(2006)\citenamefont
		{Lundstr\"om}, \citenamefont {Brodin}, \citenamefont {Lundin}, \citenamefont
		{Marklund}, \citenamefont {Bingham}, \citenamefont {Collier}, \citenamefont
		{Mendon\c{c}a},\ and\ \citenamefont {Norreys}}]{Lundstroem_2006}%
	\BibitemOpen
	\bibfield  {author} {\bibinfo {author} {\bibfnamefont {E.}~\bibnamefont
			{Lundstr\"om}}, \bibinfo {author} {\bibfnamefont {G.}~\bibnamefont {Brodin}},
		\bibinfo {author} {\bibfnamefont {J.}~\bibnamefont {Lundin}}, \bibinfo
		{author} {\bibfnamefont {M.}~\bibnamefont {Marklund}}, \bibinfo {author}
		{\bibfnamefont {R.}~\bibnamefont {Bingham}}, \bibinfo {author} {\bibfnamefont
			{J.}~\bibnamefont {Collier}}, \bibinfo {author} {\bibfnamefont {J.~T.}\
			\bibnamefont {Mendon\c{c}a}},\ and\ \bibinfo {author} {\bibfnamefont
			{P.}~\bibnamefont {Norreys}},\ }\href@noop {} {\bibfield  {journal} {\bibinfo
			{journal} {Phys. Rev. Lett.}\ }\textbf {\bibinfo {volume} {96}},\ \bibinfo
		{pages} {083602} (\bibinfo {year} {2006})}\BibitemShut {NoStop}%
	\bibitem [{\citenamefont {Lundin}\ \emph {et~al.}(2007)\citenamefont {Lundin},
		\citenamefont {Stenflo}, \citenamefont {Brodin}, \citenamefont {Marklund},\
		and\ \citenamefont {Shukla}}]{Lundin_2007}%
	\BibitemOpen
	\bibfield  {author} {\bibinfo {author} {\bibfnamefont {J.}~\bibnamefont
			{Lundin}}, \bibinfo {author} {\bibfnamefont {L.}~\bibnamefont {Stenflo}},
		\bibinfo {author} {\bibfnamefont {G.}~\bibnamefont {Brodin}}, \bibinfo
		{author} {\bibfnamefont {M.}~\bibnamefont {Marklund}},\ and\ \bibinfo
		{author} {\bibfnamefont {P.~K.}\ \bibnamefont {Shukla}},\ }\href@noop {}
	{\bibfield  {journal} {\bibinfo  {journal} {Phys. Plasmas}\ }\textbf
		{\bibinfo {volume} {14}},\ \bibinfo {pages} {064503} (\bibinfo {year}
		{2007})}\BibitemShut {NoStop}%
	\bibitem [{\citenamefont {King}, \citenamefont {Di~Piazza},\ and\ \citenamefont
		{Keitel}(2010{\natexlab{b}})}]{King_2010}%
	\BibitemOpen
	\bibfield  {author} {\bibinfo {author} {\bibfnamefont {B.}~\bibnamefont
			{King}}, \bibinfo {author} {\bibfnamefont {A.}~\bibnamefont {Di~Piazza}},\
		and\ \bibinfo {author} {\bibfnamefont {C.~H.}\ \bibnamefont {Keitel}},\
	}\href@noop {} {\bibfield  {journal} {\bibinfo  {journal} {Nature Photon.}\
		}\textbf {\bibinfo {volume} {4}},\ \bibinfo {pages} {92} (\bibinfo {year}
		{2010}{\natexlab{b}})}\BibitemShut {NoStop}%
	\bibitem [{\citenamefont {Jeong}\ \emph {et~al.}(2020)\citenamefont {Jeong},
		\citenamefont {Bulanov}, \citenamefont {Sasorov}, \citenamefont {Korn},
		\citenamefont {Koga},\ and\ \citenamefont {Bulanov}}]{Jeong_2020}%
	\BibitemOpen
	\bibfield  {author} {\bibinfo {author} {\bibfnamefont {T.~M.}\ \bibnamefont
			{Jeong}}, \bibinfo {author} {\bibfnamefont {S.~V.}\ \bibnamefont {Bulanov}},
		\bibinfo {author} {\bibfnamefont {P.~V.}\ \bibnamefont {Sasorov}}, \bibinfo
		{author} {\bibfnamefont {G.}~\bibnamefont {Korn}}, \bibinfo {author}
		{\bibfnamefont {J.~K.}\ \bibnamefont {Koga}},\ and\ \bibinfo {author}
		{\bibfnamefont {S.~S.}\ \bibnamefont {Bulanov}},\ }\href@noop {} {\bibfield
		{journal} {\bibinfo  {journal} {Phys. Rev. A}\ }\textbf {\bibinfo {volume}
			{102}},\ \bibinfo {pages} {023504} (\bibinfo {year} {2020})}\BibitemShut
	{NoStop}%
	\bibitem [{\citenamefont {Karbstein}\ \emph {et~al.}(2019)\citenamefont
		{Karbstein}, \citenamefont {Blinne}, \citenamefont {Gies},\ and\
		\citenamefont {Zepf}}]{Karbstein_2021_c}%
	\BibitemOpen
	\bibfield  {author} {\bibinfo {author} {\bibfnamefont {F.}~\bibnamefont
			{Karbstein}}, \bibinfo {author} {\bibfnamefont {A.}~\bibnamefont {Blinne}},
		\bibinfo {author} {\bibfnamefont {H.}~\bibnamefont {Gies}},\ and\ \bibinfo
		{author} {\bibfnamefont {M.}~\bibnamefont {Zepf}},\ }\href@noop {} {\bibfield
		{journal} {\bibinfo  {journal} {Phys. Rev. Lett.}\ }\textbf {\bibinfo
			{volume} {123}},\ \bibinfo {pages} {091802} (\bibinfo {year}
		{2019})}\BibitemShut {NoStop}%
	\bibitem [{\citenamefont {Dumlu}, \citenamefont {Nakamiya},\ and\ \citenamefont
		{Tanaka}(2022)}]{Dumlu_2022}%
	\BibitemOpen
	\bibfield  {author} {\bibinfo {author} {\bibfnamefont {C.~K.}\ \bibnamefont
			{Dumlu}}, \bibinfo {author} {\bibfnamefont {Y.}~\bibnamefont {Nakamiya}},\
		and\ \bibinfo {author} {\bibfnamefont {K.~A.}\ \bibnamefont {Tanaka}},\
	}\href@noop {} {\bibfield  {journal} {\bibinfo  {journal} {Phys. Rev. D}\
		}\textbf {\bibinfo {volume} {106}},\ \bibinfo {pages} {116001} (\bibinfo
		{year} {2022})}\BibitemShut {NoStop}%
	\bibitem [{\citenamefont {Berezin}\ and\ \citenamefont
		{Fedotov}(2024)}]{Berezin_2024}%
	\BibitemOpen
	\bibfield  {author} {\bibinfo {author} {\bibfnamefont {A.~V.}\ \bibnamefont
			{Berezin}}\ and\ \bibinfo {author} {\bibfnamefont {A.~M.}\ \bibnamefont
			{Fedotov}},\ }\href@noop {} {\bibfield  {journal} {\bibinfo  {journal} {Phys.
				Rev. D}\ }\textbf {\bibinfo {volume} {110}},\ \bibinfo {pages} {016009}
		(\bibinfo {year} {2024})}\BibitemShut {NoStop}%
	\bibitem [{\citenamefont {Di~Piazza}\ \emph {et~al.}(2012)\citenamefont
		{Di~Piazza}, \citenamefont {M\"{u}ller}, \citenamefont {Hatsagortsyan},\ and\
		\citenamefont {Keitel}}]{Di_Piazza_2012}%
	\BibitemOpen
	\bibfield  {author} {\bibinfo {author} {\bibfnamefont {A.}~\bibnamefont
			{Di~Piazza}}, \bibinfo {author} {\bibfnamefont {C.}~\bibnamefont
			{M\"{u}ller}}, \bibinfo {author} {\bibfnamefont {K.~Z.}\ \bibnamefont
			{Hatsagortsyan}},\ and\ \bibinfo {author} {\bibfnamefont {C.~H.}\
			\bibnamefont {Keitel}},\ }\href@noop {} {\bibfield  {journal} {\bibinfo
			{journal} {Rev. Mod. Phys.}\ }\textbf {\bibinfo {volume} {84}},\ \bibinfo
		{pages} {1177} (\bibinfo {year} {2012})}\BibitemShut {NoStop}%
	\bibitem [{\citenamefont {Dunne}(2014)}]{Dunne_2014}%
	\BibitemOpen
	\bibfield  {author} {\bibinfo {author} {\bibfnamefont {G.~V.}\ \bibnamefont
			{Dunne}},\ }\href@noop {} {\bibfield  {journal} {\bibinfo  {journal} {Eur.
				Phys. J. Special Topics}\ }\textbf {\bibinfo {volume} {223}},\ \bibinfo
		{pages} {1055} (\bibinfo {year} {2014})}\BibitemShut {NoStop}%
	\bibitem [{\citenamefont {King}\ and\ \citenamefont
		{Elkina}(2016)}]{King_2016_c}%
	\BibitemOpen
	\bibfield  {author} {\bibinfo {author} {\bibfnamefont {B.}~\bibnamefont
			{King}}\ and\ \bibinfo {author} {\bibfnamefont {N.}~\bibnamefont {Elkina}},\
	}\href@noop {} {\bibfield  {journal} {\bibinfo  {journal} {Phys. Rev. A}\
		}\textbf {\bibinfo {volume} {94}},\ \bibinfo {pages} {062102} (\bibinfo
		{year} {2016})}\BibitemShut {NoStop}%
	\bibitem [{\citenamefont {Karbstein}(2020)}]{Karbstein_2020}%
	\BibitemOpen
	\bibfield  {author} {\bibinfo {author} {\bibfnamefont {F.}~\bibnamefont
			{Karbstein}},\ }\bibfield  {title} {\enquote {\bibinfo {title} {Probing
				vacuum polarization effects with high-intensity lasers},}\ }\href@noop {}
	{\bibfield  {journal} {\bibinfo  {journal} {Particles}\ }\textbf {\bibinfo
			{volume} {3}},\ \bibinfo {pages} {39} (\bibinfo {year} {2020})}\BibitemShut
	{NoStop}%
	\bibitem [{\citenamefont {Gonoskov}\ \emph {et~al.}(2022)\citenamefont
		{Gonoskov}, \citenamefont {Blackburn}, \citenamefont {Marklund},\ and\
		\citenamefont {Bulanov}}]{Gonoskov_2022}%
	\BibitemOpen
	\bibfield  {author} {\bibinfo {author} {\bibfnamefont {A.}~\bibnamefont
			{Gonoskov}}, \bibinfo {author} {\bibfnamefont {T.~G.}\ \bibnamefont
			{Blackburn}}, \bibinfo {author} {\bibfnamefont {M.}~\bibnamefont
			{Marklund}},\ and\ \bibinfo {author} {\bibfnamefont {S.~S.}\ \bibnamefont
			{Bulanov}},\ }\href@noop {} {\bibfield  {journal} {\bibinfo  {journal} {Rev.
				Mod. Phys.}\ }\textbf {\bibinfo {volume} {94}},\ \bibinfo {pages} {045001}
		(\bibinfo {year} {2022})}\BibitemShut {NoStop}%
	\bibitem [{\citenamefont {Fedotov}\ \emph {et~al.}(2023)\citenamefont
		{Fedotov}, \citenamefont {Ilderton}, \citenamefont {Karbstein}, \citenamefont
		{King}, \citenamefont {Seipt}, \citenamefont {Taya},\ and\ \citenamefont
		{Torgrimsson}}]{Fedotov_2023}%
	\BibitemOpen
	\bibfield  {author} {\bibinfo {author} {\bibfnamefont {A.}~\bibnamefont
			{Fedotov}}, \bibinfo {author} {\bibfnamefont {A.}~\bibnamefont {Ilderton}},
		\bibinfo {author} {\bibfnamefont {F.}~\bibnamefont {Karbstein}}, \bibinfo
		{author} {\bibfnamefont {B.}~\bibnamefont {King}}, \bibinfo {author}
		{\bibfnamefont {D.}~\bibnamefont {Seipt}}, \bibinfo {author} {\bibfnamefont
			{H.}~\bibnamefont {Taya}},\ and\ \bibinfo {author} {\bibfnamefont
			{G.}~\bibnamefont {Torgrimsson}},\ }\href@noop {} {\bibfield  {journal}
		{\bibinfo  {journal} {Phys. Rep.}\ }\textbf {\bibinfo {volume} {1010}},\
		\bibinfo {pages} {1} (\bibinfo {year} {2023})}\BibitemShut {NoStop}%
	\bibitem [{\citenamefont {Mignani}\ \emph {et~al.}(2017)\citenamefont
		{Mignani}, \citenamefont {Testa}, \citenamefont {González~Caniulef},
		\citenamefont {Taverna}, \citenamefont {Turolla}, \citenamefont {Zane},\ and\
		\citenamefont {Wu}}]{MNRAS:Mignani:2017}%
	\BibitemOpen
	\bibfield  {author} {\bibinfo {author} {\bibfnamefont {R.~P.}\ \bibnamefont
			{Mignani}}, \bibinfo {author} {\bibfnamefont {V.}~\bibnamefont {Testa}},
		\bibinfo {author} {\bibfnamefont {D.}~\bibnamefont {González~Caniulef}},
		\bibinfo {author} {\bibfnamefont {R.}~\bibnamefont {Taverna}}, \bibinfo
		{author} {\bibfnamefont {R.}~\bibnamefont {Turolla}}, \bibinfo {author}
		{\bibfnamefont {S.}~\bibnamefont {Zane}},\ and\ \bibinfo {author}
		{\bibfnamefont {K.}~\bibnamefont {Wu}},\ }\bibfield  {title} {\enquote
		{\bibinfo {title} {Evidence for vacuum birefringence from the first
				optical-polarimetry measurement of the isolated neutron star {RX}
				{J1856}.5-3754},}\ }\href {https://doi.org/10.1093/mnras/stw2798} {\bibfield
		{journal} {\bibinfo  {journal} {Monthly Notices of the Royal Astronomical
				Society}\ }\textbf {\bibinfo {volume} {465}},\ \bibinfo {pages} {492}
		(\bibinfo {year} {2017})}\BibitemShut {NoStop}%
	\bibitem [{\citenamefont {Capparelli}\ \emph {et~al.}(2017)\citenamefont
		{Capparelli}, \citenamefont {Damiano}, \citenamefont {Maiani},\ and\
		\citenamefont {Polosa}}]{EPJ:Capparelli:2017}%
	\BibitemOpen
	\bibfield  {author} {\bibinfo {author} {\bibfnamefont {L.~M.}\ \bibnamefont
			{Capparelli}}, \bibinfo {author} {\bibfnamefont {A.}~\bibnamefont {Damiano}},
		\bibinfo {author} {\bibfnamefont {L.}~\bibnamefont {Maiani}},\ and\ \bibinfo
		{author} {\bibfnamefont {A.~D.}\ \bibnamefont {Polosa}},\ }\bibfield  {title}
	{\enquote {\bibinfo {title} {A note on polarized light from magnetars},}\
	}\href {https://doi.org/10.1140/epjc/s10052-017-5342-3} {\bibfield  {journal}
		{\bibinfo  {journal} {The European Physical Journal C}\ }\textbf {\bibinfo
			{volume} {77}},\ \bibinfo {pages} {754} (\bibinfo {year} {2017})}\BibitemShut
	{NoStop}%
	\bibitem [{\citenamefont {Jarlskog}\ \emph {et~al.}(1973)\citenamefont
		{Jarlskog}, \citenamefont {Jönsson}, \citenamefont {Prünster},
		\citenamefont {Schulz}, \citenamefont {Willutzki},\ and\ \citenamefont
		{Winter}}]{PRD:Jarlskog:1973}%
	\BibitemOpen
	\bibfield  {author} {\bibinfo {author} {\bibfnamefont {G.}~\bibnamefont
			{Jarlskog}}, \bibinfo {author} {\bibfnamefont {L.}~\bibnamefont {Jönsson}},
		\bibinfo {author} {\bibfnamefont {S.}~\bibnamefont {Prünster}}, \bibinfo
		{author} {\bibfnamefont {H.~D.}\ \bibnamefont {Schulz}}, \bibinfo {author}
		{\bibfnamefont {H.~J.}\ \bibnamefont {Willutzki}},\ and\ \bibinfo {author}
		{\bibfnamefont {G.~G.}\ \bibnamefont {Winter}},\ }\bibfield  {title}
	{\enquote {\bibinfo {title} {Measurement of {Delbr}{\textbackslash}"uck
				{Scattering} and {Observation} of {Photon} {Splitting} at {High}
				{Energies}},}\ }\href {https://doi.org/10.1103/PhysRevD.8.3813} {\bibfield
		{journal} {\bibinfo  {journal} {Physical Review D}\ }\textbf {\bibinfo
			{volume} {8}},\ \bibinfo {pages} {3813--3823} (\bibinfo {year}
		{1973})}\BibitemShut {NoStop}%
	\bibitem [{\citenamefont {Akhmadaliev}\ \emph {et~al.}(1998)\citenamefont
		{Akhmadaliev} \emph {et~al.}}]{Akhmadaliev:1998zz}%
	\BibitemOpen
	\bibfield  {author} {\bibinfo {author} {\bibfnamefont {S.~Z.}\ \bibnamefont
			{Akhmadaliev}}, \bibinfo {author} {\bibfnamefont {G.~Y.}\ \bibnamefont
			{Kezerashvili}}, \bibinfo {author} {\bibfnamefont {S.~G.}\ \bibnamefont
			{Klimenko}}, \bibinfo {author} {\bibfnamefont {V.~M.}\ \bibnamefont
			{Malyshev}}, \bibinfo {author} {\bibfnamefont {A.~L.}\ \bibnamefont
			{Maslennikov}}, \bibinfo {author} {\bibfnamefont {A.~M.}\ \bibnamefont
			{Milov}}, \bibinfo {author} {\bibfnamefont {A.~I.}\ \bibnamefont {Milstein}},
		\bibinfo {author} {\bibfnamefont {N.~Y.}\ \bibnamefont {Muchnoi}}, \bibinfo
		{author} {\bibfnamefont {A.~I.}\ \bibnamefont {Naumenkov}}, \bibinfo {author}
		{\bibfnamefont {V.~S.}\ \bibnamefont {Panin}}, \emph {et~al.},\ }\bibfield {title} {\enquote {\bibinfo
			{title} {{Delbruck scattering at energies of 140-450 MeV}},}\ }\href
	{https://doi.org/10.1103/PhysRevC.58.2844} {\bibfield  {journal} {\bibinfo
			{journal} {Phys. Rev. C}\ }\textbf {\bibinfo {volume} {58}},\ \bibinfo
		{pages} {2844--2850} (\bibinfo {year} {1998})}\BibitemShut {NoStop}%
	\bibitem [{\citenamefont {Aaboud}\ \emph {et~al.}(2017)\citenamefont {Aaboud}
		\emph {et~al.}}]{NP:Aaboud:2017}%
	\BibitemOpen
	\bibfield  {author} {\bibinfo {author} {\bibfnamefont {M.}~\bibnamefont
			{Aaboud}}, \bibinfo {author} {\bibfnamefont {G.}~\bibnamefont {Aad}},
		\bibinfo {author} {\bibfnamefont {B.}~\bibnamefont {Abbott}}, \bibinfo
		{author} {\bibfnamefont {J.}~\bibnamefont {Abdallah}}, \bibinfo {author}
		{\bibfnamefont {O.}~\bibnamefont {Abdinov}}, \bibinfo {author} {\bibfnamefont
			{B.}~\bibnamefont {Abeloos}}, \bibinfo {author} {\bibfnamefont {S.~H.}\
			\bibnamefont {Abidi}}, \bibinfo {author} {\bibfnamefont {O.~S.}\ \bibnamefont
			{AbouZeid}}, \bibinfo {author} {\bibfnamefont {N.~L.}\ \bibnamefont
			{Abraham}}, \bibinfo {author} {\bibfnamefont {H.}~\bibnamefont {Abramowicz}},
		\emph {et~al.},\ }\bibfield  {title} {\enquote {\bibinfo {title}
			{Evidence for light-by-light scattering in heavy-ion collisions with the
				{ATLAS} detector at the {LHC}},}\ }\href {https://doi.org/10.1038/nphys4208}
	{\bibfield  {journal} {\bibinfo  {journal} {Nature Physics}\ }\textbf
		{\bibinfo {volume} {13}},\ \bibinfo {pages} {852--858} (\bibinfo {year}
		{2017})}\BibitemShut {NoStop}%
	\bibitem [{\citenamefont {Sirunyan}\ \emph {et~al.}(2019)\citenamefont
		{Sirunyan} \emph {et~al.}}]{CMS:2018erd}%
	\BibitemOpen
	\bibfield  {author} {\bibinfo {author} {\bibfnamefont {A.~M.}\ \bibnamefont
			{Sirunyan}}, \bibinfo {author} {\bibfnamefont {A.}~\bibnamefont {Tumasyan}},
		\bibinfo {author} {\bibfnamefont {W.}~\bibnamefont {Adam}}, \bibinfo {author}
		{\bibfnamefont {F.}~\bibnamefont {Ambrogi}}, \bibinfo {author} {\bibfnamefont
			{E.}~\bibnamefont {Asilar}}, \bibinfo {author} {\bibfnamefont
			{T.}~\bibnamefont {Bergauer}}, \bibinfo {author} {\bibfnamefont
			{J.}~\bibnamefont {Brandstetter}}, \bibinfo {author} {\bibfnamefont
			{M.}~\bibnamefont {Dragicevic}}, \bibinfo {author} {\bibfnamefont
			{J.}~\bibnamefont {Erö}}, \bibinfo {author} {\bibfnamefont {A.}~\bibnamefont
			{Escalante Del~Valle}}, \emph {et~al.},\ }\bibfield
	{title} {\enquote {\bibinfo {title} {{Evidence for light-by-light scattering
					and searches for axion-like particles in ultraperipheral PbPb collisions at
					$\sqrt{s_\mathrm{NN}} =$ 5.02 TeV}},}\ }\href
	{https://doi.org/10.1016/j.physletb.2019.134826} {\bibfield  {journal}
		{\bibinfo  {journal} {Phys. Lett. B}\ }\textbf {\bibinfo {volume} {797}},\
		\bibinfo {pages} {134826} (\bibinfo {year} {2019})},\ \Eprint
	{https://arxiv.org/abs/1810.04602} {arXiv:1810.04602 [hep-ex]} \BibitemShut
	{NoStop}%
	\bibitem [{\citenamefont {Aad}\ \emph {et~al.}(2019)\citenamefont {Aad} \emph
		{et~al.}}]{ATLAS:2019azn}%
	\BibitemOpen
	\bibfield  {author} {\bibinfo {author} {\bibnamefont {{ATLAS
					Collaboration}}}, \bibinfo {author} {\bibfnamefont {G.}~\bibnamefont {Aad}},
		\bibinfo {author} {\bibfnamefont {B.}~\bibnamefont {Abbott}}, \bibinfo
		{author} {\bibfnamefont {D.}~\bibnamefont {Abbott}}, \bibinfo {author}
		{\bibfnamefont {O.}~\bibnamefont {Abdinov}}, \bibinfo {author} {\bibfnamefont
			{A.}~\bibnamefont {Abed~Abud}}, \bibinfo {author} {\bibfnamefont
			{K.}~\bibnamefont {Abeling}}, \bibinfo {author} {\bibfnamefont
			{D.}~\bibnamefont {Abhayasinghe}}, \bibinfo {author} {\bibfnamefont
			{S.}~\bibnamefont {Abidi}}, \bibinfo {author} {\bibfnamefont
			{O.}~\bibnamefont {AbouZeid}}, \emph {et~al.},\ }\bibfield
	{title} {\enquote {\bibinfo {title} {{Observation of light-by-light
					scattering in ultraperipheral Pb+Pb collisions with the ATLAS detector}},}\
	}\href {https://doi.org/10.1103/PhysRevLett.123.052001} {\bibfield  {journal}
		{\bibinfo  {journal} {Phys. Rev. Lett.}\ }\textbf {\bibinfo {volume} {123}},\
		\bibinfo {pages} {052001} (\bibinfo {year} {2019})},\ \Eprint
	{https://arxiv.org/abs/1904.03536} {arXiv:1904.03536 [hep-ex]} \BibitemShut
	{NoStop}%
	\bibitem [{\citenamefont {Berestetskii}, \citenamefont {Lifshitz},\ and\
		\citenamefont {Pitaevskii}(1982)}]{Landau_b_4_1982}%
	\BibitemOpen
	\bibfield  {author} {\bibinfo {author} {\bibfnamefont {V.~B.}\ \bibnamefont
			{Berestetskii}}, \bibinfo {author} {\bibfnamefont {E.~M.}\ \bibnamefont
			{Lifshitz}},\ and\ \bibinfo {author} {\bibfnamefont {L.~P.}\ \bibnamefont
			{Pitaevskii}},\ }\href@noop {} {\emph {\bibinfo {title} {Quantum
				Electrodynamics}}}\ (\bibinfo  {publisher} {Elsevier Butterworth-Heinemann,
		Oxford},\ \bibinfo {year} {1982})\BibitemShut {NoStop}%
	\bibitem [{\citenamefont {Radier}\ \emph {et~al.}(2022)\citenamefont {Radier},
		\citenamefont {Chalus}, \citenamefont {Charbonneau}, \citenamefont
		{Thambirajah}, \citenamefont {Deschamps}, \citenamefont {David},
		\citenamefont {Barbe}, \citenamefont {Etter}, \citenamefont {Matras},
		\citenamefont {Ricaud}, \citenamefont {Leroux}, \citenamefont {Richard},
		\citenamefont {Lureau}, \citenamefont {Baleanu}, \citenamefont {Banici},
		\citenamefont {Gradinariu}, \citenamefont {Caldararu}, \citenamefont
		{Capiteanu}, \citenamefont {Naziru}, \citenamefont {Diaconescu},
		\citenamefont {Iancu}, \citenamefont {Dabu}, \citenamefont {Ursescu},
		\citenamefont {Dancus}, \citenamefont {Ur}, \citenamefont {Tanaka},\ and\
		\citenamefont {Zamfir}}]{HPLSE:Radier:2022}%
	\BibitemOpen
	\bibfield  {author} {\bibinfo {author} {\bibfnamefont {C.}~\bibnamefont
			{Radier}}, \bibinfo {author} {\bibfnamefont {O.}~\bibnamefont {Chalus}},
		\bibinfo {author} {\bibfnamefont {M.}~\bibnamefont {Charbonneau}}, \bibinfo
		{author} {\bibfnamefont {S.}~\bibnamefont {Thambirajah}}, \bibinfo {author}
		{\bibfnamefont {G.}~\bibnamefont {Deschamps}}, \bibinfo {author}
		{\bibfnamefont {S.}~\bibnamefont {David}}, \bibinfo {author} {\bibfnamefont
			{J.}~\bibnamefont {Barbe}}, \bibinfo {author} {\bibfnamefont
			{E.}~\bibnamefont {Etter}}, \bibinfo {author} {\bibfnamefont
			{G.}~\bibnamefont {Matras}}, \bibinfo {author} {\bibfnamefont
			{S.}~\bibnamefont {Ricaud}}, \bibinfo {author} {\bibfnamefont
			{V.}~\bibnamefont {Leroux}}, \bibinfo {author} {\bibfnamefont
			{C.}~\bibnamefont {Richard}}, \bibinfo {author} {\bibfnamefont
			{F.}~\bibnamefont {Lureau}}, \bibinfo {author} {\bibfnamefont
			{A.}~\bibnamefont {Baleanu}}, \bibinfo {author} {\bibfnamefont
			{R.}~\bibnamefont {Banici}}, \bibinfo {author} {\bibfnamefont
			{A.}~\bibnamefont {Gradinariu}}, \bibinfo {author} {\bibfnamefont
			{C.}~\bibnamefont {Caldararu}}, \bibinfo {author} {\bibfnamefont
			{C.}~\bibnamefont {Capiteanu}}, \bibinfo {author} {\bibfnamefont
			{A.}~\bibnamefont {Naziru}}, \bibinfo {author} {\bibfnamefont
			{B.}~\bibnamefont {Diaconescu}}, \bibinfo {author} {\bibfnamefont
			{V.}~\bibnamefont {Iancu}}, \bibinfo {author} {\bibfnamefont
			{R.}~\bibnamefont {Dabu}}, \bibinfo {author} {\bibfnamefont {D.}~\bibnamefont
			{Ursescu}}, \bibinfo {author} {\bibfnamefont {I.}~\bibnamefont {Dancus}},
		\bibinfo {author} {\bibfnamefont {C.~A.}\ \bibnamefont {Ur}}, \bibinfo
		{author} {\bibfnamefont {K.~A.}\ \bibnamefont {Tanaka}},\ and\ \bibinfo
		{author} {\bibfnamefont {N.~V.}\ \bibnamefont {Zamfir}},\ }\bibfield  {title}
	{\enquote {\bibinfo {title} {10 {PW} peak power femtosecond laser pulses at
				{ELI}-{NP}},}\ }\href {https://doi.org/10.1017/hpl.2022.11} {\bibfield
		{journal} {\bibinfo  {journal} {High Power Laser Science and Engineering}\
		}\textbf {\bibinfo {volume} {10}},\ \bibinfo {pages} {e21} (\bibinfo {year}
		{2022})}\BibitemShut {NoStop}%
	\bibitem [{\citenamefont {King}\ and\ \citenamefont
		{Keitel}(2012)}]{NJP:King:2012}%
	\BibitemOpen
	\bibfield  {author} {\bibinfo {author} {\bibfnamefont {B.}~\bibnamefont
			{King}}\ and\ \bibinfo {author} {\bibfnamefont {C.~H.}\ \bibnamefont
			{Keitel}},\ }\bibfield  {title} {\enquote {\bibinfo {title} {Photon-photon
				scattering in collisions of intense laser pulses},}\ }\href
	{https://doi.org/10.1088/1367-2630/14/10/103002} {\bibfield  {journal}
		{\bibinfo  {journal} {New Journal of Physics}\ }\textbf {\bibinfo {volume}
			{14}},\ \bibinfo {pages} {103002} (\bibinfo {year} {2012})}\BibitemShut
	{NoStop}%
	\bibitem [{\citenamefont {Bernard}\ \emph {et~al.}(2000)\citenamefont
		{Bernard}, \citenamefont {Moulin}, \citenamefont {Amiranoff}, \citenamefont
		{Braun}, \citenamefont {Chambaret}, \citenamefont {Darpentigny},
		\citenamefont {Grillon}, \citenamefont {Ranc},\ and\ \citenamefont
		{Perrone}}]{EPJD:Bernard:2000}%
	\BibitemOpen
	\bibfield  {author} {\bibinfo {author} {\bibfnamefont {D.}~\bibnamefont
			{Bernard}}, \bibinfo {author} {\bibfnamefont {F.}~\bibnamefont {Moulin}},
		\bibinfo {author} {\bibfnamefont {F.}~\bibnamefont {Amiranoff}}, \bibinfo
		{author} {\bibfnamefont {A.}~\bibnamefont {Braun}}, \bibinfo {author}
		{\bibfnamefont {J.}~\bibnamefont {Chambaret}}, \bibinfo {author}
		{\bibfnamefont {G.}~\bibnamefont {Darpentigny}}, \bibinfo {author}
		{\bibfnamefont {G.}~\bibnamefont {Grillon}}, \bibinfo {author} {\bibfnamefont
			{S.}~\bibnamefont {Ranc}},\ and\ \bibinfo {author} {\bibfnamefont
			{F.}~\bibnamefont {Perrone}},\ }\bibfield  {title} {\enquote {\bibinfo
			{title} {Search for stimulated photon-photon scattering in vacuum},}\ }\href
	{https://doi.org/10.1007/s100530050535} {\bibfield  {journal} {\bibinfo
			{journal} {The European Physical Journal D - Atomic, Molecular, Optical and
				Plasma Physics}\ }\textbf {\bibinfo {volume} {10}},\ \bibinfo {pages} {141}
		(\bibinfo {year} {2000})}\BibitemShut {NoStop}%
	\bibitem [{\citenamefont {Heinzl}, \citenamefont {King},\ and\ \citenamefont
		{Liu}(2024{\natexlab{a}})}]{Heinzl:2024cia}%
	\BibitemOpen
	\bibfield  {author} {\bibinfo {author} {\bibfnamefont {T.}~\bibnamefont
			{Heinzl}}, \bibinfo {author} {\bibfnamefont {B.}~\bibnamefont {King}},\ and\
		\bibinfo {author} {\bibfnamefont {D.}~\bibnamefont {Liu}},\ }\bibfield
	{title} {\enquote {\bibinfo {title} {{Coherent enhancement of QED
					cross-sections in electromagnetic backgrounds}},}\ }\href@noop {} {\
		(\bibinfo {year} {2024}{\natexlab{a}})},\ \Eprint
	{https://arxiv.org/abs/2412.10574} {arXiv:2412.10574 [hep-ph]} \BibitemShut
	{NoStop}%
	\bibitem [{\citenamefont {Lundström}\ \emph {et~al.}(2006)\citenamefont
		{Lundström}, \citenamefont {Brodin}, \citenamefont {Lundin}, \citenamefont
		{Marklund}, \citenamefont {Bingham}, \citenamefont {Collier}, \citenamefont
		{Mendonça},\ and\ \citenamefont {Norreys}}]{PRL:Lundstrom:2006}%
	\BibitemOpen
	\bibfield  {author} {\bibinfo {author} {\bibfnamefont {E.}~\bibnamefont
			{Lundström}}, \bibinfo {author} {\bibfnamefont {G.}~\bibnamefont {Brodin}},
		\bibinfo {author} {\bibfnamefont {J.}~\bibnamefont {Lundin}}, \bibinfo
		{author} {\bibfnamefont {M.}~\bibnamefont {Marklund}}, \bibinfo {author}
		{\bibfnamefont {R.}~\bibnamefont {Bingham}}, \bibinfo {author} {\bibfnamefont
			{J.}~\bibnamefont {Collier}}, \bibinfo {author} {\bibfnamefont {J.~T.}\
			\bibnamefont {Mendonça}},\ and\ \bibinfo {author} {\bibfnamefont
			{P.}~\bibnamefont {Norreys}},\ }\bibfield  {title} {\enquote {\bibinfo
			{title} {Using {High}-{Power} {Lasers} for {Detection} of {Elastic}
				{Photon}-{Photon} {Scattering}},}\ }\href
	{https://doi.org/10.1103/PhysRevLett.96.083602} {\bibfield  {journal}
		{\bibinfo  {journal} {Physical Review Letters}\ }\textbf {\bibinfo {volume}
			{96}},\ \bibinfo {pages} {083602} (\bibinfo {year} {2006})},\ \bibinfo {note}
	{publisher: American Physical Society}\BibitemShut {NoStop}%
	\bibitem [{\citenamefont {Lundin}\ \emph {et~al.}(2006)\citenamefont {Lundin},
		\citenamefont {Marklund}, \citenamefont {Lundström}, \citenamefont {Brodin},
		\citenamefont {Collier}, \citenamefont {Bingham}, \citenamefont {Mendonça},\
		and\ \citenamefont {Norreys}}]{PRA:Lundin:2006}%
	\BibitemOpen
	\bibfield  {author} {\bibinfo {author} {\bibfnamefont {J.}~\bibnamefont
			{Lundin}}, \bibinfo {author} {\bibfnamefont {M.}~\bibnamefont {Marklund}},
		\bibinfo {author} {\bibfnamefont {E.}~\bibnamefont {Lundström}}, \bibinfo
		{author} {\bibfnamefont {G.}~\bibnamefont {Brodin}}, \bibinfo {author}
		{\bibfnamefont {J.}~\bibnamefont {Collier}}, \bibinfo {author} {\bibfnamefont
			{R.}~\bibnamefont {Bingham}}, \bibinfo {author} {\bibfnamefont {J.~T.}\
			\bibnamefont {Mendonça}},\ and\ \bibinfo {author} {\bibfnamefont
			{P.}~\bibnamefont {Norreys}},\ }\bibfield  {title} {\enquote {\bibinfo
			{title} {Analysis of four-wave mixing of high-power lasers for the detection
				of elastic photon-photon scattering},}\ }\href
	{https://doi.org/10.1103/PhysRevA.74.043821} {\bibfield  {journal} {\bibinfo
			{journal} {Physical Review A}\ }\textbf {\bibinfo {volume} {74}},\ \bibinfo
		{pages} {043821} (\bibinfo {year} {2006})}\BibitemShut {NoStop}%
	\bibitem [{\citenamefont {King}, \citenamefont {Hu},\ and\ \citenamefont
		{Shen}(2018{\natexlab{b}})}]{PRA:King:2018}%
	\BibitemOpen
	\bibfield  {author} {\bibinfo {author} {\bibfnamefont {B.}~\bibnamefont
			{King}}, \bibinfo {author} {\bibfnamefont {H.}~\bibnamefont {Hu}},\ and\
		\bibinfo {author} {\bibfnamefont {B.}~\bibnamefont {Shen}},\ }\bibfield
	{title} {\enquote {\bibinfo {title} {Three-pulse photon-photon scattering},}\
	}\href {https://doi.org/10.1103/PhysRevA.98.023817} {\bibfield  {journal}
		{\bibinfo  {journal} {Physical Review A}\ }\textbf {\bibinfo {volume} {98}},\
		\bibinfo {pages} {023817} (\bibinfo {year} {2018}{\natexlab{b}})}\BibitemShut
	{NoStop}%
	\bibitem [{\citenamefont {Gies}\ \emph {et~al.}(2018)\citenamefont {Gies},
		\citenamefont {Karbstein}, \citenamefont {Kohlf\"urst},\ and\ \citenamefont
		{Seegert}}]{PRD:Gies:2018b}%
	\BibitemOpen
	\bibfield  {author} {\bibinfo {author} {\bibfnamefont {H.}~\bibnamefont
			{Gies}}, \bibinfo {author} {\bibfnamefont {F.}~\bibnamefont {Karbstein}},
		\bibinfo {author} {\bibfnamefont {C.}~\bibnamefont {Kohlf\"urst}},\ and\
		\bibinfo {author} {\bibfnamefont {N.}~\bibnamefont {Seegert}},\ }\bibfield
	{title} {\enquote {\bibinfo {title} {Photon-photon scattering at the
				high-intensity frontier},}\ }\href
	{https://doi.org/10.1103/PhysRevD.97.076002} {\bibfield  {journal} {\bibinfo
			{journal} {Phys. Rev. D}\ }\textbf {\bibinfo {volume} {97}},\ \bibinfo
		{pages} {076002} (\bibinfo {year} {2018})}\BibitemShut {NoStop}%
	\bibitem [{\citenamefont {Euler}\ and\ \citenamefont
		{Kockel}(1935)}]{Euler_1935}%
	\BibitemOpen
	\bibfield  {author} {\bibinfo {author} {\bibfnamefont {H.}~\bibnamefont
			{Euler}}\ and\ \bibinfo {author} {\bibfnamefont {B.}~\bibnamefont {Kockel}},\
	}\bibfield  {title} {\enquote {\bibinfo {title} {Über die {Streuung} von
				{Licht} an {Licht} nach der {Diracschen} {Theorie}},}\ }\href
	{https://doi.org/10.1007/BF01493898} {\bibfield  {journal} {\bibinfo
			{journal} {Naturwissenschaften}\ }\textbf {\bibinfo {volume} {23}},\ \bibinfo
		{pages} {246} (\bibinfo {year} {1935})}\BibitemShut {NoStop}%
	\bibitem [{\citenamefont {Mckenna}\ and\ \citenamefont
		{Platzman}(1963)}]{PR:McKenna:1963}%
	\BibitemOpen
	\bibfield  {author} {\bibinfo {author} {\bibfnamefont {J.}~\bibnamefont
			{Mckenna}}\ and\ \bibinfo {author} {\bibfnamefont {P.~M.}\ \bibnamefont
			{Platzman}},\ }\bibfield  {title} {\enquote {\bibinfo {title} {Nonlinear
				{Interaction} of {Light} in a {Vacuum}},}\ }\href
	{https://doi.org/10.1103/PhysRev.129.2354} {\bibfield  {journal} {\bibinfo
			{journal} {Physical Review}\ }\textbf {\bibinfo {volume} {129}},\ \bibinfo
		{pages} {2354} (\bibinfo {year} {1963})}\BibitemShut {NoStop}%
	\bibitem [{\citenamefont {Inc.}(2022)}]{vendor:Mathworks:MATLAB}%
	\BibitemOpen
	\bibfield  {author} {\bibinfo {author} {\bibfnamefont {T.~M.}\ \bibnamefont
			{Inc.}},\ }\href {https://www.mathworks.com} {\enquote {\bibinfo {title}
			{Matlab version: 9.13.0 (r2022b)},}\ } (\bibinfo {year} {2022})\BibitemShut
	{NoStop}%
	\bibitem [{\citenamefont {Parker}\ \emph {et~al.}(2018)\citenamefont {Parker},
		\citenamefont {Danson}, \citenamefont {Egan}, \citenamefont {Elsmere},
		\citenamefont {Girling}, \citenamefont {Harvey}, \citenamefont {Hillier},
		\citenamefont {Hussey}, \citenamefont {Masoero}, \citenamefont {McLoughlin},
		\citenamefont {Penman}, \citenamefont {Treadwell}, \citenamefont {Winter},\
		and\ \citenamefont {Hopps}}]{HPLSE:Parker:2018}%
	\BibitemOpen
	\bibfield  {author} {\bibinfo {author} {\bibfnamefont {S.}~\bibnamefont
			{Parker}}, \bibinfo {author} {\bibfnamefont {C.}~\bibnamefont {Danson}},
		\bibinfo {author} {\bibfnamefont {D.}~\bibnamefont {Egan}}, \bibinfo {author}
		{\bibfnamefont {S.}~\bibnamefont {Elsmere}}, \bibinfo {author} {\bibfnamefont
			{M.}~\bibnamefont {Girling}}, \bibinfo {author} {\bibfnamefont
			{E.}~\bibnamefont {Harvey}}, \bibinfo {author} {\bibfnamefont
			{D.}~\bibnamefont {Hillier}}, \bibinfo {author} {\bibfnamefont
			{D.}~\bibnamefont {Hussey}}, \bibinfo {author} {\bibfnamefont
			{S.}~\bibnamefont {Masoero}}, \bibinfo {author} {\bibfnamefont
			{J.}~\bibnamefont {McLoughlin}}, \bibinfo {author} {\bibfnamefont
			{R.}~\bibnamefont {Penman}}, \bibinfo {author} {\bibfnamefont
			{P.}~\bibnamefont {Treadwell}}, \bibinfo {author} {\bibfnamefont
			{D.}~\bibnamefont {Winter}},\ and\ \bibinfo {author} {\bibfnamefont
			{N.}~\bibnamefont {Hopps}},\ }\bibfield  {title} {\enquote {\bibinfo {title}
			{400 {TW} operation of {Orion} at ultra-high contrast},}\ }\href
	{https://doi.org/10.1017/hpl.2018.44} {\bibfield  {journal} {\bibinfo
			{journal} {High Power Laser Science and Engineering}\ }\textbf {\bibinfo
			{volume} {6}},\ \bibinfo {pages} {e47} (\bibinfo {year} {2018})}\BibitemShut
	{NoStop}%
	\bibitem [{\citenamefont {Doyle}\ \emph {et~al.}(2022)\citenamefont {Doyle},
		\citenamefont {Khademi}, \citenamefont {Hilz}, \citenamefont {Sävert},
		\citenamefont {Schäfer}, \citenamefont {Schreiber},\ and\ \citenamefont
		{Zepf}}]{NJP:Doyle:2022}%
	\BibitemOpen
	\bibfield  {author} {\bibinfo {author} {\bibfnamefont {L.}~\bibnamefont
			{Doyle}}, \bibinfo {author} {\bibfnamefont {P.}~\bibnamefont {Khademi}},
		\bibinfo {author} {\bibfnamefont {P.}~\bibnamefont {Hilz}}, \bibinfo {author}
		{\bibfnamefont {A.}~\bibnamefont {Sävert}}, \bibinfo {author} {\bibfnamefont
			{G.}~\bibnamefont {Schäfer}}, \bibinfo {author} {\bibfnamefont
			{J.}~\bibnamefont {Schreiber}},\ and\ \bibinfo {author} {\bibfnamefont
			{M.}~\bibnamefont {Zepf}},\ }\bibfield  {title} {\enquote {\bibinfo {title}
			{Experimental estimates of the photon background in a potential
				light-by-light scattering study},}\ }\href
	{https://doi.org/10.1088/1367-2630/ac4ad3} {\bibfield  {journal} {\bibinfo
			{journal} {New Journal of Physics}\ }\textbf {\bibinfo {volume} {24}},\
		\bibinfo {pages} {025003} (\bibinfo {year} {2022})}\BibitemShut {NoStop}%
	\bibitem [{\citenamefont {MacLeod}(2007)}]{MS:Macleod:2007}%
	\BibitemOpen
	\bibfield  {author} {\bibinfo {author} {\bibfnamefont {A.}~\bibnamefont
			{MacLeod}},\ }\emph {\bibinfo {title} {Measuring the gain of a
			photomultiplier tube}},\ \href@noop {} {Master's thesis},\ \bibinfo  {school}
	{McGill University} (\bibinfo {year} {2007})\BibitemShut {NoStop}%
	\bibitem [{\citenamefont {Gol’tsman}\ \emph {et~al.}(2001)\citenamefont
		{Gol’tsman}, \citenamefont {Okunev}, \citenamefont {Chulkova},
		\citenamefont {Lipatov}, \citenamefont {Semenov}, \citenamefont {Smirnov},
		\citenamefont {Voronov}, \citenamefont {Dzardanov}, \citenamefont
		{Williams},\ and\ \citenamefont {Sobolewski}}]{APL:Goltsman:2001}%
	\BibitemOpen
	\bibfield  {author} {\bibinfo {author} {\bibfnamefont {G.~N.}\ \bibnamefont
			{Gol’tsman}}, \bibinfo {author} {\bibfnamefont {O.}~\bibnamefont {Okunev}},
		\bibinfo {author} {\bibfnamefont {G.}~\bibnamefont {Chulkova}}, \bibinfo
		{author} {\bibfnamefont {A.}~\bibnamefont {Lipatov}}, \bibinfo {author}
		{\bibfnamefont {A.}~\bibnamefont {Semenov}}, \bibinfo {author} {\bibfnamefont
			{K.}~\bibnamefont {Smirnov}}, \bibinfo {author} {\bibfnamefont
			{B.}~\bibnamefont {Voronov}}, \bibinfo {author} {\bibfnamefont
			{A.}~\bibnamefont {Dzardanov}}, \bibinfo {author} {\bibfnamefont
			{C.}~\bibnamefont {Williams}},\ and\ \bibinfo {author} {\bibfnamefont
			{R.}~\bibnamefont {Sobolewski}},\ }\bibfield  {title} {\enquote {\bibinfo
			{title} {{Picosecond superconducting single-photon optical detector}},}\
	}\href {https://doi.org/10.1063/1.1388868} {\bibfield  {journal} {\bibinfo
			{journal} {Applied Physics Letters}\ }\textbf {\bibinfo {volume} {79}},\
		\bibinfo {pages} {705} (\bibinfo {year} {2001})}\BibitemShut {NoStop}%
	\bibitem [{\citenamefont {Cova}\ \emph {et~al.}(1996)\citenamefont {Cova},
		\citenamefont {Ghioni}, \citenamefont {Lacaita}, \citenamefont {Samori},\
		and\ \citenamefont {Zappa}}]{Optica:Cova:1996}%
	\BibitemOpen
	\bibfield  {author} {\bibinfo {author} {\bibfnamefont {S.}~\bibnamefont
			{Cova}}, \bibinfo {author} {\bibfnamefont {M.}~\bibnamefont {Ghioni}},
		\bibinfo {author} {\bibfnamefont {A.}~\bibnamefont {Lacaita}}, \bibinfo
		{author} {\bibfnamefont {C.}~\bibnamefont {Samori}},\ and\ \bibinfo {author}
		{\bibfnamefont {F.}~\bibnamefont {Zappa}},\ }\bibfield  {title} {\enquote
		{\bibinfo {title} {Avalanche photodiodes and quenching circuits for
				single-photon detection},}\ }\href {https://doi.org/10.1364/AO.35.001956}
	{\bibfield  {journal} {\bibinfo  {journal} {Appl. Opt.}\ }\textbf {\bibinfo
			{volume} {35}},\ \bibinfo {pages} {1956--1976} (\bibinfo {year}
		{1996})}\BibitemShut {NoStop}%
	\bibitem [{\citenamefont {Quantum}(2024)}]{SingleQuantum:SNSPD}%
	\BibitemOpen
	\bibfield  {author} {\bibinfo {author} {\bibfnamefont {S.}~\bibnamefont
			{Quantum}},\ }\href {https://www.singlequantum.com/technology/snspd/}
	{\enquote {\bibinfo {title} {{Superconducing Nanowire Single Photon
					Detectors}},}\ }\bibinfo {howpublished} {see
		\url{https://www.singlequantum.com/technology/snspd/}} (\bibinfo {year}
	{2024})\BibitemShut {NoStop}%
	\bibitem [{\citenamefont {Zhou}\ \emph {et~al.}(2018)\citenamefont {Zhou},
		\citenamefont {Han}, \citenamefont {Lv}, \citenamefont {Li}, \citenamefont
		{Lu}, \citenamefont {Wang}, \citenamefont {Song}, \citenamefont {Tan},
		\citenamefont {Liang}, \citenamefont {Feng},\ and\ \citenamefont
		{Cai}}]{IEEE:Zhou:2018}%
	\BibitemOpen
	\bibfield  {author} {\bibinfo {author} {\bibfnamefont {X.}~\bibnamefont
			{Zhou}}, \bibinfo {author} {\bibfnamefont {T.}~\bibnamefont {Han}}, \bibinfo
		{author} {\bibfnamefont {Y.}~\bibnamefont {Lv}}, \bibinfo {author}
		{\bibfnamefont {J.}~\bibnamefont {Li}}, \bibinfo {author} {\bibfnamefont
			{W.}~\bibnamefont {Lu}}, \bibinfo {author} {\bibfnamefont {Y.}~\bibnamefont
			{Wang}}, \bibinfo {author} {\bibfnamefont {X.}~\bibnamefont {Song}}, \bibinfo
		{author} {\bibfnamefont {X.}~\bibnamefont {Tan}}, \bibinfo {author}
		{\bibfnamefont {S.}~\bibnamefont {Liang}}, \bibinfo {author} {\bibfnamefont
			{Z.}~\bibnamefont {Feng}},\ and\ \bibinfo {author} {\bibfnamefont
			{S.}~\bibnamefont {Cai}},\ }\bibfield  {title} {\enquote {\bibinfo {title}
			{Large-{Area} {4H}-{SiC} {Ultraviolet} {Avalanche} {Photodiodes} {Based} on
				{Variable}-{Temperature} {Reflow} {Technique}},}\ }\href
	{https://doi.org/10.1109/LED.2018.2871798} {\bibfield  {journal} {\bibinfo
			{journal} {IEEE Electron Device Letters}\ }\textbf {\bibinfo {volume} {39}},\
		\bibinfo {pages} {1724--1727} (\bibinfo {year} {2018})},\ \bibinfo {note}
	{conference Name: IEEE Electron Device Letters}\BibitemShut {NoStop}%
	\bibitem [{\citenamefont {Karbstein}\ \emph
		{et~al.}(2022{\natexlab{b}})\citenamefont {Karbstein}, \citenamefont
		{Ullmann}, \citenamefont {Mosman},\ and\ \citenamefont
		{Zepf}}]{Karbstein:2022uwf}%
	\BibitemOpen
	\bibfield  {author} {\bibinfo {author} {\bibfnamefont {F.}~\bibnamefont
			{Karbstein}}, \bibinfo {author} {\bibfnamefont {D.}~\bibnamefont {Ullmann}},
		\bibinfo {author} {\bibfnamefont {E.~A.}\ \bibnamefont {Mosman}},\ and\
		\bibinfo {author} {\bibfnamefont {M.}~\bibnamefont {Zepf}},\ }\bibfield
	{title} {\enquote {\bibinfo {title} {{Direct Accessibility of the Fundamental
					Constants Governing Light-by-Light Scattering}},}\ }\href
	{https://doi.org/10.1103/PhysRevLett.129.061802} {\bibfield  {journal}
		{\bibinfo  {journal} {Phys. Rev. Lett.}\ }\textbf {\bibinfo {volume} {129}},\
		\bibinfo {pages} {061802} (\bibinfo {year} {2022}{\natexlab{b}})},\ \Eprint
	{https://arxiv.org/abs/2207.09866} {arXiv:2207.09866 [hep-ph]} \BibitemShut
	{NoStop}%
	\bibitem [{\citenamefont {MacLeod}\ and\ \citenamefont
		{King}(2024)}]{PRA:Macleod:2024}%
	\BibitemOpen
	\bibfield  {author} {\bibinfo {author} {\bibfnamefont {A.~J.}\ \bibnamefont
			{MacLeod}}\ and\ \bibinfo {author} {\bibfnamefont {B.}~\bibnamefont {King}},\
	}\bibfield  {title} {\enquote {\bibinfo {title} {Fundamental constants from
				photon-photon scattering in three-beam collisions},}\ }\href
	{https://doi.org/10.1103/PhysRevA.110.032216} {\bibfield  {journal} {\bibinfo
			{journal} {Physical Review A}\ }\textbf {\bibinfo {volume} {110}},\ \bibinfo
		{pages} {032216} (\bibinfo {year} {2024})}\BibitemShut {NoStop}%
	\bibitem [{\citenamefont {Schütze}\ \emph {et~al.}(2024)\citenamefont
		{Schütze}, \citenamefont {Doyle}, \citenamefont {Schreiber}, \citenamefont
		{Zepf},\ and\ \citenamefont {Karbstein}}]{PRD:Schutze:2024}%
	\BibitemOpen
	\bibfield  {author} {\bibinfo {author} {\bibfnamefont {F.}~\bibnamefont
			{Schütze}}, \bibinfo {author} {\bibfnamefont {L.}~\bibnamefont {Doyle}},
		\bibinfo {author} {\bibfnamefont {J.}~\bibnamefont {Schreiber}}, \bibinfo
		{author} {\bibfnamefont {M.}~\bibnamefont {Zepf}},\ and\ \bibinfo {author}
		{\bibfnamefont {F.}~\bibnamefont {Karbstein}},\ }\bibfield  {title} {\enquote
		{\bibinfo {title} {Dark-field setup for the measurement of light-by-light
				scattering with high-intensity lasers},}\ }\href
	{https://doi.org/10.1103/PhysRevD.109.096009} {\bibfield  {journal} {\bibinfo
			{journal} {Physical Review D}\ }\textbf {\bibinfo {volume} {109}},\ \bibinfo
		{pages} {096009} (\bibinfo {year} {2024})}\BibitemShut {NoStop}%
	\bibitem [{Orc(2024)}]{OrcaQUEST2}%
	\BibitemOpen
	\href@noop {} {\emph {\bibinfo {title} {ORCA-Quest2 qCMOS camera}}},\
	\bibinfo {organization} {Hamamatsu} (\bibinfo {year} {2024})\BibitemShut
	{NoStop}%
	\bibitem [{\citenamefont {Janesick}\ \emph {et~al.}(1990)\citenamefont
		{Janesick}, \citenamefont {Elliott}, \citenamefont {Dingiziam}, \citenamefont
		{Bredthauer}, \citenamefont {Chandler}, \citenamefont {Westphal},\ and\
		\citenamefont {Gunn}}]{SPIE:Janesick:1990}%
	\BibitemOpen
	\bibfield  {author} {\bibinfo {author} {\bibfnamefont {J.~R.}\ \bibnamefont
			{Janesick}}, \bibinfo {author} {\bibfnamefont {T.~S.}\ \bibnamefont
			{Elliott}}, \bibinfo {author} {\bibfnamefont {A.}~\bibnamefont {Dingiziam}},
		\bibinfo {author} {\bibfnamefont {R.~A.}\ \bibnamefont {Bredthauer}},
		\bibinfo {author} {\bibfnamefont {C.~E.}\ \bibnamefont {Chandler}}, \bibinfo
		{author} {\bibfnamefont {J.~A.}\ \bibnamefont {Westphal}},\ and\ \bibinfo
		{author} {\bibfnamefont {J.~E.}\ \bibnamefont {Gunn}},\ }\bibfield  {title}
	{\enquote {\bibinfo {title} {New advancements in charge-coupled device
				technology: subelectron noise and 4096 x 4096 pixel {CCDs}},}\ }in\ \href
	{https://doi.org/10.1117/12.19452} {\emph {\bibinfo {booktitle}
			{Charge-{Coupled} {Devices} and {Solid} {State} {Optical} {Sensors}}}},\
	Vol.\ \bibinfo {volume} {1242}\ (\bibinfo  {publisher} {SPIE},\ \bibinfo
	{year} {1990})\ pp.\ \bibinfo {pages} {223--237}\BibitemShut {NoStop}%
	\bibitem [{\citenamefont {Tiffenberg}\ \emph {et~al.}(2017)\citenamefont
		{Tiffenberg}, \citenamefont {Sofo-Haro}, \citenamefont {Drlica-Wagner},
		\citenamefont {Essig}, \citenamefont {Guardincerri}, \citenamefont {Holland},
		\citenamefont {Volansky},\ and\ \citenamefont {Yu}}]{PRL:Tiffenberg:2017}%
	\BibitemOpen
	\bibfield  {author} {\bibinfo {author} {\bibfnamefont {J.}~\bibnamefont
			{Tiffenberg}}, \bibinfo {author} {\bibfnamefont {M.}~\bibnamefont
			{Sofo-Haro}}, \bibinfo {author} {\bibfnamefont {A.}~\bibnamefont
			{Drlica-Wagner}}, \bibinfo {author} {\bibfnamefont {R.}~\bibnamefont
			{Essig}}, \bibinfo {author} {\bibfnamefont {Y.}~\bibnamefont {Guardincerri}},
		\bibinfo {author} {\bibfnamefont {S.}~\bibnamefont {Holland}}, \bibinfo
		{author} {\bibfnamefont {T.}~\bibnamefont {Volansky}},\ and\ \bibinfo
		{author} {\bibfnamefont {T.-T.}\ \bibnamefont {Yu}},\ }\bibfield  {title}
	{\enquote {\bibinfo {title} {Single-{Electron} and {Single}-{Photon}
				{Sensitivity} with a {Silicon} {Skipper} {CCD}},}\ }\href
	{https://doi.org/10.1103/PhysRevLett.119.131802} {\bibfield  {journal}
		{\bibinfo  {journal} {Physical Review Letters}\ }\textbf {\bibinfo {volume}
			{119}},\ \bibinfo {pages} {131802} (\bibinfo {year} {2017})}\BibitemShut
	{NoStop}%
	\bibitem [{\citenamefont {Born}\ and\ \citenamefont
		{Infeld}(1934)}]{Born:1934gh}%
	\BibitemOpen
	\bibfield  {author} {\bibinfo {author} {\bibfnamefont {M.}~\bibnamefont
			{Born}}\ and\ \bibinfo {author} {\bibfnamefont {L.}~\bibnamefont {Infeld}},\
	}\bibfield  {title} {\enquote {\bibinfo {title} {{Foundations of the new
					field theory}},}\ }\href {https://doi.org/10.1098/rspa.1934.0059} {\bibfield
		{journal} {\bibinfo  {journal} {Proc. Roy. Soc. Lond. A}\ }\textbf {\bibinfo
			{volume} {144}},\ \bibinfo {pages} {425--451} (\bibinfo {year}
		{1934})}\BibitemShut {NoStop}%
	\bibitem [{\citenamefont {Fradkin}\ and\ \citenamefont
		{Tseytlin}(1985)}]{FRADKIN1985123}%
	\BibitemOpen
	\bibfield  {author} {\bibinfo {author} {\bibfnamefont {E.}~\bibnamefont
			{Fradkin}}\ and\ \bibinfo {author} {\bibfnamefont {A.}~\bibnamefont
			{Tseytlin}},\ }\bibfield  {title} {\enquote {\bibinfo {title} {Non-linear
				electrodynamics from quantized strings},}\ }\href
	{https://doi.org/https://doi.org/10.1016/0370-2693(85)90205-9} {\bibfield
		{journal} {\bibinfo  {journal} {Physics Letters B}\ }\textbf {\bibinfo
			{volume} {163}},\ \bibinfo {pages} {123--130} (\bibinfo {year}
		{1985})}\BibitemShut {NoStop}%
	\bibitem [{\citenamefont {Ellis}, \citenamefont {Mavromatos},\ and\
		\citenamefont {You}(2017)}]{Ellis:2017edi}%
	\BibitemOpen
	\bibfield  {author} {\bibinfo {author} {\bibfnamefont {J.}~\bibnamefont
			{Ellis}}, \bibinfo {author} {\bibfnamefont {N.~E.}\ \bibnamefont
			{Mavromatos}},\ and\ \bibinfo {author} {\bibfnamefont {T.}~\bibnamefont
			{You}},\ }\bibfield  {title} {\enquote {\bibinfo {title} {{Light-by-Light
					Scattering Constraint on Born-Infeld Theory}},}\ }\href
	{https://doi.org/10.1103/PhysRevLett.118.261802} {\bibfield  {journal}
		{\bibinfo  {journal} {Phys. Rev. Lett.}\ }\textbf {\bibinfo {volume} {118}},\
		\bibinfo {pages} {261802} (\bibinfo {year} {2017})},\ \Eprint
	{https://arxiv.org/abs/1703.08450} {arXiv:1703.08450 [hep-ph]} \BibitemShut
	{NoStop}%
	\bibitem [{\citenamefont {Ellis}\ \emph {et~al.}(2022)\citenamefont {Ellis},
		\citenamefont {Mavromatos}, \citenamefont {Roloff},\ and\ \citenamefont
		{You}}]{Ellis:2022uxv}%
	\BibitemOpen
	\bibfield  {author} {\bibinfo {author} {\bibfnamefont {J.}~\bibnamefont
			{Ellis}}, \bibinfo {author} {\bibfnamefont {N.~E.}\ \bibnamefont
			{Mavromatos}}, \bibinfo {author} {\bibfnamefont {P.}~\bibnamefont {Roloff}},\
		and\ \bibinfo {author} {\bibfnamefont {T.}~\bibnamefont {You}},\ }\bibfield
	{title} {\enquote {\bibinfo {title} {{Light-by-light scattering at future
					$e^+e^-$ colliders}},}\ }\href
	{https://doi.org/10.1140/epjc/s10052-022-10565-w} {\bibfield  {journal}
		{\bibinfo  {journal} {Eur. Phys. J. C}\ }\textbf {\bibinfo {volume} {82}},\
		\bibinfo {pages} {634} (\bibinfo {year} {2022})},\ \Eprint
	{https://arxiv.org/abs/2203.17111} {arXiv:2203.17111 [hep-ph]} \BibitemShut
	{NoStop}%
	\bibitem [{\citenamefont {Fouch\'e}, \citenamefont {Battesti},\ and\
		\citenamefont {Rizzo}(2016)}]{Fouche:2016qqj}%
	\BibitemOpen
	\bibfield  {author} {\bibinfo {author} {\bibfnamefont {M.}~\bibnamefont
			{Fouch\'e}}, \bibinfo {author} {\bibfnamefont {R.}~\bibnamefont {Battesti}},\
		and\ \bibinfo {author} {\bibfnamefont {C.}~\bibnamefont {Rizzo}},\ }\bibfield
	{title} {\enquote {\bibinfo {title} {{Limits on nonlinear
					electrodynamics}},}\ }\href {https://doi.org/10.1103/PhysRevD.93.093020}
	{\bibfield  {journal} {\bibinfo  {journal} {Phys. Rev. D}\ }\textbf {\bibinfo
			{volume} {93}},\ \bibinfo {pages} {093020} (\bibinfo {year} {2016})},\
	\bibinfo {note} {[Erratum: Phys.Rev.D 95, 099902 (2017)]},\ \Eprint
	{https://arxiv.org/abs/1605.04102} {arXiv:1605.04102 [physics.optics]}
	\BibitemShut {NoStop}%
	\bibitem [{\citenamefont {Macleod}\ and\ \citenamefont
		{King}(2024)}]{Macleod:2024jxl}%
	\BibitemOpen
	\bibfield  {author} {\bibinfo {author} {\bibfnamefont {A.~J.}\ \bibnamefont
			{Macleod}}\ and\ \bibinfo {author} {\bibfnamefont {B.}~\bibnamefont {King}},\
	}\bibfield  {title} {\enquote {\bibinfo {title} {{Fundamental constants from
					photon-photon scattering in three-beam collisions}},}\ }\href
	{https://doi.org/10.1103/PhysRevA.110.032216} {\bibfield  {journal} {\bibinfo
			{journal} {Phys. Rev. A}\ }\textbf {\bibinfo {volume} {110}},\ \bibinfo
		{pages} {032216} (\bibinfo {year} {2024})},\ \Eprint
	{https://arxiv.org/abs/2406.10342} {arXiv:2406.10342 [hep-ph]} \BibitemShut
	{NoStop}%
	\bibitem [{\citenamefont {Heinzl}, \citenamefont {King},\ and\ \citenamefont
		{Liu}(2024{\natexlab{b}})}]{king24toappear}%
	\BibitemOpen
	\bibfield  {author} {\bibinfo {author} {\bibfnamefont {T.}~\bibnamefont
			{Heinzl}}, \bibinfo {author} {\bibfnamefont {B.}~\bibnamefont {King}},\ and\
		\bibinfo {author} {\bibfnamefont {D.}~\bibnamefont {Liu}},\ }\href@noop {} {}
	(\bibinfo {year} {2024}{\natexlab{b}})\BibitemShut {NoStop}%
	\bibitem [{\citenamefont {Davila}, \citenamefont {Schubert},\ and\
		\citenamefont {Trejo}(2014)}]{Davila:2013wba}%
	\BibitemOpen
	\bibfield  {author} {\bibinfo {author} {\bibfnamefont {J.~M.}\ \bibnamefont
			{Davila}}, \bibinfo {author} {\bibfnamefont {C.}~\bibnamefont {Schubert}},\
		and\ \bibinfo {author} {\bibfnamefont {M.~A.}\ \bibnamefont {Trejo}},\
	}\bibfield  {title} {\enquote {\bibinfo {title} {{Photonic processes in
					Born-Infeld theory}},}\ }\href {https://doi.org/10.1142/S0217751X14501747}
	{\bibfield  {journal} {\bibinfo  {journal} {Int. J. Mod. Phys. A}\ }\textbf
		{\bibinfo {volume} {29}},\ \bibinfo {pages} {1450174} (\bibinfo {year}
		{2014})},\ \Eprint {https://arxiv.org/abs/1310.8410} {arXiv:1310.8410
		[hep-ph]} \BibitemShut {NoStop}%
	\bibitem [{\citenamefont {Cowan}\ \emph {et~al.}(2011)\citenamefont {Cowan},
		\citenamefont {Cranmer}, \citenamefont {Gross},\ and\ \citenamefont
		{Vitells}}]{Cowan:2010js}%
	\BibitemOpen
	\bibfield  {author} {\bibinfo {author} {\bibfnamefont {G.}~\bibnamefont
			{Cowan}}, \bibinfo {author} {\bibfnamefont {K.}~\bibnamefont {Cranmer}},
		\bibinfo {author} {\bibfnamefont {E.}~\bibnamefont {Gross}},\ and\ \bibinfo
		{author} {\bibfnamefont {O.}~\bibnamefont {Vitells}},\ }\bibfield  {title}
	{\enquote {\bibinfo {title} {{Asymptotic formulae for likelihood-based tests
					of new physics}},}\ }\href {https://doi.org/10.1140/epjc/s10052-011-1554-0}
	{\bibfield  {journal} {\bibinfo  {journal} {Eur. Phys. J. C}\ }\textbf
		{\bibinfo {volume} {71}},\ \bibinfo {pages} {1554} (\bibinfo {year}
		{2011})},\ \bibinfo {note} {[Erratum: Eur.Phys.J.C 73, 2501 (2013)]},\
	\Eprint {https://arxiv.org/abs/1007.1727} {arXiv:1007.1727 [physics.data-an]}
	\BibitemShut {NoStop}%
	\bibitem [{\citenamefont {Blinne}\ \emph {et~al.}(2019)\citenamefont {Blinne},
		\citenamefont {Gies}, \citenamefont {Karbstein}, \citenamefont {Kohlfürst},\
		and\ \citenamefont {Zepf}}]{PRD:Blinne:2019}%
	\BibitemOpen
	\bibfield  {author} {\bibinfo {author} {\bibfnamefont {A.}~\bibnamefont
			{Blinne}}, \bibinfo {author} {\bibfnamefont {H.}~\bibnamefont {Gies}},
		\bibinfo {author} {\bibfnamefont {F.}~\bibnamefont {Karbstein}}, \bibinfo
		{author} {\bibfnamefont {C.}~\bibnamefont {Kohlfürst}},\ and\ \bibinfo
		{author} {\bibfnamefont {M.}~\bibnamefont {Zepf}},\ }\bibfield  {title}
	{\enquote {\bibinfo {title} {All-optical signatures of quantum vacuum
				nonlinearities in generic laser fields},}\ }\href
	{https://doi.org/10.1103/PhysRevD.99.016006} {\bibfield  {journal} {\bibinfo
			{journal} {Physical Review D}\ }\textbf {\bibinfo {volume} {99}},\ \bibinfo
		{pages} {016006} (\bibinfo {year} {2019})}\BibitemShut {NoStop}%
	\bibitem [{\citenamefont {Bromage}\ \emph {et~al.}(2021)\citenamefont
		{Bromage}, \citenamefont {Bahk}, \citenamefont {Bedzyk}, \citenamefont
		{Begishev}, \citenamefont {Bucht}, \citenamefont {Dorrer}, \citenamefont
		{Feng}, \citenamefont {Jeon}, \citenamefont {Mileham}, \citenamefont
		{Roides}, \citenamefont {Shaughnessy}, \citenamefont {Shoup}, \citenamefont
		{Spilatro}, \citenamefont {Webb}, \citenamefont {Weiner},\ and\ \citenamefont
		{Zuegel}}]{HPLSE:Bromage:2021}%
	\BibitemOpen
	\bibfield  {author} {\bibinfo {author} {\bibfnamefont {J.}~\bibnamefont
			{Bromage}}, \bibinfo {author} {\bibfnamefont {S.-W.}\ \bibnamefont {Bahk}},
		\bibinfo {author} {\bibfnamefont {M.}~\bibnamefont {Bedzyk}}, \bibinfo
		{author} {\bibfnamefont {I.~A.}\ \bibnamefont {Begishev}}, \bibinfo {author}
		{\bibfnamefont {S.}~\bibnamefont {Bucht}}, \bibinfo {author} {\bibfnamefont
			{C.}~\bibnamefont {Dorrer}}, \bibinfo {author} {\bibfnamefont
			{C.}~\bibnamefont {Feng}}, \bibinfo {author} {\bibfnamefont {C.}~\bibnamefont
			{Jeon}}, \bibinfo {author} {\bibfnamefont {C.}~\bibnamefont {Mileham}},
		\bibinfo {author} {\bibfnamefont {R.~G.}\ \bibnamefont {Roides}}, \bibinfo
		{author} {\bibfnamefont {K.}~\bibnamefont {Shaughnessy}}, \bibinfo {author}
		{\bibfnamefont {M.~J.}\ \bibnamefont {Shoup}}, \bibinfo {author}
		{\bibfnamefont {M.}~\bibnamefont {Spilatro}}, \bibinfo {author}
		{\bibfnamefont {B.}~\bibnamefont {Webb}}, \bibinfo {author} {\bibfnamefont
			{D.}~\bibnamefont {Weiner}},\ and\ \bibinfo {author} {\bibfnamefont {J.~D.}\
			\bibnamefont {Zuegel}},\ }\bibfield  {title} {\enquote {\bibinfo {title}
			{{MTW}-{OPAL}: a technology development platform for ultra-intense optical
				parametric chirped-pulse amplification systems},}\ }\href
	{https://doi.org/10.1017/hpl.2021.45} {\bibfield  {journal} {\bibinfo
			{journal} {High Power Laser Science and Engineering}\ }\textbf {\bibinfo
			{volume} {9}} (\bibinfo {year} {2021}),\ 10.1017/hpl.2021.45}\BibitemShut
	{NoStop}%
	\bibitem [{\citenamefont {Barut}(1980)}]{Barut_b_1980}%
	\BibitemOpen
	\bibfield  {author} {\bibinfo {author} {\bibfnamefont {A.~O.}\ \bibnamefont
			{Barut}},\ }\href@noop {} {\emph {\bibinfo {title} {Electrodynamics and
				Classical Theory of Fields and Particles}}}\ (\bibinfo  {publisher} {Dover,
		New York},\ \bibinfo {year} {1980})\BibitemShut {NoStop}%
	\bibitem [{\citenamefont {Salamin}(2007)}]{APB:Salamin:2007}%
	\BibitemOpen
	\bibfield  {author} {\bibinfo {author} {\bibfnamefont {Y.}~\bibnamefont
			{Salamin}},\ }\bibfield  {title} {\enquote {\bibinfo {title} {Fields of a
				{Gaussian} beam beyond the paraxial approximation},}\ }\href
	{https://doi.org/10.1007/s00340-006-2442-4} {\bibfield  {journal} {\bibinfo
			{journal} {Applied Physics B}\ }\textbf {\bibinfo {volume} {86}},\ \bibinfo
		{pages} {319} (\bibinfo {year} {2007})}\BibitemShut {NoStop}%
	\bibitem [{\citenamefont {Gusynin}\ and\ \citenamefont
		{Shovkovy}(1999)}]{JMP:Gusynin:1999}%
	\BibitemOpen
	\bibfield  {author} {\bibinfo {author} {\bibfnamefont {V.~P.}\ \bibnamefont
			{Gusynin}}\ and\ \bibinfo {author} {\bibfnamefont {I.~A.}\ \bibnamefont
			{Shovkovy}},\ }\bibfield  {title} {\enquote {\bibinfo {title} {Derivative
				{Expansion} of the {Effective} {Action} for {QED} in 2+1 and 3+1
				dimensions},}\ }\href {https://doi.org/10.1063/1.533037} {\bibfield
		{journal} {\bibinfo  {journal} {Journal of Mathematical Physics}\ }\textbf
		{\bibinfo {volume} {40}},\ \bibinfo {pages} {5406} (\bibinfo {year}
		{1999})}\BibitemShut {NoStop}%
\end{thebibliography}
%

\end{document}